\documentclass[
    reprint,
    preprintnumbers,
    superscriptaddress,
    nofootinbib,
     amsmath,amssymb,
     aps,
     prd,
    floatfix, longbibliography
    ]{revtex4-2}

    \usepackage{graphicx}
    \usepackage[caption=false]{subfig}
    \usepackage[usenames,dvipsnames]{xcolor} 
    \usepackage{pgfplots}
    \usepackage[utf8]{inputenc}
    \usepackage{color}
    \usepackage{hyperref}
    \usepackage[normalem]{ulem} 
    \usepackage{physics}
    \usepackage{enumitem}
    \usepackage{comment}
    \usepackage{bm}
    \usepackage{booktabs}
\hypersetup{
  breaklinks = true,
  colorlinks   = true, 
  urlcolor     = Blue, 
  linkcolor    = Blue, 
  citecolor   = Blue 
}
    \allowdisplaybreaks[1]
    \graphicspath{{figures/}}

\usepackage{float}

\newcommand{\oxford}{Astrophysics, University of Oxford, DWB, Keble Road, Oxford OX1 3RH, United Kingdom}

\newcommand{\splitatcommas}[1]{%
  \begingroup
  \begingroup\lccode`~=`, \lowercase{\endgroup
    \edef~{\mathchar\the\mathcode`, \penalty0 \noexpand\hspace{0pt plus 1em}}%
  }\mathcode`,="8000 #1%
  \endgroup
}

\begin{document}

\title{Cosmological constraints on Galileon dark energy with broken shift symmetry}

\author{William J. Wolf}
\email{william.wolf@stx.ox.ac.uk}
\affiliation{\oxford}
\author{Pedro G. Ferreira}
\email{pedro.ferreira@physics.ox.ac.uk}
\affiliation{\oxford}
\author{Carlos Garc\'ia-Garc\'ia}
\email{carlos.garcia-garcia@physics.ox.ac.uk}
\affiliation{\oxford}

\begin{abstract}
Current cosmological data seem to show that dark energy is evolving in time and that it possibly crossed the phantom divide in the past. So far the only theories that lead to such a behavior involve a non-trivial coupling between dark energy, in the form of a scalar field, and the gravitational or matter sector. We show that there is another possibility involving both a non-trivial kinetic sector in a cubic Galileon theory and a scalar field potential that breaks the Galileon shift symmetry, which can lead to a similar phenomenology on large scales. We perform a full Bayesian analysis using the latest cosmological data, including DESI DR2 BAO measurements, type Ia SNe measurements from DESY5, Union3, and Pantheon+, and CMB data from Planck and ACT. We find that it is statistically strongly favored over a Universe dominated by a cosmological constant (with a Bayes factor of $\log B\simeq 6.5$). Yet, as with other non-minimally coupled theories, it has severe ancillary gravitational effects. These can be mitigated to some extent, but as with other viable theories, the penalty is ever more elaborate scalar field models of dark energy.
\end{abstract}

\maketitle


\section{Introduction}\label{sec:introduction}

The latest constraints on the expansion history of the Universe, from a combination of data such as Baryonic Accoustic Oscillations (BAO), Supernovae (SNe), and the Cosmic Microwave Background (CMB), seem to favour a time evolving dark energy over the standard $\Lambda$-Cold Dark Matter (CDM) model of cosmology where dark energy is driven by the cosmological constant $\Lambda$ \cite{DESI:2025zgx}. 

While these findings are still subject to scrutiny, they now represent a $\simeq 4\sigma$ tension, and have sparked a tremendous amount of interest in the recent physics literature. Topics of interest include assessing various theoretical proposals for dynamical dark energy involving scalar fields and their implications for new physics, such as quintessence \cite{Wolf:2023uno, Wolf:2024eph, Shlivko:2024llw, Tada:2024znt, Payeur:2024dnq, Gialamas:2025pwv, Lodha:2025qbg, Luu:2025fgw, Mishra:2025goj, Bhattacharya:2024hep, Wang:2024hwd, Gomez-Valent:2025mfl, Cline:2025sbt, Lin:2025gne, Goh:2025upc, Hossain:2025grx}, modified gravity \cite{Wolf:2024stt, Wolf:2025jed, Ye:2024ywg, Chudaykin:2024gol, Ye:2025ulq, Pan:2025psn, Goldstein:2025epp, Cai:2025mas, Adam:2025kve, Lu:2025gki, Cataneo:2025vae}, and an interacting dark sector \cite{Khoury:2025txd,  Chakraborty:2025syu, vanderWesthuizen:2025vcb, vanderWesthuizen:2025mnw, Shah:2025ayl, Braglia:2025gdo, Chen:2025ywv, vanderWesthuizen:2025rip, Andriot:2025los, Li:2024qso}. Much attention has also been paid to whether or not the cosmological inference is biased depending on different ways of parameterizing dark energy \cite{Wolf:2025jlc, Shlivko:2025fgv, Giare:2024gpk, Giani:2025hhs, Dinda:2025iaq, Wu:2025wyk, Yang:2025oax, Lee:2025pzo, Li:2025vuh}, model agnostic reconstructions \cite{Dinda:2024ktd, DESI:2024aqx, Jiang:2024xnu, Berti:2025phi, Li:2025ops}, and deconstructing the data and assumptions that have gone into the analysis \cite{Efstathiou_2024, Efstathiou:2025tie, Carloni:2024zpl, DES:2025tir, Cortes:2025joz, Wang:2024dka, Huang:2025som, Gialamas:2024lyw, Dinda:2025svh, Ghosh:2024kyd, Bansal:2025ipo, Chan-GyungPark:2024mlx, Chan-GyungPark:2024brx, Qiang:2025cxp, Chaudhary:2025pcc, Camarena:2025upt, Luongo:2024fww, Luongo:2024zhc, Sousa-Neto:2025gpj, Chaudhary:2025vzy, Afroz:2025iwo, Toomey:2025xyo, Adi:2025hyj, RoyChoudhury:2024wri, RoyChoudhury:2025dhe}.

As has been widely discussed (e.g., \cite{Linder:2025zxb, Wolf:2024stt, Wolf:2025jed, Lodha:2025qbg, Berti:2025phi, Ye:2024ywg, Ozulker:2025ehg, Scherer:2025esj, Gomez-Valent:2025mfl, Keeley:2025rlg, Liu:2025bss}), assuming the cosmological data is correct and fit for analysis, one of the most curious persistent features in the combined data is a preference for a phantom crossing around $z\approx 0.5$, indicating cosmic acceleration that is stronger than expected with $\Lambda$ before this time, but weaker than $\Lambda$ at more recent times. 
In this paper, we build on \cite{Wolf:2024eph, Wolf:2024stt, Wolf:2025jed}, and further explore the consequences of assuming that dark energy is driven by a single scalar field. If we assume dark energy is driven by a single scalar field, it can be described by a very general action \cite{Park:2010cw},
\begin{equation}\label{eq:fullaction}
S=\int d^4 x\sqrt{-g}\left[\frac{M^2_{\rm P}}{2}F(\varphi)R+K(\varphi)X 
-V(\varphi)\right] +\mathcal{S}_{\mathrm{m}},
\end{equation}
where $R$ is the Ricci scalar, $M_{\mathrm{P}}$ is the Planck mass, $X=-\partial_\mu\varphi\partial^\mu\varphi/2$ and $\mathcal{S}_{\mathrm{m}}$ is the action for matter fields. The form of Eq.~\eqref{eq:fullaction} is motivated by effective field theory (EFT), where we are concerned with a single scalar degree of freedom. This scalar degree of freedom can have a kinetic and potential energy, as well as couple to gravity. This is represented by the functions $K(\varphi)$, $V(\varphi)$, and $F(\varphi)$ which are in principle free functions of the scalar field $\varphi$. The EFT philosophy provides a systematic approach where one considers all the operators which are allowed by locality and symmetry, and then constructs a hierarchy of terms which are gradually suppressed by inverse powers of the energy scale. The phenomenology associated with dark energy occurs on the largest scales and in the far infrared. Furthermore, observations suggest that if dark energy is dynamical it has only very recently begun thawing which suggests a small field excursion ($\Delta \varphi \ll M_{\mathrm{P}}$) over this small portion of cosmic history. Thus, the scalar field operators most relevant to dark energy will be the lowest order operators in an EFT expansion.

The simplest, non-trivial possibility (setting $F=K=1$) consists of a minimally coupled, canonical scalar field with a potential -- quintessence \cite{Wolf:2023uno, Wolf:2024eph, Caldwell:1997ii, Tsujikawa:2013fta, Ratra:1987rm, Copeland:2006wr}. Yet, quintessence does not seem to be a good description of the cosmological data \cite{Wolf:2024eph, Wolf:2025jed}. One can generalize dark energy by introducing a non-minimal coupling to gravity ($F\neq 1$, while keeping the kinetic term canonical, $K=1$) to find that this is a much better description of the cosmological data concerning the expansion history. Nevertheless, the non-minimal coupling introduces a host of other problems in gravitational physics \cite{Wolf:2025jed, Wolf:2024stt}. 

In this paper, we continue this systematic exploration (see \cite{Wolf:2023uno, Wolf:2024eph, Wolf:2024stt, Wolf:2025jlc, Wolf:2025jed} for analysis of the previously mentioned cases). Here, we include a potential and revert to minimal coupling between the scalar field and the Ricci curvature ($F=1$), but now allow for a non-minimal kinetic sector. We will show that introducing such non-trivial structures, again, results in a much better statistical description of the cosmological data. Yet, as in the non-minimal case, it leads to ancillary gravitational effects on other scales that must be dealt with.

Sections \ref{sec:DE} and \ref{sec:phenom} situate scalar field dark energy with a modified kinetic term within the class of Galileon gravitational theories. Section \ref{sec:evidence} describes the cosmological data and statistical methodology used for comparing dark energy models and computes the posterior constraints and phenomenological quantities of interest, demonstrating that this dark energy model is a far better statistical description of the cosmological data than $\Lambda$CDM or thawing quintessence and performs similar to the non-minimally coupled model in \cite{Wolf:2025jed} as it similarly allows for phantom crossing behavior.
Section \ref{sec:grav_consequences} explores the gravitational consequences.
Section \ref{sec:conclusion} concludes.

\section{Dynamical dark energy}\label{sec:DE}
In a homogeneous, isotropic Universe which satisfies the Einstein field equations, the scale factor of the Universe, $a(t)$, obeys the Friedmann equations 
\begin{equation}
H^2=\frac{1}{3 M_{\mathrm{P}}^2} \rho_{\mathrm{tot}}, \quad \dot{H}=-\frac{1}{2 M_{\mathrm{P}}^2}\left(\rho_{\mathrm{tot}}+p_{\mathrm{tot}}\right),
\end{equation}
where $H\equiv \dot{a}/a$, ${\dot a}=da/dt$ and $\rho_{\mathrm{tot}}$ and $p_{\mathrm{tot}}$ refer to the energy density and pressure of all the energy species present: radiation, matter, and dark energy. Constraints on the expansion history indicate that the universe is in a period of accelerated expansion. Assuming the Einstein field equations, one then infers that the Universe is dominated by dark energy at late times, whose effects on the dynamics of the Universe are completely described by its equation of state, defined as the ratio
\begin{equation}
w(a) \equiv \frac{p_{\rm DE}}{\rho_{\rm DE}}.
\end{equation}
The simplest candidates for dark energy are the cosmological constant $\Lambda$ or a scalar field $\varphi$. It has become customary, 
however, to assess the expansion history data in a less model dependent way, parameterizing the dark energy equation of state in terms of two parameters, 
\begin{equation}
    w(a)\simeq w_0+w_a(1-a),
\end{equation}
with $w_0$ being the value of the equation of state today and $w_a$ capturing any time variation in the equation of state \cite{Linder:2002et, Chevallier:2000qy}. Current constraints on the equation of state seem to disfavour a cosmological constant $(w_0, w_a)=(-1, 0)$ at $\simeq 4\sigma$ and indicate that the time evolution $w_a$ is large and negative. 

Previously, assuming the Einstein field equations and  Eq.~\eqref{eq:fullaction}, we have considered \cite{Wolf:2023uno, Wolf:2024eph, Wolf:2025jlc}
\begin{align}
    F(\varphi) &= 1, \quad
    K(\varphi) =1, \quad
    V(\varphi) \simeq V_0 + \frac{1}{2}m^2 \varphi^2.
    \label{eq:minimal_functions}
\end{align}
This corresponds to a general model of thawing quintessence minimally coupled to gravity where the coefficient $m^2$ can be positive (``slow-roll'') or negative (``hilltop''). As previously discussed, thawing quintessence is not a good description of the data. When one projects the thawing quintessence priors onto the $(w_0, w_a)$ parameter space (using the same errors as the cosmological data), they are disjoint with the data posteriors; these theories cannot produce significant enough temporal variation in the equation of state (captured by the parameter $w_a$) in the regimes where the data has constraining power. This is further born out in a number of statistical measures where thawing quintessence  only offers a moderate (at best) improvement over $\Lambda$CDM, or no improvement at all, depending on the exact combination of data used in the analysis \cite{Wolf:2024eph, Wolf:2025jlc, Wolf:2025jed}. 

We can move beyond the Einstein field equations and consider a generalized  quintessence model where
\begin{align}\label{eq:nonminimal_functions}
    F(\varphi) &\simeq 1-\xi\varphi^2, \quad
    K(\varphi) =1, \\
    V(\varphi) & \simeq V_0 + \beta\varphi + \frac{1}{2}m^2\varphi^2.
    \nonumber
\end{align}
Introducing the non-minimal coupling $\xi$, modifies the Einstein field equations in such a way that one infers an increase of the energy density of the dark energy field in the past -- i.e.\ a  {\it phantom} equation of state that violates the null energy condition and corresponds to $ w< -1$ -- before it begins to thaw to cross the phantom divide at recent times \cite{Wolf:2024stt, Wolf:2025jed, Ye:2024ywg, Gannouji:2006jm, Boisseau:2000pr, Caldwell:1999ew, Torres:2002pe}. Such dynamical evolution is a far better fit and description of the cosmological data as these predictions overlap with the data posteriors when this theory's priors are projected onto the $(w_0, w_a)$ plane. As this theory allows for dark energy to go phantom, dark energy can both evolve sharply over intermediate redshift regimes where the data has much of its constraining power, without overshooting the inferred value of the equation of state today which the data suggests has not evolved to be too far away from $-1$ (i.e.\ $w_0\simeq-0.75$). Calculating the Bayesian evidence, $\Delta \chi^2$, and other statistical measures then confirm that this model is a significantly better fit to the data that $\Lambda$CDM or thawing quintessence. 

There is a problem, of course: a scalar field with a non-minimal coupling is strongly at odds with local fifth force tests and necessitates introducing a screening mechanism to protect the theory on local scales. Given the undesirable gravitational consequences and the necessity to introduce screening leads us down an ever more complex and \textit{ad hoc} path, it makes sense to continue a systematic exploration of Eq.~\eqref{eq:fullaction} by considering whether there are viable models {\it without} a non-minimal coupling between the scalar field and the Ricci scalar, but which have higher order terms in the kinetic sector.

Theories with a modified kinetic sector have been extensively studied in the past. An appealing aspect of these theories is that it is possible to assume an additional symmetry of the action:  shift symmetry, where the action is invariant under transformations of the form $\varphi \rightarrow \varphi+c+c_\mu x^\mu$,  known as Galileon symmetry \cite{Deffayet:2009wt, Deffayet:2010qz, Nicolis:2008in}. Such theories have a number of attractive, theoretical, properties, but also some downsides. A notable downside is that they lead to effective equations of state which are phantom throughout the evolution of the Universe \cite{Traykova:2021hbr, Deffayet:2010qz} (see Appendix \ref{appendix:shiftsymmetric} for a brief discussion) which is in striking contradiction with current observations. 

In this paper we pursue the strategy of considering the simplest theories with non-minimal kinetic terms (from the point of view of EFT) which may be observationally viable.  We thus consider
\begin{align}
    F(\varphi) &= 1, \quad
    K(\varphi) \simeq \alpha + \frac{\beta}{(\Lambda_3)^3} \square\varphi, \quad
    V(\varphi) \simeq V_0 + \frac{1}{2}m^2 \varphi^2.
    \label{eq:kinetic_functions}
\end{align}
The kinetic sector has been modified to include a general coefficient $\alpha$ for the standard kinetic term $X$ in addition to a coefficient $\beta$ multiplying X$\square \varphi$, making this part of the cubic Galileon/kinetic braiding class of gravitational theories, which have been studied in a variety of cosmological contexts \cite{Deffayet:2010qz, Nicolis:2008in, DeFelice:2010pv, Traykova:2021hbr, Ye:2024kus, Zumalacarregui:2020cjh, Brahma:2020eqd, Barreira:2012kk, Barreira:2014jha, Leloup:2019fas, Renk:2017rzu, Peirone:2019aua, Frusciante:2019puu, DeFelice:2011bh, Bartlett:2020tjd, Barreira:2013jma, Tsujikawa:2025wca, Samaddar:2025xak, Deffayet:2009wt}. $\Lambda_3$ is an energy scale defined as $\Lambda_3=(M_{\mathrm{P}}H_0^2)^{1/3}$. The $X\Box\varphi$ term arises naturally in the EFT framework as it is an allowed operator. If one imposes shift symmetry, this restricts the potential term to be zero or constant. However, scalar field operators in the potential function are of course also naturally present in a general EFT framework.
As we shall see, the presence of a massive potential breaks this shift symmetry and can lead to viable dark energy (see \cite{Tsujikawa:2025wca} for a similar construct). This theory can also be translated into the familiar Horndeski language for scalar-tensor theories (see Appendix \ref{appendix:horndeski} for a brief discussion).

If the kinetic term is going to drive the accelerated expansion, we need $\alpha<0$, as with a standard kinetic term the theory is driven towards a trivial attractor solution \cite{Zumalacarregui:2020cjh}.  We also have a quadratic potential with a massive scalar field where $V_0$ is in units of $M_{\rm P}^2 \tilde{H}^2_0$, $m^2$ is in units of $\tilde{H}^2_0$, both $\alpha$ and $\beta$ are dimensionless, and $M_{\rm P} =1$. In order to ensure consistency in the normalization of this model, we follow \cite{Traykova:2021hbr} and fix $\tilde{H}^2_0$ to a fiducial value $\tilde{H}_0 = 67.5$\,km\,s$^{-1}$Mpc$^{-1}$ in defining the model coefficients.
The equation of motion for this dark energy model is\footnote{See e.g.\ \cite{DeFelice:2011bh, Bellini:2014fua} for the most general equation of motion in Horndeski gravity. One can then specify the Horndeski functions corresponding to the theory under discussion here, which are given by $G_2 = \alpha X - V(\varphi)$, $G_3 = -\frac{1}{\Lambda_3^3} \beta X$, $G_4=\frac{1}{2}M_{\mathrm{P}}$, and $G_5 = 0$, to derive the equation of motion for this dark energy model.} 
\begin{equation}
\alpha\left(\ddot{\varphi}+3 H \dot{\varphi}\right)-\frac{3 \beta}{\Lambda^3_3} \dot{\varphi}\left(3 H^2 \dot{\varphi}+\dot{H} \dot{\varphi}+2 H \ddot{\varphi}\right)=-V^{\prime}(\varphi),
\end{equation}
where we recover the familiar Klein-Gordon equation in a Friedmann background for $\alpha=1$ and $\beta =0$. 
Note that, when the kinetic term $\alpha < 0$ and has the wrong sign, this effectively inverts the sign of the mass term in the potential such that the standard positive mass potential now has a ``negative'' mass.

This model has been implemented in \texttt{hi\_class} \cite{hi_class1, hi_class2}  in order to solve both the background cosmological equations of motion as well as to evolve linear cosmological perturbations and perform stability checks to ensure that ghost and gradient instabilities are avoided. We fix $\beta=-1$ (as this coefficient just affects the normalization of the scalar field \cite{Traykova:2021hbr, Barreira:2013jma}) and we use a shooting method to tune $V_0$ in order to satisfy the Friedmann equations.

\section{Phenomenology of Galileon gravity with a potential}\label{sec:phenom}

Let us first discuss the shift symmetric version of the theory given in Eq.~\eqref{eq:kinetic_functions} without a potential ($V(\varphi) =0$) which is familiar in the literature due to its ability to provide self-accelerating solutions to the Universe without a cosmological constant or potential. In order for this theory to provide an accelerating universe, it requires a ``negative'' kinetic energy, meaning that $\alpha < 0$ and differs from the standard kinetic term in field theories. Of course, this has the potential to create catastrophic ghost and gradient instabilities; however, the theory is stable for certain parameter choices. The ``no ghost'' and ``no gradient'' conditions are
\begin{align}\label{eq:stability1}
 D\equiv\alpha_K + \frac{3}{2} \alpha_B^2>0, \quad
 c_s^2 > 0,
\end{align}
where $\alpha_B$ is the braiding parameter, $\alpha_K$ is the kineticity parameter, and $c^2_s$ is the speed of sound for scalar perturbations \cite{Bellini:2014fua}. These parameters are defined to be
\begin{eqnarray}\label{eq:alphas}
\alpha_B &=& -\frac{2 \dot{\varphi} X}{H M_{\mathrm{P}}^2}\frac{\beta}{(\Lambda_3)^3}, \nonumber \\ 
 \alpha_K &=& \frac{2 X}{H^2 M_{\mathrm{P}}^2} \left( \alpha - 6 \dot{\varphi} H \frac{\beta}{(\Lambda_3)^3} \right), \nonumber \\
 c^2_s&=&- \frac{(2-\alpha_B)({\dot H}-\frac{1}{2}H^2\alpha_B)-H{\dot \alpha}_B+\rho_{\rm M}+P_{\rm M}}{H^2D},\nonumber \\
\end{eqnarray}
In this theory there is no running of the Planck mass and gravitational waves propagate at the speed of light (i.e.\ $\alpha_{\rm M} = \alpha_T = 0$). As one can see from Eqs.~\eqref{eq:stability1} and \eqref{eq:alphas}, when the kinetic term has the wrong sign, the inclusion of the kinetic braiding term is necessary to stabilise the theory by allowing $D>0$.

Without a potential (i.e.\ when the theory has a shift symmetry) there is an attractor solution $\dot{\varphi} H \rightarrow \zeta$, where $\zeta$ is a constant. These solutions lead to a phantom equation $w_{\mathrm{\varphi}} < -1$ of state the approaches a de Sitter state from below \cite{Deffayet:2010qz, Traykova:2021hbr}. Phantom dark energy leads to an expansion rate that is greater than non-phantom dark energy and the cosmological data does provide evidence for a larger inferred $H(z)$ than $\Lambda$CDM can produce in the past. However, as the cosmological data is solidly in the thawing region at more recent times, this indicates that $H(z)$ must transition to be weaker than inferred in $\Lambda$CDM. As these theories cannot cross the phantom divide from below, they are thus observationally ruled out (see Fig.\ \ref{Fig:shiftsym_wedge} in the appendix).

\begin{figure*}[t]
   \centering
   \includegraphics[width=0.48\textwidth]{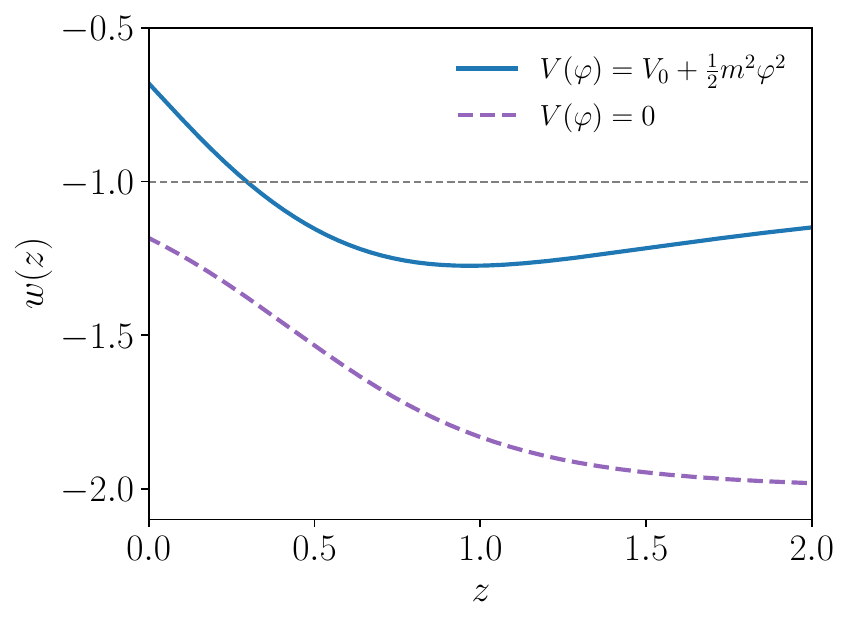}
   \hfill
   \includegraphics[width=0.48\textwidth]{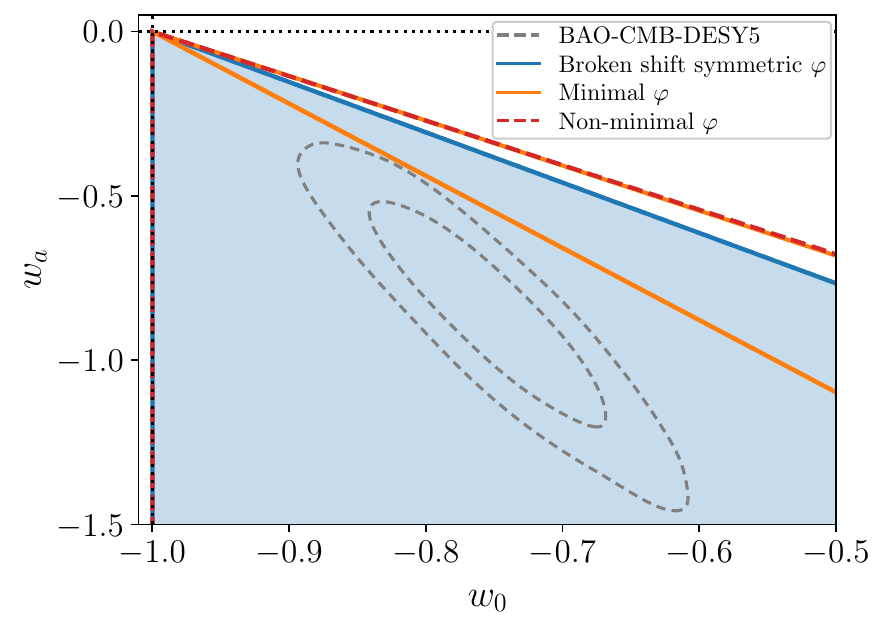}
   \vskip -0.1in
   \caption{(Left) Equation of state evolution $w(z)$ for a cubic Galileon model with and without a potential. (Right) Phase-space wedge structure for various dark energy models where their theory priors are compared to the data posteriors.}
   \label{fig:wz_combined}
\end{figure*}

With a potential $V(\varphi)$, it is possible for these models to evolve from $w_{\varphi} < -1$ to $w_{\varphi} > -1$ at later times, potentially offering a viable model of dark energy. The scalar field energy density and pressure are
\begin{equation}
\begin{aligned}
p_\varphi & =\frac{\alpha}{2}\dot{\varphi}^2 +\frac{\beta}{(\Lambda_3)^3}\dot{\varphi}^2 \ddot{\varphi} -V(\varphi), \\
\rho_\varphi & =\frac{\alpha}{2}\dot{\varphi}^2-\frac{3 \beta}{(\Lambda_3)^3}H \dot{\varphi}^3 + V(\varphi).
\end{aligned}
\end{equation}
To get an intuition for the behavior of $w_{\varphi}$ in this model, first consider the case where $V(\varphi)=0$ and scalar field is on its attractor solution. We have that $\dot{\varphi} = \zeta/H$ and $\ddot{\varphi} = -\zeta \dot{H}/H^2$, where $\dot{H} = -\frac{3}{2} H^2 \left(1 + w_{\rm eff}\right)$ and $w_{\mathrm{eff}}$ is the effective equation of state of the background. This allows us to write $p_\varphi = 1+3(\beta/\alpha) (w_{\mathrm{eff}}+1)$ and $\rho_\varphi = 1-6(\beta/\alpha)$. Solving the Friedman equations for this model and our choice of scale and normalization for the field for a cosmology with $\Omega_{\varphi} \simeq 0.7$ and $H_0 \simeq 67$ km s$^{-1}$ Mpc$^{-1}$ results in $\beta/\alpha \simeq 1/3$, which allows us to estimate that the dark energy equation of state will evolve from $w_\varphi \simeq -2 \rightarrow -1$ as the Universe moves from matter domination to dark energy domination, approaching the de Sitter state from below as previously discussed (see the left pannel of Fig.~\ref{fig:wz_combined}). 

When the potential is present, $V(\varphi)\neq0$, the field still begins on the same attractor solution at early times but receives a contribution from the potential which raises the value of $w_\varphi$. Eventually, the field rolls off the attractor solution and $\dot{\varphi} H$ begins to evolve. The presence of the potential allows $w_\varphi$ to cross the phantom divide from below, which resembles the behavior of the non-minimally coupled models in Eq.~\eqref{eq:minimal_functions} \cite{Wolf:2024stt, Wolf:2025jed, Ye:2024ywg}. This is illustrated in Fig.~\ref{fig:wz_combined} where we have plotted two representative kinetic braiding models, one with and one without a potential.

In this work, we focus on the latest expansion history data. Before undertaking a direct comparison of this model with the full complement of cosmological data relating to the expansion history, we first project the model's ``theory priors'' into the $(w_0, w_a)$ parameter space in order to get an initial assessment of whether or not it can offer a good description of the data. To do so, we follow the prescription detailed in \cite{Wolf:2024eph, Wolf:2025jlc, Garcia-Garcia:2019cvr, Traykova:2021hbr}. In brief, we generate the exact observables (or in some cases compressed versions of the observables) that this model would predict for the observables probed in surveys of interest. That is, we generate the model's predictions over a wide range of parameter choices for $D_A(z)$, $D_L(z)$, etc at the redshifts measured in BAO, SNe, and CMB probes, using DESI BAO DR2 measurements, as well as compressed versions of the DESY5 SNe and Planck CMB data (again in the manner described in detail in \cite{Wolf:2024eph, Wolf:2025jlc}). Then, using the known errors and uncertainties of the data, we 
fit $(w_0, w_a)$ to these predictions through
\begin{equation}\label{hfit}
H^2(a)=H_0^2\left[\Omega_{\mathrm{m}} a^{-3}+\left(1-\Omega_{\mathrm{m}}\right) e^{3 w_a (a-1)} a^{-3(1+w_0+w_a)}\right],
\end{equation}
as all of the relevant observables are directly sensitive to $H(z)$.
For the purpose of this exercise, we consider $\alpha \in [-20, 0]$, $m^2 \in [-80, 0]$, $\varphi_i \in [0,1]$, $H_0 \in [60, 75]$, and $\Omega_{\mathrm{m}} \in [0.25, 0.40]$ in generating the theory priors.

The results are depicted in the right panel of Fig.~\ref{fig:wz_combined}, where one can see that the theory priors share a complete overlap with the data constraints on $(w_0, w_a)$ and occupies the same region of parameter space as the non-minimally coupled model from \cite{Wolf:2024stt, Wolf:2025jed}.
This indicates that this model is a good description of the data as it shares significant overlap with the posterior constraints when both the theory priors and cosmological data posteriors are interpreted in terms of $(w_0, w_a)$ parameters. This is also not terribly surprising considering the phenomenology of the equation of state for the non-minimal and kinetic sectors of Eq.~\eqref{eq:fullaction} (with a potential) are remarkably similar.  While both the non-minimally coupled model and the broken shift symmetric model can remain in the phantom regime today for certain parameter choices (i.e.\ $w_0 < -1$), most of the volume of the theoretical priors is located in the thawing regime. As the phantom regime is ruled out, we focus on the thawing regime due to current observational constraints and in Fig.~\ref{fig:wz_combined} we depict the projection of the theory priors as running up to the phantom boundary. See the appendix~\ref{appendix:shiftsymmetric} for a comparison of these models with pure shift symmetric models which are confined to the phantom regime.
We turn now to a full statistical analysis. 

\section{Cosmological Data and Constraints}\label{sec:evidence}

In order to assess this dark energy model's ability to describe cosmological data, we use similar dataset combinations to those used in the DESI DR2 2025 analysis \cite{DESI:2025zgx}; DESI 2025 BAO measurements, the SNe Ia samples from DES-Y5, Pantheon+, and Union3 \cite{Scolnic:2021amr, Rubin:2023ovl, DES:2024jxu}, the TTTEEE CMB measurements from Planck 2018 \cite{Planck:2019nip, Planck:2018vyg},  and the CMB lensing from ACT DR6 \cite{ACT:2023dou, ACT:2023kun}. 

DESI 2025 BAO measurements consist of a set of six measurements of the angular diameter distance, $D_M$, the Hubble parameter, $D_H = c/H$, and the angle-averaged quantity $D_V = (z D_M^2 D_H)^3$, relative to the sound horizon, obtained from measurements of the BAO scale. The SNe data is made of light curves from different SNe Ia samples which allows us to measure the luminosity distance $D_L$. The CMB data consists of angular power spectra. From Planck, we use the auto- and cross-correlation of the CMB temperature and $E$-mode polarization fields. We use the low-$\ell$ power spectra, obtained with the ``Commander'' component separation algorithm, in the range $2 \leq \ell \leq 29$ and the high-$\ell$ Planck PR3 \texttt{plik} likelihood covering $30 \leq \ell \leq 2508$ for the temperature auto-correlation ($TT$) and $30 \leq \ell \leq 1996$ for the $TE$ and $EE$ components \cite{Aghanim:2019ame}. Finally, we use ACT DR6 as provided by the official likelihood, consisting of the ACT CMB reconstructed lensing power spectra between the scales $40 < L < 763$. 

For BAO, SNe, and CMB data, we use the likelihoods as implemented in \texttt{Cobaya} \cite{Cobaya}. We then use both the nested sampler \texttt{polychord} \cite{Handley:2015fda} and a Metropolis-Hastings sampler \cite{Lewis:2002ah} to derive the parameter posterior distributions and to compare the models statistically.
We sample the cosmological and dark energy model parameters in Table~\ref{Table: priors} and impose uniform priors, while keeping the neutrino mass fixed to \(\sum m_\nu = 0.06\,\mathrm{eV}\). As explained in the following paragraph, the number of parameters and shape of the degeneracies make this a challenging model to sample efficiently. Thus, in order to assess convergence when deploying Monte-Carlo methods, we adopt the Gelmin-Rubic statistic, requiring $R < 0.05$. 

\begin{table}[t]
\centering
\begin{tabular}{|l|c|}
\hline
\multicolumn{2}{|c|}{\textbf{Cosmological Parameters}} \\
\hline
$\Omega_b h^2$ & $\mathcal{U}(0.020, 0.025)$ \\
$\Omega_m$ & $\mathcal{U}(0.25, 0.40)$ \\
$H_0$ [km/s/Mpc] & $\mathcal{U}(60, 80)$ \\
$\ln(10^{10} A_s)$ & $\mathcal{U}(2.95, 3.15)$ \\
$n_s$ & $\mathcal{U}(0.9, 1.05)$ \\
$\tau$ & $\mathcal{U}(0.02, 0.1)$ \\
\hline
\multicolumn{2}{|c|}{\textbf{Dark Energy Parameters}} \\
\hline
$\alpha$ & $\mathcal{U}(-10, 0.0)$ \\
$m^2$ & $\mathcal{U}(-80, 0)$ \\
$A \equiv \phi_i |m^2|^{1.41}$ & $\mathcal{U}(0, 10)$ \\
\hline
\end{tabular}
\caption{Uniform prior distributions for cosmological and dark energy parameters. 
$\mathcal{U}(a, b)$ denotes a uniform distribution over $[a, b]$.}
\label{Table: priors}
\end{table}

When sampling this dark energy model, as mentioned before, we fix $\beta = -1$, as this corresponds to a well-known rescaling of the Galileon model, as well as $\dot{\varphi}_i$ to a small, non-zero value, as the early scalar field evolution rapidly converges to the well-known attractor solution. Thus, we vary $\alpha$, $m^2$, and $\varphi_i$, while tuning $V_0$ with a shooting algorithm to ensure that the Friedmann equations are satisfied for the choices of cosmological parameters. In sampling this model, due to the shape of the degeneracies in $\varphi_i$, we found it far more efficient to change variables from $\varphi_i$ to $A=\phi_i |m^2|^{1.41}$ and vary $A$ instead.

Finally, we
compute the theory predictions for cosmological distances and power spectra with \texttt{hi\_class} in order to compare the dark energy theory with the aforementioned datasets, calculating $\Delta \chi^2$ statistic with respect to $\Lambda$CDM. The $\chi^2$ is defined by $\chi^2=-2 \log \mathcal{L}$,
where $ \log \mathcal{L}$ is the log likelihood.  However, there are significantly more parameters in this model than the $\Lambda$CDM model, so it will be informative to examine more refined statistical measures such as the Bayesian evidence and the Akaike Information Criterion (AIC), both of which account for and penalize the introduction of new parameters. 
The Bayesian evidence is given by,
\begin{equation}
    \log \mathcal{Z} = \log \int \mathcal{L}(D |\theta, M) P(\theta | M) \, d\theta,
\end{equation}
where $\mathcal{L}(D | \theta, M)$ is the likelihood,  $P(\theta| M)$ is the prior, and $\theta$ the sampled parameters. We report the difference with respect to $\Lambda$CDM using the following convention, $\Delta \chi^2_{\varphi \Lambda} = \chi^2_\varphi - \chi^2_\Lambda$ and $\log B_{\varphi \Lambda } = \log \mathcal{Z}_{\varphi} - \log\mathcal{Z}_{\Lambda}$, meaning that the more favoured model with respect to the $\Delta \chi^2$ fit is given by negative values while the more favoured model with respect to the Bayesian evidence is given by positive values. A Bayes factor $\log B \geq 5$ is considered to be strong evidence for one model over the other, while $5\geq \log B \geq 2.5$ is moderate evidence and $\log B \leq 2.5$ is inconclusive \cite{Jeffreys1939, Trotta:2005ar, Trotta:2008qt}. We also report the AIC as $\Delta {\rm AIC_{\varphi \Lambda}}$, with AIC being defined as
$
 {\rm AIC}=2\Delta k+\chi^2    
$,
where $\Delta k$ is the difference in the number of parameters between the models and imposes a penalty on the model with more parameters. Conventionally, a $\Delta \rm{AIC}  \leq -10 $ is considered strong evidence,  $\Delta \rm{AIC}  \simeq -5$ considered moderate evidence, and closer to zero is considered to be statistically indistinguishable \cite{Burnham2002, Liddle:2007fy}.

\begin{table}[t]
    \centering
    \renewcommand{\arraystretch}{1.3}
    \begin{tabular}{@{}lccc@{}}
        \hline
        \hline
        & Pantheon+ & Union3 & DESY5 \\
        \hline
        $\Delta \chi^2_{\varphi \Lambda }$ & $-11.9$ & $-17.5$ & $ -22.8$ \\
        $\Delta \mathrm{AIC}_{\varphi \Lambda }$ & $-5.9$ & $-11.5$ & $ -16.8$ \\
        $\log B_{\varphi \Lambda }$ & $-0.03$ & $5.01$ & $6.52$ \\
        \hline
        \end{tabular}
    \caption{Statistical comparison of the cubic Galileon/kinetic braiding dark energy model relative to $\Lambda$CDM, combining BAO and CMB data with three different SNe datasets. $\log B$ denotes the log Bayes factor in favor of the extended model, and $\Delta\chi^2$ is the improvement in best-fit $\chi^2$. The Bayes factors $\log B$ have an uncertainty of $\simeq \pm 0.6$.}
    \label{Table: bayes_chi2_single_model}
\end{table}

This dark energy model offers a substantial improvement in terms of its ability to fit the cosmological expansion history data as can be seen from examining its $\Delta \chi^2$ statistics in Table \ref{Table: bayes_chi2_single_model}, on par with both the parametric $(w_0, w_a)$ model and the non-minimally coupled model of \cite{Wolf:2025jed, Wolf:2024stt}. This Galileon model has three additional varied free parameters when compared with $\Lambda$CDM, given by $\{\alpha, m^2, \varphi_{i}\}$ (both $\Lambda$ and $V_0$ are fixed by the other cosmological and model parameters). Consequently, $\Delta k_{\varphi \Lambda} = 3$. 

As has been consistently seen across the recent literature, the statistical evidence for dynamical dark energy when using the Pantheon+ SNe sample is far less significant than when the other SNe samples are used. This is born out both with respect to raw $\Delta \chi^2$ fit as well as more sophisticated statistical measures such as information criteria and Bayesian evidence which penalize the introduction of new parameters. Indeed, $\Lambda$CDM and dynamical dark energy are roughly indistinguishable statistically with the Pantheon+ SNe  \cite{Lodha:2025qbg, Wolf:2025jed, Wolf:2024eph}, which remains the case with the dark energy model considered here. The dependence of our conclusions on which SNe sample is used is particularly interesting. In particular, \cite{Efstathiou_2024} has prominently argued that supernovae systematics may be driving some of these results and correcting these systematics brings DESY5 in closer alignment with the older Pantheon+ analysis. However, \cite{DES:2025tir} has argued that both Union3 and DESY5 utilize improved modeling frameworks and newer samples, while also using completely different analysis pipelines from each other. There remain a number of open questions here that will no doubt continue to be the subject of intense scrutiny, but for the time being it is clear that the more recent data and analyses offered by Union3 and DESY5 provide significantly stronger evidence for dynamical dark energy.

Finally, it is worth noting that this model, while clearly outperforming $\Lambda$CDM in a manner comparable to the parametric $(w_0, w_a)$ model, in every measure slightly underperforms the non-minimally coupled dark energy model in \cite{Wolf:2025jed, Wolf:2024stt}. In particular, it is a slightly worse overall fit to the data but has similar numbers of parameters; and so, we would naturally expect it to yield slightly lesser statistical evidence in other measures. For reference, the non-minimally coupled model has $\log B \simeq 7.3$ and a $\Delta \chi^2 \simeq -23.6$ (see Table 1 in \cite{Wolf:2025jed} for more details), and has the same number of free parameters as the cubic Galileon model and so would have $\Delta \rm{AIC_{\varphi \Lambda }}=-17.6$. However, the broken shift symmetric/cubic Galileon model still produces very significant improvements over $\Lambda$CDM or standard thawing quintessence particularly when the DESY5 SNe data is used.

\begin{figure}[t]
   \centering
    {%
       \includegraphics[width=\columnwidth]{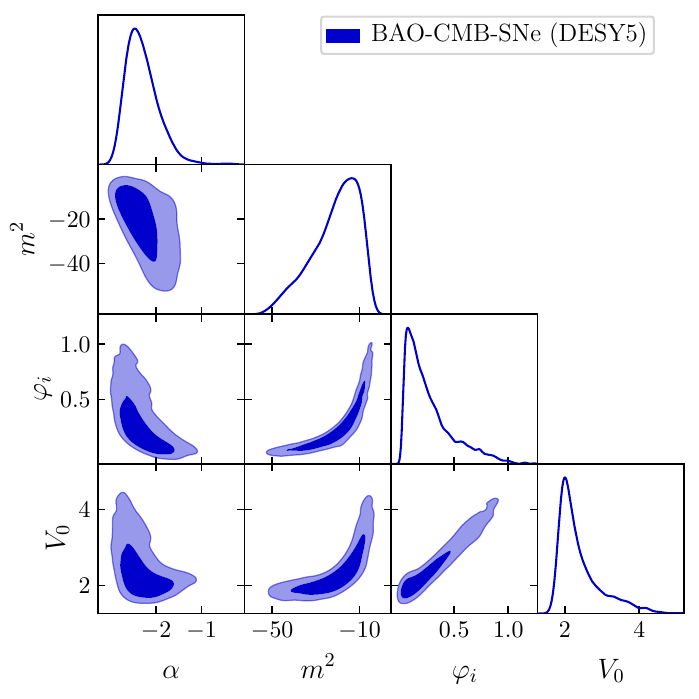}
    }
    \vskip -0.3in
    \caption{68\% and 95\% C.L. posterior distributions for the cubic Galileon/kinetic braiding dark energy model parameters with a broken shift symmetry. }\label{Fig:Posteriors}
\end{figure}

\begin{figure}[t]
   \centering
    {%
       \includegraphics[width=\columnwidth]{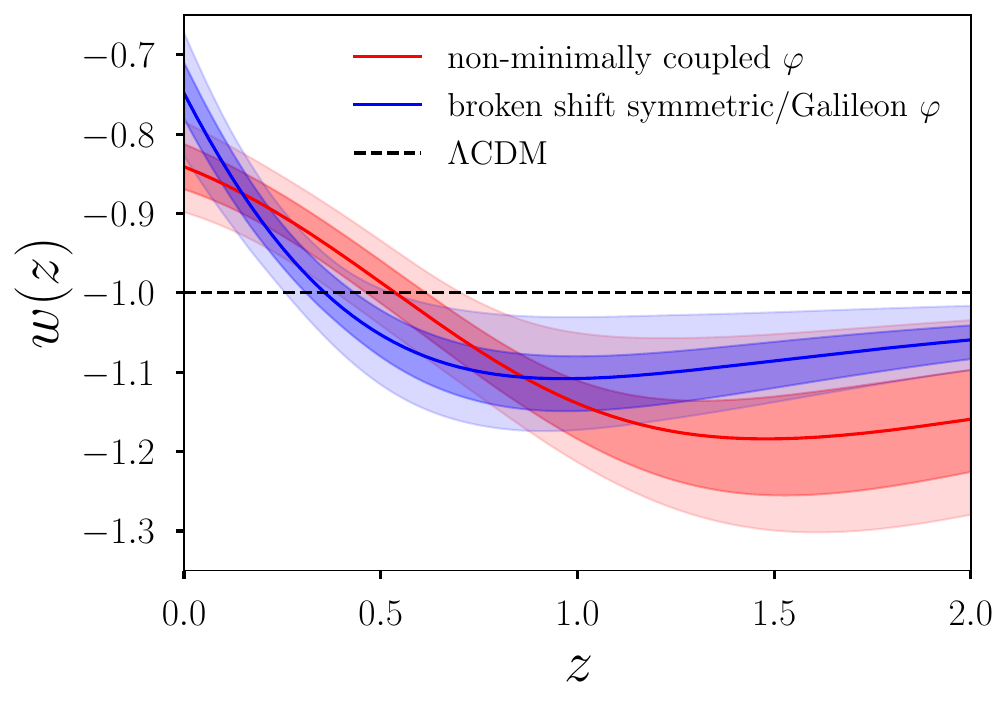}
    }
    \vskip -0.3in
    \caption{Posterior constraints on $w(z)$ for the non-minimally coupled dark energy model \cite{Wolf:2025jed} and the cubic Galileon dark energy model considered here. Both dark energy constructs produce similar equations of state which explains their similar success in fitting the expansion history data. }\label{Fig:wPosteriors}
\end{figure}

Given that the data combination with the strongest evidence for dynamical dark energy is that which uses the DESY5 SNe sample, this is of particular interest. The posterior constraints when using this SNe sample in combination with CMB and BAO on the dark energy model parameters from Eq.~\eqref{eq:kinetic_functions} are given in Fig.~\ref{Fig:Posteriors}, with the 1$\sigma$ regions $\alpha = -2.31^{+0.25}_{-0.41}$, 
$m^2 = -21.0^{+14}_{-6.0}$, 
$A = 4.4^{+1.3}_{-1.0}$, 
$V_0 = 2.36^{+0.16}_{-0.61}$. As with the non-minimally coupled model of \cite{Wolf:2024stt, Wolf:2025jed} or with parametric $(w_0, w_a)$ models, we see here the same features where the cosmological data favours an equation of state which is phantom in the past, but crosses the phantom divide and begins to thaw around $z\simeq 0.5$. The posterior constraints on the equation of state for this model are given in Fig.~\eqref{Fig:wPosteriors}. $w(z=0) = -0.749^{+0.039}_{-0.032}$, placing the equation of state today at approximately 7$\sigma$ away from the cosmological constant value. Despite the substantially different microphysics at play between this dark energy model (driven by a modified kinetic sector plus a potential) and other kinds of exotic interactions (such as the direct coupling between the scalar and gravity in \cite{Wolf:2024stt, Wolf:2025jed, Ye:2024ywg}), they produce remarkably similar equations of state. Given that they are, for all intents and purposes, indistinguishable statistically as discussed in the paragraph above, this adds further substantiation to the argument that cosmological data will not be able to uniquely pin down the fundamental action for dark energy and can only give us limited insight into dark energy microphysics \cite{Ferreira:2025fpn, Wolf:2023uno}. Any dark energy theory producing an equation of state like the ones seen here will reproduce the ability to fit this collection of cosmological data as one can easily construct any number of distinct dark energy proposals by utilizing a different form of the scalar field potential. Furthermore, the small field excursions ($\Delta \varphi \ll M_{\mathrm{P}}$) relevant to dynamical dark energy indicate that one can simply Taylor expand the potential and find that many other potentials will agree to an excellent approximation with the massive potential considered here.

The cosmological data used here constrains primarily the expansion history of the universe. However, this model will impact the strength of gravity and possibly lead to signatures in the growth of cosmic structure. Thus, while no growth data was used in above analysis (as we await new growth measurements from DESI along with its covariance matrix with the BAO measurements), we can compute the constraints on quantities that will affect growth and be of interest as new data becomes available. Given that exotic gravitational interactions (whether coming from non-minimal couplings or non-canonical kinetic interactions) will lead to deviations from the growth of cosmic structure expected from general relativity (GR), such data (depending on its quality) may provide important discriminating power between more exotic scalar field interactions and minimal, canonical scalar fields or $\Lambda$. We now turn to analyzing these additional gravitational consequences.

\section{Ancillary Gravitational Consequences}\label{sec:grav_consequences}

\begin{figure*}[t]
   \centering
   \includegraphics[width=0.48\textwidth]{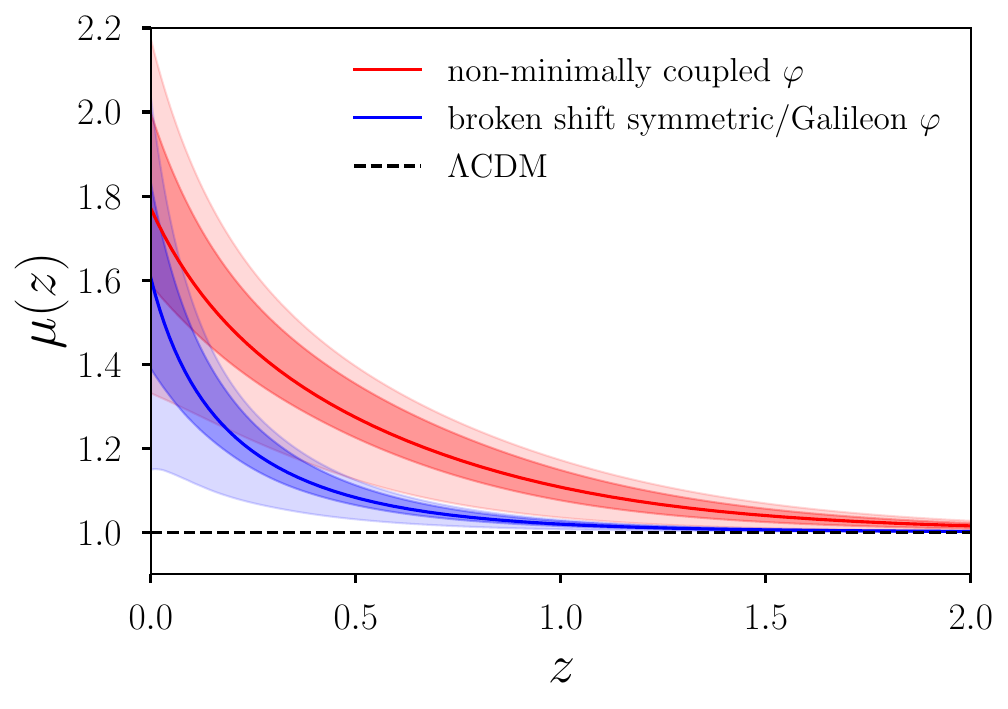}
   \hfill
   \includegraphics[width=0.48\textwidth]{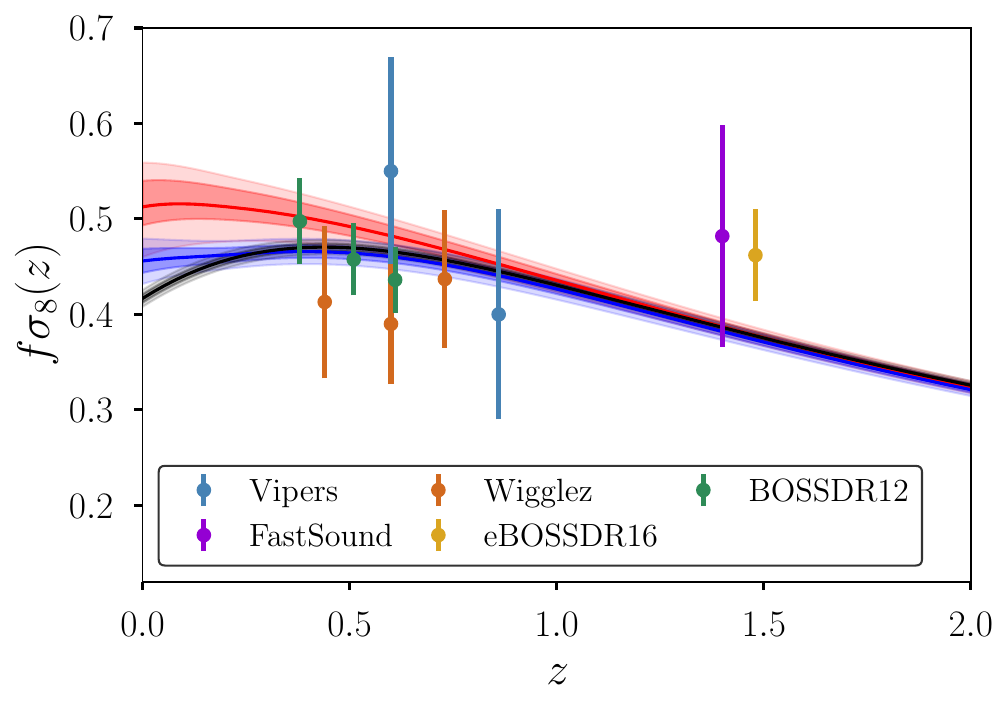}
   \vskip -0.1in
   \caption{Posterior constraints on $\mu(z)$ (Left) and f$\sigma_8$ (Right) for the non-minimally coupled dark energy model \cite{Wolf:2025jed} and cubic Galileon dark energy model considered here. While both models predict similar phenomenology for the dark energy equation of state and fit the expansion history data remarkably well, the models predict very different modified gravity effects which will show up in their predictions for growth.}
   \label{fig:mu_combined}
\end{figure*}

While this theory has no explicit coupling between the scalar field and the Einstein-Hilbert term, it does, nevertheless lead to non-trivial gravitational effects.
The presence of a braiding term, captured at the linear level by $\alpha_B$, indicates that this theory will affect the strength of gravity on cosmological and clustering scales. In the quasi-static regime, which is useful not only for local constraints but also for (cosmological) gravitational clustering, the Newton-Poisson equation can be written as
\begin{equation}
    \nabla^2 \Phi=4\pi G_0 \mu(z) {\bar \rho}_{\rm M}\delta_{\rm M},
\end{equation}
where $\Phi$ is the Newtonian potential, ${\bar \rho}_{\rm M}$ and $\delta_{\rm M}$ are the background and fractional density perturbation of matter, and $\mu(z)$ captures the time evolution of the {\it effective} Newton's constant on the scales of gravitational clustering (with $\mu(z)=1$ we recover GR).
For the theory we are considering we have 
\begin{equation}
\mu(z)=1+\frac{\alpha_B^2}{2c_s^2\left(\alpha_K+\frac{3}{2} \alpha_B^2\right)}
\end{equation}
in the quasi-static approximation \cite{Bellini:2014fua, Shah:2025vnt, Ishak:2024jhs}.
From examining Eq.~\eqref{eq:alphas}, we can see that some of the additional terms can give $\mathcal{O}(1)$ contributions, leading us to expect departures from GR. In Fig.\ \ref{fig:mu_combined}, we show the expected behavior of $\mu(z)$ derived from the posterior analysis of the previous section. One can see that $\mu(z)>1$ and is growing at late times. We compare it with $\mu(z)$ from the non-minimal models studied in
\cite{Wolf:2024stt, Wolf:2025jed} and we can see that, for much of cosmic time, it is closer to one, but it changes more rapidly at very late times. This means that we expect its cumulative effect in these (broken) shift symmetric models to be smaller than in the more conventional minimally coupled models. 

On a large scale, the time-varying Newton's constant will affect the growth rate of structure, $f \equiv d\ln\delta_M/d\ln a$. We have that the evolution equation for the growth rate is given by
\begin{eqnarray}
\frac{df}{d\ln a}+f^2+\left(1+\frac{d \ln(aH)}{d\ln a}\right)f=\frac{3}{2} \mu(a)\Omega_{\rm M}(a) \nonumber
\end{eqnarray}
where $\Omega_M(a)$ is the fractional energy density of matter dependent on time. We can clearly see that $\mu(z)>1$ will enhance the growth of structure. This is manifest in the right hand panel of Fig.\ \ref{fig:mu_combined}, where there is enhanced growth at late times. But, as foreshadowed above, the cumulative effect is much smaller than for the standard non-minimal theories we considered and so the modifications to growth are smaller than in these theories and only manifest themselves substantially for $z<0.3$. 

For local constraints, what matters is $\mu(z=0)$. From the left hand panel, we see that $\mu(z=0)\sim 1.6\pm 0.2$, on par with what we find with the non-minimally coupled models, although about $10-20\%$ smaller. This means that, as it stands, such a model is strongly inconsistent with the fifth force constraints, where we have constraints of $\mu(0)-1 < 10^{-3}$ on laboratory scales, or even tighter on astrophysical scales \cite{Uzan:2024ded, Will:2014kxa, Ke:2021jtj, Bassi2022}. These constraints would rule such a model out.

Nevertheless, it is well known that such models with a cubic Galileon, have Vainshtein screening which, in certain conditions will protect a gravitational system from the fifth forces. This mechanism has been studied in great detail \cite{Babichev:2013usa}; for a point-like (or spherically symmetric)  source of mass $M$, one can define a Vainshtein radius
\begin{eqnarray}
r_V=\left[\frac{M}{M_{\rm P}}\frac{1}{(\Lambda_3)^3}\right]^{1/3} \nonumber
\end{eqnarray}
which leads to a supression of the fifth force, $F_5$, relative to the Newtonian force, $F_N$, on scales $r<r_V$ of the form
\begin{eqnarray}
    \frac{F_5}{F_N}\simeq \left(\frac{r}{r_V}\right)^{2/3} \nonumber
\end{eqnarray}
Thus, it would seem that this theory already has ``built in'' a mechanism for evading fifth force constraints. For example, the Milky Way will have a Vainshtein screening scale $r_V\sim$ few Megaparsecs which means its dynamics will be fully screened.

A more careful look at the details of this theory throws up well known concerns. The strong coupling scale -- the energy scale above which one cannot use the action presented here and new degrees of freedom need to be considered -- is set by $\Lambda_3\sim  10^{-13} \ {\rm eV}$ which corresponds to a length scale $\lambda\sim 10^3$ {\rm km}. Such properties of these theories (which include that of a massive graviton) are well known -- what one thinks of as a microscopic theory breaks down on macroscopic scale. To be able to work with such a theory as a truly predictive microscopic theory (which is, in the end, what we are looking for) we will need to add additional degrees of freedom and terms to the action. 

A further concern is whether the Vainshtein screening mechanism is indeed present with broken shift symmetry. If the mass term in the low energy EFT arises from integrating out new, massive degrees of freedom which are present at higher energy, it has been argued \cite{Burrage_2021} that a self-consistent analysis shows that the Vainshtein mechanism will not  emerge in the presence of all the higher order operators which must be taken into account. While the argument made in \cite{Burrage_2021} is specific to one particular way of generating a mass term for the scalar field -- as yet we do not have a ``no-go theorem'' for the Vainshtein mechanism in massive Galileon theories -- it is indicative of a possible, severe problem that may jeopardize such theories.

Given the significant overlap between the phenomenology of non-minimally coupled models and the Galileon models considered here in terms of their equations of state (as well as the aforementioned similarities in terms of their ability to affect the strength of gravity on other scales), it is also worth briefly mentioning an effect that could potentially definitively discriminate between these classes of theories. The gravitational wave (GW) luminosity distance probed by bright sirens is defined as $d_L^{\mathrm{gw}}(z)= (M_{\mathrm{eff}}(0) /M_{\mathrm{eff}}(z))  d_L^{\mathrm{em}}$ \cite{LISACosmologyWorkingGroup:2019mwx},
where $d_L^{\mathrm{em}}$ is the luminosity distance associated the electromagnetic (EM) counterpart and $M^2_{\rm eff}/2 = G_4(\varphi)$ is the effective Planck mass associated with the Horndeski function $G_4$ which multiplies the Ricci scalar R. The non-minimally coupled models, which do have an $\alpha_{\mathrm{M}} \equiv \mathrm{d} \log M_{\mathrm{eff}}^2/\mathrm{~d} \log a \neq 0$, will be sensitive to such measurements \cite{LISACosmologyWorkingGroup:2019mwx, Lagos:2019kds, Wolf:2019hun}, whereas the Galileon models considered here have $\alpha_{\mathrm{M}}=0$ and will predict the same GW luminosity distance as GR.

Finally, an additional concern is the effect of the theory on the largest scales. We have looked at how the fifth force will affect the growth of structure. One probe that can access this effect on the largest scales is the Integrated Sachs Wolfe, $\Delta T/T|_{\rm ISW}$, in the CMB temperature anisotropies and how it correlates with a tracer of the matter field (like the galaxy distribution). Indeed  we have that $
    \Delta T/T|_{\rm ISW}=2\int d\tau\partial_\tau \Phi
$ where $\tau$ is conformal time, $\Phi$ is the gravitational potential and the integral is from last scattering until today. It has been shown that cubic Galileons are grossly inconsistent with measured $\Delta T/T|_{\rm ISW}$ \cite{Renk:2017rzu} and it is conceivable the broken shift symmetric version will inherit some of the same problems here although, as has been shown in \cite{Kable:2021yws} it is possible to construct theories which are compatible with the ISW measurement. A comparison with current data would be required to make a more definitive statement on whether this is, indeed, a problem.

\section{Discussion}\label{sec:conclusion}

In this paper we have continued on our campaign to understand what the current constraints on the expansion of the Universe imply for scalar field dark energy. We emphasise that we are agnostic about whether the measurements are correct or not, or whether the {\it interpretation} of the measurements are correct. We are assuming they are and exploring the consequences. We also, crucially, emphasize that we are {\it not} proposing models but are trying to, in as systematic a manner as possible, to understand whether the simplest scalar fields models are viable and consistent with  observations. 

To guide us, in our systematic approach, we are using the philosophy of effective field theory, considering the terms in the action which should be relevant at a given energy scales. We are, furthermore, aided by the fact that the relevant dynamical range of the scalar field which is detectable by observations is small, $\Delta \varphi/M_{\rm Pl}<1$. This means that we can consider polynomial expansions of the free functions and only keep the leading order terms. As a result, we effectively constrain a small number of coefficients and are able to make robust conclusions for what is, in principle, a broad class of models.

The broken shift symmetric model (or broken Galileon model) we have considered here is an alternative to the non-minimally coupled theories we have considered before. While there is no direct coupling to the Einstein-Hilbert term, it still leads to new gravitational effects which affect the expansion of the Universe and can be interpreted as a modification in the strength of Newtonian gravity on clustering scales. For the cubic-Galileon model (which is shift symmetric) the equation of state, $w(z)<-1$ throughout and does not reproduce the observed features found in current data. The addition of a potential mitigates this and leads to a ``phantom crossing'', much like with the non-minimal models we have considered before. 

Thus it would seem that, on cosmological scales, one requires, yet again, a theory with non-trivial gravitational consequences. And, as is well known, this will have ancillary consequences: it will affect the growth rate of structure and will lead to observable fifth forces on laboratory and Solar System scales. Naively, given the precision of current constraints on fifth forces, we could rule out this theory. Yet, remarkably, this theory naturally possesses a screening mechanism which might shield local measurements from the effect of fifth forces. The Vainshtein mechanism at play here is well known and, while effective, raises concerns -- it has a strong coupling scale at very low energies meaning that a more complete theory is required if we are to claim any success
at constructing a microphysical theory of dark energy.

From the point of view of an effective field theory, this may be overkill. Note that we are trying to come up with a microscopic theory which will explain the accelerated expansion we see today but we do not need it to be valid on all scales. Indeed, asking for it to be valid at Gigaparsec scales and on kilometre scales (or less) may be too tall an order. This is, in fact, the philosophy behind the effective field theory approach -- to have a description which is valid in a well defined range of scales, not {\it all} scales. But if that is the case, one might question why one is pursuing a microphysical explanation for dark energy at all. We have a perfectly good explanation of dark energy, in terms of an equation of state, and the bulk properties of a ``fluid''. Such a description mirrors how we describe standard model physics on cosmological scales -- not in terms of the Dirac equation or Yang-Mills fields but in terms of baryon densities and photon phase space distributions. 

Thus we find ourselves in a quandary. We can give in and forgo the aim of constructing a fundamental dark energy theory. We are then left in a situation where we will never fully determine what is the microphysical theory behind one of the most striking phenomenon in cosmology. If we  try to pursue microphysical proposals, we need to do so in as general a way as possible (as we have done here and in the past). But we are finding that, in the case of scalar field dark energy, one has to consider ever more complex proposals that can accommodate not only the expansion of the Universe but also ancillary gravitational effects on smaller scales -- indeed, that possibility has been explored in the case of broken shift symmetric models in \cite{Tsujikawa:2025wca} and for more general scalar field actions in \cite{Yao:2025wlx}. This is disheartening but also encouraging. Disheartening because it would seem like we are going down a well trodden path of adding epicycles to make a theory work -- we are entering the Ptolomaeic Era of dark energy. Encouraging because some of these theories may lead to new observable effects that can be tested by non-cosmological means -- in the laboratory or on astrophysical scales -- leading to a completely new window on dark energy.

\section*{Acknowledgements}
PGF acknowledges support from STFC and the Beecroft Trust. CGG acknowledges support from the Beecroft Trust. WJW acknowledges support from St. Cross College, University of Oxford.

\appendix
\section{Link to Horndeski Gravity}
\label{appendix:horndeski}

Although the scope of this paper is not the study of specific models that might fit the data well but to infer the microphysical properties of dark energy by building the most general action in an EFT-like approach, it is possible to link our results to the well-known Horndeski Gravity. Given its familiarity and that most of the Galileon's literature is expressed in this context, we proceed to show the connection of our EFT-like actions and Horndeski Gravity.

Horndeski gravity represents the most general scalar-tensor theory of gravity with second order equations of motion \cite{Horndeski:1974wa, Charmousis:2011bf} and is given by the action,
\begin{equation}
S[g_{\mu\nu},\phi]
=
\int d^4x \sqrt{-g}
\sum_{i=2}^{5} \frac{1}{8\pi G_N}\,\mathcal{L}_i[g_{\mu\nu},\phi]
 .
\end{equation}
where the various components of the action are defined as
\begin{align}
\mathcal{L}_2 &= G_2(\phi, X), \tag{A2} \\[6pt]
\mathcal{L}_3 &= -G_3(\phi, X)\Box\phi, \tag{A3} \\[6pt]
\mathcal{L}_4 &= G_4(\phi, X)R
+ G_{4X}(\phi, X)
\left[
(\Box\phi)^2 - \phi_{;\mu\nu}\phi^{;\mu\nu}
\right], \tag{A4} \\[6pt]
\mathcal{L}_5 &= G_5(\phi, X)G_{\mu\nu}\phi^{;\mu\nu}
- \frac{1}{6} G_{5X}(\phi, X)
\left[
(\Box\phi)^3
\right. \notag\\
&\quad\left.
+ 2\,\phi_{;\mu}{}^{\nu}\phi_{;\nu}{}^{\alpha}\phi_{;\alpha}{}^{\mu}
- 3\,\phi_{;\mu\nu}\phi^{;\mu\nu}\Box\phi
\right]. \tag{A5}
\end{align}

The dark energy model given by Eq.~\eqref{eq:kinetic_functions}which is the main subject of this paper can be naturally expressed in the Horndeski language,
\begin{equation}
\begin{aligned}
G_2 &= \alpha X - \left( V_0 + \frac{1}{2} m^2 \varphi^2 \right),
&\quad
G_3 &= -\frac{\beta X}{\Lambda_3^3},
\\[6pt]
G_4 &= \frac{1}{2} M_{\mathrm{P}}^2,
&\quad
G_5 &= 0 .
\end{aligned}
\tag{A6}
\end{equation}
Similarly, one can write a closely related, very general shift symmetric model (the subject of Appendix \ref{appendix:shiftsymmetric}) as, 
\begin{equation}
\begin{aligned}
G_2 &= \alpha_1 X + \frac{\alpha_2}{\Lambda_2^4} X^2 - V_0,
&\quad
G_3 &= -\frac{1}{\Lambda_3^3}
\left(
\beta_1 X + \frac{\beta_2}{\Lambda_2^4} X^2
\right),
\\[6pt]
G_4 &= \frac{1}{2} M_P^2,
&\quad
G_5 &= 0, 
\end{aligned}
\tag{A7}
\end{equation}
where the only difference is that we have dropped the $m^2\varphi^2$ term in the potential that breaks the shift symmetry and we have included the next set of higher order terms in the Lagrangian. As the Horndeski framework has been extensively studied, one can then use these definitions to easily compute the scalar field equation of motion, the $\alpha_B$ and $\alpha_K$ parameters, and the strength of gravity $\mu(z)$ from the expressions in \cite{Bellini:2014fua}.

\section{Shift Symmetric Model}\label{appendix:shiftsymmetric}

A general symmetric theory of dark energy is given by \cite{Traykova:2021hbr}:

\begin{equation}\label{eq:shift_sym}
\begin{split}
S = \int d^4x \sqrt{-g}\Bigg[ \frac{M^2_{\rm P}}{2}R 
&+ \left(\alpha_1 + \frac{\beta_1}{(\Lambda_3)^3}\Box\varphi\right)X \\
&+ \left(\frac{\alpha_2}{(\Lambda_4)^4} 
   + \frac{\beta_2}{(\Lambda_3)^3 (\Lambda_4)^4}\Box\varphi\right)X^2 \Bigg],
\end{split}
\end{equation}
where the $\alpha_i$ and $\beta_i$ are dimensionless constants and $\Lambda_4=(H^2_0M^2_{\rm P})^{1/4}$. We note that this includes all the lowest order terms according to EFT of theories with shift symmetry. In \cite{Traykova:2021hbr}, the authors explored the phenomenology of such theories in great detail and found that the equation of state was phantom {\it throughout the history of the Universe}. 

\begin{figure}[t]
   \centering
    {%
       \includegraphics[width=\columnwidth]{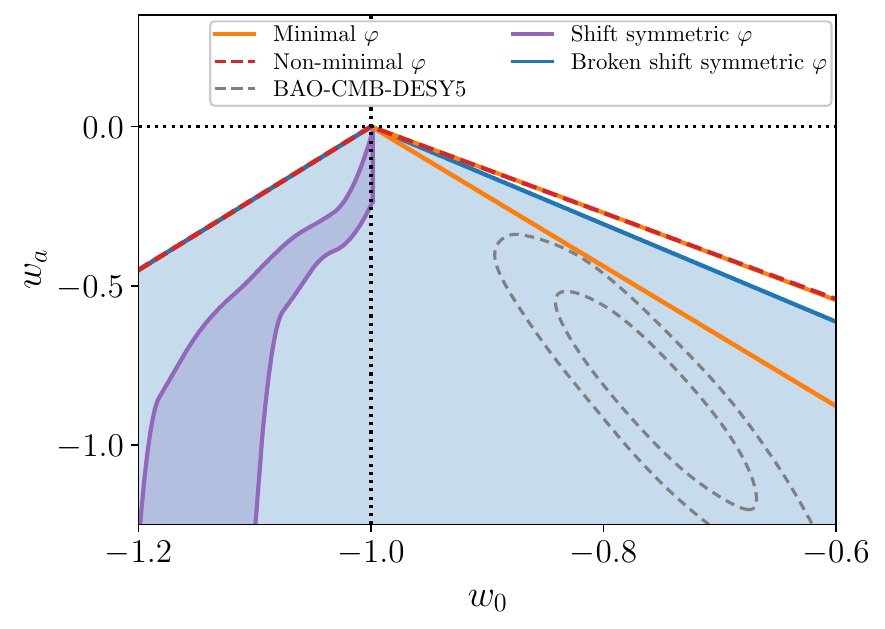}
    }
    \vskip -0.3in
    \caption{Theory priors of various dark energy models as in Fig.~\ref{fig:wz_combined}, but now including the pure shift symmetric version of the Galileon theory.}\label{Fig:shiftsym_wedge}
\end{figure}

\begin{figure*}[t]
   \centering
   \includegraphics[width=0.48\textwidth]{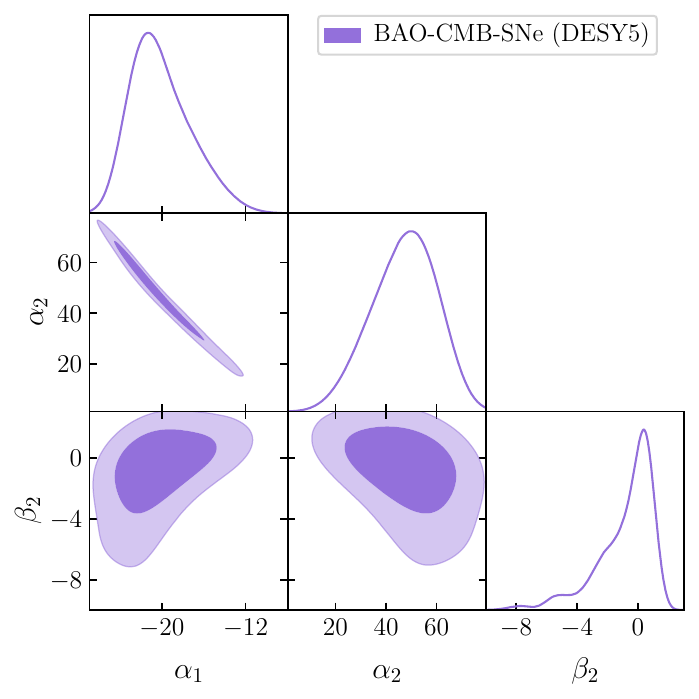}
   \hfill
   \includegraphics[width=0.48\textwidth]{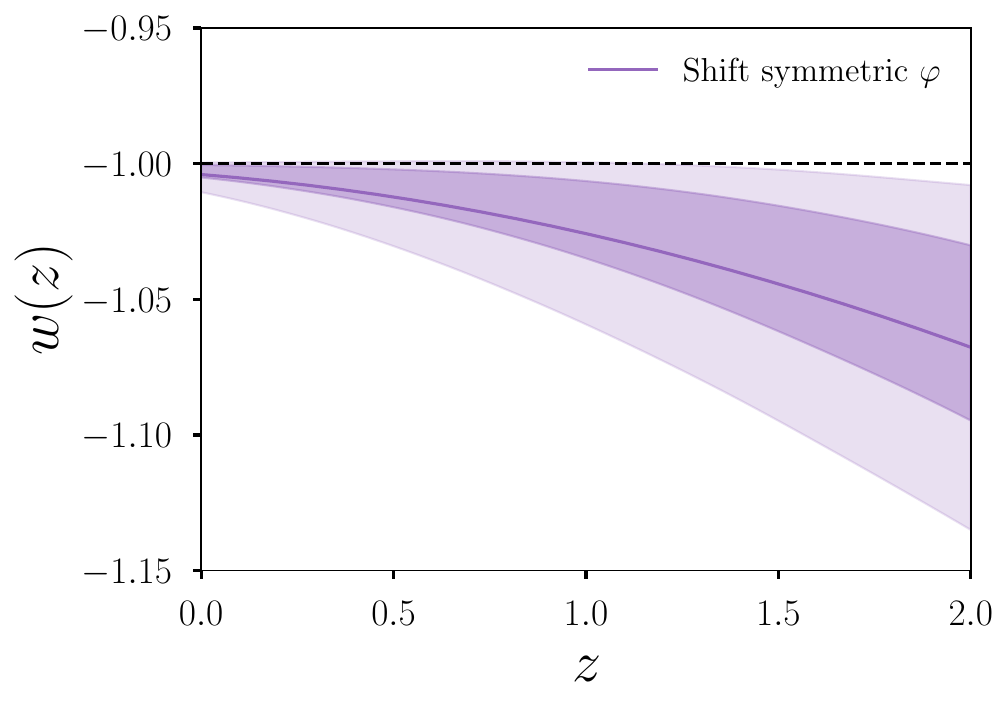}
   \vskip -0.1in
   \caption{(Left) 68\% and 95\% C.L. posterior distributions for the shift symmetric dark energy model parameters.  (Right) Posterior constraints on $w(z)$ for the shift symmetric dark energy model.}
   \label{fig:appendix_posteriors}
\end{figure*}

It is useful to translate what this means into where the equation of state parameters, ($w_0$,$w_a$), lie on the plane. To do so, we sample the ranges $\alpha_1 \in [-30, 0]$, $\alpha_2 \in [-100, 100]$, and $\beta_2 \in [-100, 100]$ for the dark energy parameters (corresponding roughly to the distributions found in \cite{Traykova:2021hbr}), and as before $H_0 \in [60, 75]$, and $\Omega_{\mathrm{m}} \in [0.25, 0.40]$ for the cosmological background parameters. We then project them into the $(w_0, w_a)$ parameter space according to the same procedure discussed earlier whereby we fit the predictions from the dark energy model to $(w_0, w_a)$ using the known errors and uncertainty of the relevant datasets \cite{Wolf:2024eph, Wolf:2025jlc}. In Fig.\ \ref{Fig:shiftsym_wedge}, we find this projection and we can see that it is completely disjunct with the values of the parameters preferred by the current data and that the theory entirely projects onto $w_0<-1$ region of parameter space.  
In this figure, we also see that both the non-minimally coupled and the broken shift symmetric/cubic Galileon theory with a potential can also occupy parameter space in the phantom region if, for certain parameter choices, the field has not thawed enough to cross the phantom divide.

Interestingly, we see that the theory projection into the $(w_0, w_a)$ plane shows that the theory is slightly closer to the data posteriors on $(w_0, w_a)$ than $\Lambda$CDM at $(-1, 0)$ as it migrates down the $w_a$ axis that indicates temporal evolution. To investigate further, we obtain the posterior constraints and quantify the $\Delta \chi^2$ fit using the DESY5 SNe samples in addition to the CMB and BAO data. We have that $\Delta \chi_{\varphi \Lambda}^2 \simeq -7$, which corresponds to the modest improvement we anticipated by examining the shift symmetric parameter space from in Fig.\ \ref{Fig:shiftsym_wedge} closest to the $(w_0, w_a)$ data constraints. However, the introduction of two additional varied free parameters (in this case, $\alpha_1$ is the parameter that is tuned/fixed to satisfy the Friedmann equation) results in $\Delta \mathrm{AIC}_{\varphi \Lambda} \simeq -3$. The posterior constraints on the model parameters and the posterior constraints on $w(\varphi)$ are given in Fig.\ \ref{fig:appendix_posteriors}, where it can be seen that the shift symmetric model remains in the phantom region throughout its entire history as it asymptotes to $w(\varphi) \rightarrow -1$.

\begin{figure}[t]
   \centering
    {%
       \includegraphics[width=\columnwidth]{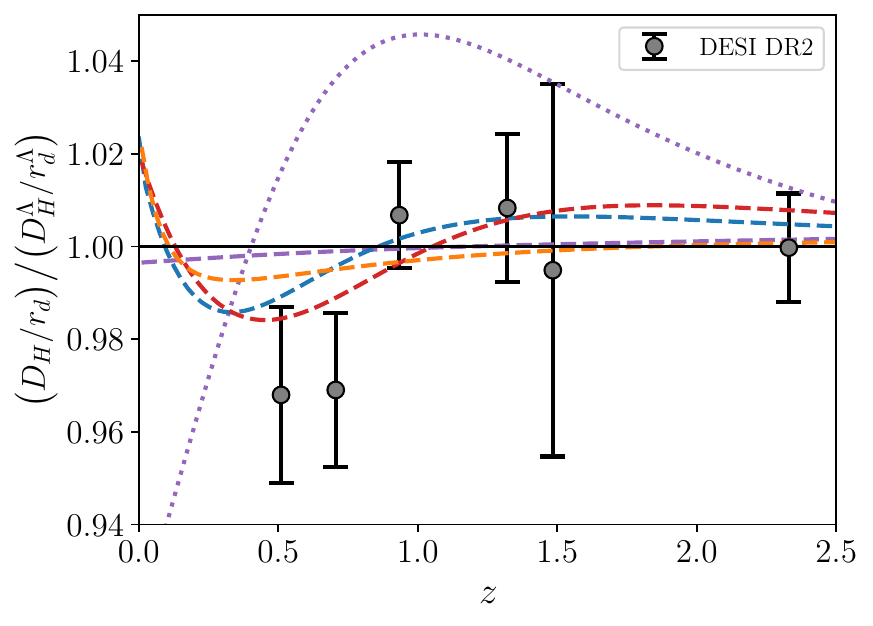}
    }
    \vskip -0.3in
    \caption{Predicted $D_H$ for the best fit dark energy models normalized to the $\Lambda$CDM best fit (CMB+BAO+DESY5). (blue) broken shift symmetric Galileon/kinetic braiding scalar field (red) non-minimally coupled scalar field (yellow) minimally coupled quintessence (purple dashed) general shift symmetric theory (purple dotted) shift symmetric with $\alpha_2=\beta_2=V=0$ as in Fig.\ \ref{fig:wz_combined}. }\label{Fig:DESIvModels}
\end{figure}

Finally, it is worth briefly commenting on the difference in behavior of the equation of state between the shift symmetric version of \eqref{eq:kinetic_functions} with $V(\varphi)=0$ depicted in Fig.\ \ref{fig:wz_combined} and the more general shift symmetric theory presented in Eq.\ \eqref{eq:shift_sym}. That equation of state remains much deeper in the phantom regime and evolves more rapidly, meaning that it would find itself in the bottom far left area of the shift symmetric theory priors in Fig.\ \ref{Fig:shiftsym_wedge}, which is very far from the data constraints. To confirm this, we can also run an MCMC fixing $\alpha_2 = \beta_2 =0$, which corresponds to the model considered earlier in Eq.\ \eqref{eq:kinetic_functions} with  $V(\varphi) =0$. This is incredibly inconsistent with the data, with $\Delta \chi_{\varphi \Lambda}^2 \simeq 240$. The addition of $\alpha_2$ and $\beta_2$ introduces considerably more dynamical freedom and allows the equation of state to evolve dynamically in a manner that is much closer $w(\varphi) \simeq -1$. Consequently, it can offer a best fit statistical description of the data that is closer to $\Lambda$CDM than the most basic shift symmetric theory with only $\alpha_1$ and $\beta_1$. However, as we have seen, purely shift symmetric theories are a far worse description of the cosmological expansion history data than quintessence, non-minimally coupled fields, and Galileon/kinetic braiding gravity with a broken shift symmetry. 

In Fig.\ \ref{Fig:DESIvModels}, one can see the predicted $D_H =c/H$ for each of the models represented on Fig.\ \ref{Fig:shiftsym_wedge} for their best fit parameters to the combined CMB, BAO, and SNe (DESY5) data (all normalized to the best fit $\Lambda$CDM model). That is, the Galileon model with the broken shift symmetry (blue) as expected closely tracks the non-minimal model (red) which, as we have seen, shares overlapping theoretical priors and very similar behavior for the equation of state. These are also clearly the two best descriptions of the data as we have seen from various statistical analyses and their theoretical priors. One also sees that minimally coupled thawing quintessence (yellow) does not describe the data as well. Coming then to the purely (unbroken) shift symmetric cases considered here, one sees the best fit from the general shift symmetric theory (Eq.\ \eqref{eq:shift_sym}) and the more restricted shift symmetric theory ($\alpha_2=\beta_2=V=0$; see Fig.\ \ref{fig:wz_combined} for a representative equation of state). As anticipated in the preceding discussion concerning theoretical priors, the best fit general shift symmetric theory (purple dashed) tracks $\Lambda$CDM closely with minimal dynamics. However, the restricted shift symmetric theory (purple dotted) which maintains substantial phantom behavior throughout its evolution is grossly inconsistent with the data.

\newpage
\bibliography{refs}

\begin{thebibliography}{148}%
\makeatletter
\providecommand \@ifxundefined [1]{%
 \@ifx{#1\undefined}
}%
\providecommand \@ifnum [1]{%
 \ifnum #1\expandafter \@firstoftwo
 \else \expandafter \@secondoftwo
 \fi
}%
\providecommand \@ifx [1]{%
 \ifx #1\expandafter \@firstoftwo
 \else \expandafter \@secondoftwo
 \fi
}%
\providecommand \natexlab [1]{#1}%
\providecommand \enquote  [1]{``#1''}%
\providecommand \bibnamefont  [1]{#1}%
\providecommand \bibfnamefont [1]{#1}%
\providecommand \citenamefont [1]{#1}%
\providecommand \href@noop [0]{\@secondoftwo}%
\providecommand \href [0]{\begingroup \@sanitize@url \@href}%
\providecommand \@href[1]{\@@startlink{#1}\@@href}%
\providecommand \@@href[1]{\endgroup#1\@@endlink}%
\providecommand \@sanitize@url [0]{\catcode `\\12\catcode `\$12\catcode `\&12\catcode `\#12\catcode `\^12\catcode `\_12\catcode `\%12\relax}%
\providecommand \@@startlink[1]{}%
\providecommand \@@endlink[0]{}%
\providecommand \url  [0]{\begingroup\@sanitize@url \@url }%
\providecommand \@url [1]{\endgroup\@href {#1}{\urlprefix }}%
\providecommand \urlprefix  [0]{URL }%
\providecommand \Eprint [0]{\href }%
\providecommand \doibase [0]{https://doi.org/}%
\providecommand \selectlanguage [0]{\@gobble}%
\providecommand \bibinfo  [0]{\@secondoftwo}%
\providecommand \bibfield  [0]{\@secondoftwo}%
\providecommand \translation [1]{[#1]}%
\providecommand \BibitemOpen [0]{}%
\providecommand \bibitemStop [0]{}%
\providecommand \bibitemNoStop [0]{.\EOS\space}%
\providecommand \EOS [0]{\spacefactor3000\relax}%
\providecommand \BibitemShut  [1]{\csname bibitem#1\endcsname}%
\let\auto@bib@innerbib\@empty
\bibitem [{\citenamefont {Abdul~Karim}\ \emph {et~al.}(2025)\citenamefont {Abdul~Karim} \emph {et~al.}}]{DESI:2025zgx}%
  \BibitemOpen
  \bibfield  {author} {\bibinfo {author} {\bibfnamefont {M.}~\bibnamefont {Abdul~Karim}} \emph {et~al.} (\bibinfo {collaboration} {DESI}),\ }\bibfield  {title} {\bibinfo {title} {{DESI DR2 Results II: Measurements of Baryon Acoustic Oscillations and Cosmological Constraints}},\ }\href@noop {} {\  (\bibinfo {year} {2025})},\ \Eprint {https://arxiv.org/abs/2503.14738} {arXiv:2503.14738 [astro-ph.CO]} \BibitemShut {NoStop}%
\bibitem [{\citenamefont {Wolf}\ and\ \citenamefont {Ferreira}(2023)}]{Wolf:2023uno}%
  \BibitemOpen
  \bibfield  {author} {\bibinfo {author} {\bibfnamefont {W.~J.}\ \bibnamefont {Wolf}}\ and\ \bibinfo {author} {\bibfnamefont {P.~G.}\ \bibnamefont {Ferreira}},\ }\bibfield  {title} {\bibinfo {title} {{Underdetermination of dark energy}},\ }\href {https://doi.org/10.1103/PhysRevD.108.103519} {\bibfield  {journal} {\bibinfo  {journal} {Phys. Rev. D}\ }\textbf {\bibinfo {volume} {108}},\ \bibinfo {pages} {103519} (\bibinfo {year} {2023})},\ \Eprint {https://arxiv.org/abs/2310.07482} {arXiv:2310.07482 [astro-ph.CO]} \BibitemShut {NoStop}%
\bibitem [{\citenamefont {Wolf}\ \emph {et~al.}(2024)\citenamefont {Wolf}, \citenamefont {Garc\'\i{}a-Garc\'\i{}a}, \citenamefont {Bartlett},\ and\ \citenamefont {Ferreira}}]{Wolf:2024eph}%
  \BibitemOpen
  \bibfield  {author} {\bibinfo {author} {\bibfnamefont {W.~J.}\ \bibnamefont {Wolf}}, \bibinfo {author} {\bibfnamefont {C.}~\bibnamefont {Garc\'\i{}a-Garc\'\i{}a}}, \bibinfo {author} {\bibfnamefont {D.~J.}\ \bibnamefont {Bartlett}},\ and\ \bibinfo {author} {\bibfnamefont {P.~G.}\ \bibnamefont {Ferreira}},\ }\bibfield  {title} {\bibinfo {title} {{Scant evidence for thawing quintessence}},\ }\href {https://doi.org/10.1103/PhysRevD.110.083528} {\bibfield  {journal} {\bibinfo  {journal} {Phys. Rev. D}\ }\textbf {\bibinfo {volume} {110}},\ \bibinfo {pages} {083528} (\bibinfo {year} {2024})},\ \Eprint {https://arxiv.org/abs/2408.17318} {arXiv:2408.17318 [astro-ph.CO]} \BibitemShut {NoStop}%
\bibitem [{\citenamefont {Shlivko}\ and\ \citenamefont {Steinhardt}(2024)}]{Shlivko:2024llw}%
  \BibitemOpen
  \bibfield  {author} {\bibinfo {author} {\bibfnamefont {D.}~\bibnamefont {Shlivko}}\ and\ \bibinfo {author} {\bibfnamefont {P.~J.}\ \bibnamefont {Steinhardt}},\ }\bibfield  {title} {\bibinfo {title} {{Assessing observational constraints on dark energy}},\ }\href {https://doi.org/10.1016/j.physletb.2024.138826} {\bibfield  {journal} {\bibinfo  {journal} {Phys. Lett. B}\ }\textbf {\bibinfo {volume} {855}},\ \bibinfo {pages} {138826} (\bibinfo {year} {2024})},\ \Eprint {https://arxiv.org/abs/2405.03933} {arXiv:2405.03933 [astro-ph.CO]} \BibitemShut {NoStop}%
\bibitem [{\citenamefont {Tada}\ and\ \citenamefont {Terada}(2024)}]{Tada:2024znt}%
  \BibitemOpen
  \bibfield  {author} {\bibinfo {author} {\bibfnamefont {Y.}~\bibnamefont {Tada}}\ and\ \bibinfo {author} {\bibfnamefont {T.}~\bibnamefont {Terada}},\ }\bibfield  {title} {\bibinfo {title} {{Quintessential interpretation of the evolving dark energy in light of DESI observations}},\ }\href {https://doi.org/10.1103/PhysRevD.109.L121305} {\bibfield  {journal} {\bibinfo  {journal} {Phys. Rev. D}\ }\textbf {\bibinfo {volume} {109}},\ \bibinfo {pages} {L121305} (\bibinfo {year} {2024})},\ \Eprint {https://arxiv.org/abs/2404.05722} {arXiv:2404.05722 [astro-ph.CO]} \BibitemShut {NoStop}%
\bibitem [{\citenamefont {Payeur}\ \emph {et~al.}(2025)\citenamefont {Payeur}, \citenamefont {McDonough},\ and\ \citenamefont {Brandenberger}}]{Payeur:2024dnq}%
  \BibitemOpen
  \bibfield  {author} {\bibinfo {author} {\bibfnamefont {G.}~\bibnamefont {Payeur}}, \bibinfo {author} {\bibfnamefont {E.}~\bibnamefont {McDonough}},\ and\ \bibinfo {author} {\bibfnamefont {R.}~\bibnamefont {Brandenberger}},\ }\bibfield  {title} {\bibinfo {title} {{Do observations prefer thawing quintessence?}},\ }\href {https://doi.org/10.1103/bggr-61nr} {\bibfield  {journal} {\bibinfo  {journal} {Phys. Rev. D}\ }\textbf {\bibinfo {volume} {111}},\ \bibinfo {pages} {123541} (\bibinfo {year} {2025})},\ \Eprint {https://arxiv.org/abs/2411.13637} {arXiv:2411.13637 [astro-ph.CO]} \BibitemShut {NoStop}%
\bibitem [{\citenamefont {Gialamas}\ \emph {et~al.}(2025{\natexlab{a}})\citenamefont {Gialamas}, \citenamefont {H{\"u}tsi}, \citenamefont {Raidal}, \citenamefont {Urrutia}, \citenamefont {Vasar},\ and\ \citenamefont {Veerm{\"a}e}}]{Gialamas:2025pwv}%
  \BibitemOpen
  \bibfield  {author} {\bibinfo {author} {\bibfnamefont {I.~D.}\ \bibnamefont {Gialamas}}, \bibinfo {author} {\bibfnamefont {G.}~\bibnamefont {H{\"u}tsi}}, \bibinfo {author} {\bibfnamefont {M.}~\bibnamefont {Raidal}}, \bibinfo {author} {\bibfnamefont {J.}~\bibnamefont {Urrutia}}, \bibinfo {author} {\bibfnamefont {M.}~\bibnamefont {Vasar}},\ and\ \bibinfo {author} {\bibfnamefont {H.}~\bibnamefont {Veerm{\"a}e}},\ }\bibfield  {title} {\bibinfo {title} {{Quintessence and phantoms in light of DESI 2025}},\ }\href@noop {} {\  (\bibinfo {year} {2025}{\natexlab{a}})},\ \Eprint {https://arxiv.org/abs/2506.21542} {arXiv:2506.21542 [astro-ph.CO]} \BibitemShut {NoStop}%
\bibitem [{\citenamefont {Lodha}\ \emph {et~al.}(2025)\citenamefont {Lodha} \emph {et~al.}}]{Lodha:2025qbg}%
  \BibitemOpen
  \bibfield  {author} {\bibinfo {author} {\bibfnamefont {K.}~\bibnamefont {Lodha}} \emph {et~al.},\ }\bibfield  {title} {\bibinfo {title} {{Extended Dark Energy analysis using DESI DR2 BAO measurements}},\ }\href@noop {} {\  (\bibinfo {year} {2025})},\ \Eprint {https://arxiv.org/abs/2503.14743} {arXiv:2503.14743 [astro-ph.CO]} \BibitemShut {NoStop}%
\bibitem [{\citenamefont {Luu}\ \emph {et~al.}(2025)\citenamefont {Luu}, \citenamefont {Qiu},\ and\ \citenamefont {Tye}}]{Luu:2025fgw}%
  \BibitemOpen
  \bibfield  {author} {\bibinfo {author} {\bibfnamefont {H.~N.}\ \bibnamefont {Luu}}, \bibinfo {author} {\bibfnamefont {Y.-C.}\ \bibnamefont {Qiu}},\ and\ \bibinfo {author} {\bibfnamefont {S.~H.~H.}\ \bibnamefont {Tye}},\ }\bibfield  {title} {\bibinfo {title} {{Dynamical dark energy from an ultralight axion}},\ }\href {https://doi.org/10.1103/3mpg-24d2} {\bibfield  {journal} {\bibinfo  {journal} {Phys. Rev. D}\ }\textbf {\bibinfo {volume} {112}},\ \bibinfo {pages} {023524} (\bibinfo {year} {2025})},\ \Eprint {https://arxiv.org/abs/2503.18120} {arXiv:2503.18120 [hep-ph]} \BibitemShut {NoStop}%
\bibitem [{\citenamefont {Mishra}\ \emph {et~al.}(2025)\citenamefont {Mishra}, \citenamefont {Matthewson}, \citenamefont {Sahni}, \citenamefont {Shafieloo},\ and\ \citenamefont {Shtanov}}]{Mishra:2025goj}%
  \BibitemOpen
  \bibfield  {author} {\bibinfo {author} {\bibfnamefont {S.~S.}\ \bibnamefont {Mishra}}, \bibinfo {author} {\bibfnamefont {W.~L.}\ \bibnamefont {Matthewson}}, \bibinfo {author} {\bibfnamefont {V.}~\bibnamefont {Sahni}}, \bibinfo {author} {\bibfnamefont {A.}~\bibnamefont {Shafieloo}},\ and\ \bibinfo {author} {\bibfnamefont {Y.}~\bibnamefont {Shtanov}},\ }\bibfield  {title} {\bibinfo {title} {{Braneworld Dark Energy in light of DESI DR2}},\ }\href@noop {} {\  (\bibinfo {year} {2025})},\ \Eprint {https://arxiv.org/abs/2507.07193} {arXiv:2507.07193 [astro-ph.CO]} \BibitemShut {NoStop}%
\bibitem [{\citenamefont {Bhattacharya}\ \emph {et~al.}(2024)\citenamefont {Bhattacharya}, \citenamefont {Borghetto}, \citenamefont {Malhotra}, \citenamefont {Parameswaran}, \citenamefont {Tasinato},\ and\ \citenamefont {Zavala}}]{Bhattacharya:2024hep}%
  \BibitemOpen
  \bibfield  {author} {\bibinfo {author} {\bibfnamefont {S.}~\bibnamefont {Bhattacharya}}, \bibinfo {author} {\bibfnamefont {G.}~\bibnamefont {Borghetto}}, \bibinfo {author} {\bibfnamefont {A.}~\bibnamefont {Malhotra}}, \bibinfo {author} {\bibfnamefont {S.}~\bibnamefont {Parameswaran}}, \bibinfo {author} {\bibfnamefont {G.}~\bibnamefont {Tasinato}},\ and\ \bibinfo {author} {\bibfnamefont {I.}~\bibnamefont {Zavala}},\ }\bibfield  {title} {\bibinfo {title} {{Cosmological constraints on curved quintessence}},\ }\href {https://doi.org/10.1088/1475-7516/2024/09/073} {\bibfield  {journal} {\bibinfo  {journal} {JCAP}\ }\textbf {\bibinfo {volume} {09}},\ \bibinfo {pages} {073}},\ \Eprint {https://arxiv.org/abs/2405.17396} {arXiv:2405.17396 [astro-ph.CO]} \BibitemShut {NoStop}%
\bibitem [{\citenamefont {Wang}\ \emph {et~al.}(2024)\citenamefont {Wang}, \citenamefont {Peng},\ and\ \citenamefont {Piao}}]{Wang:2024hwd}%
  \BibitemOpen
  \bibfield  {author} {\bibinfo {author} {\bibfnamefont {H.}~\bibnamefont {Wang}}, \bibinfo {author} {\bibfnamefont {Z.-Y.}\ \bibnamefont {Peng}},\ and\ \bibinfo {author} {\bibfnamefont {Y.-S.}\ \bibnamefont {Piao}},\ }\bibfield  {title} {\bibinfo {title} {{Can recent DESI BAO measurements accommodate a negative cosmological constant?}},\ }\href@noop {} {\  (\bibinfo {year} {2024})},\ \Eprint {https://arxiv.org/abs/2406.03395} {arXiv:2406.03395 [astro-ph.CO]} \BibitemShut {NoStop}%
\bibitem [{\citenamefont {G{\'o}mez-Valent}\ and\ \citenamefont {Gonz{\'a}lez-Fuentes}(2025)}]{Gomez-Valent:2025mfl}%
  \BibitemOpen
  \bibfield  {author} {\bibinfo {author} {\bibfnamefont {A.}~\bibnamefont {G{\'o}mez-Valent}}\ and\ \bibinfo {author} {\bibfnamefont {A.}~\bibnamefont {Gonz{\'a}lez-Fuentes}},\ }\bibfield  {title} {\bibinfo {title} {{Effective Phantom Divide Crossing with Standard and Negative Quintessence}},\ }\href@noop {} {\  (\bibinfo {year} {2025})},\ \Eprint {https://arxiv.org/abs/2508.00621} {arXiv:2508.00621 [astro-ph.CO]} \BibitemShut {NoStop}%
\bibitem [{\citenamefont {Cline}\ and\ \citenamefont {Muralidharan}(2025)}]{Cline:2025sbt}%
  \BibitemOpen
  \bibfield  {author} {\bibinfo {author} {\bibfnamefont {J.~M.}\ \bibnamefont {Cline}}\ and\ \bibinfo {author} {\bibfnamefont {V.}~\bibnamefont {Muralidharan}},\ }\bibfield  {title} {\bibinfo {title} {{Simple quintessence models in light of DESI-BAO observations}},\ }\href@noop {} {\  (\bibinfo {year} {2025})},\ \Eprint {https://arxiv.org/abs/2506.13047} {arXiv:2506.13047 [astro-ph.CO]} \BibitemShut {NoStop}%
\bibitem [{\citenamefont {Lin}\ \emph {et~al.}(2025)\citenamefont {Lin}, \citenamefont {Visinelli},\ and\ \citenamefont {Yanagida}}]{Lin:2025gne}%
  \BibitemOpen
  \bibfield  {author} {\bibinfo {author} {\bibfnamefont {W.}~\bibnamefont {Lin}}, \bibinfo {author} {\bibfnamefont {L.}~\bibnamefont {Visinelli}},\ and\ \bibinfo {author} {\bibfnamefont {T.~T.}\ \bibnamefont {Yanagida}},\ }\bibfield  {title} {\bibinfo {title} {{Testing Quintessence Axion Dark Energy with Recent Cosmological Results}},\ }\href@noop {} {\  (\bibinfo {year} {2025})},\ \Eprint {https://arxiv.org/abs/2504.17638} {arXiv:2504.17638 [astro-ph.CO]} \BibitemShut {NoStop}%
\bibitem [{\citenamefont {Goh}\ and\ \citenamefont {Taylor}(2025)}]{Goh:2025upc}%
  \BibitemOpen
  \bibfield  {author} {\bibinfo {author} {\bibfnamefont {L.~W.~K.}\ \bibnamefont {Goh}}\ and\ \bibinfo {author} {\bibfnamefont {A.~N.}\ \bibnamefont {Taylor}},\ }\bibfield  {title} {\bibinfo {title} {{Phantom Crossing with Quintom Models}},\ }\href@noop {} {\  (\bibinfo {year} {2025})},\ \Eprint {https://arxiv.org/abs/2509.12335} {arXiv:2509.12335 [astro-ph.CO]} \BibitemShut {NoStop}%
\bibitem [{\citenamefont {Hossain}\ and\ \citenamefont {Maqsood}(2025)}]{Hossain:2025grx}%
  \BibitemOpen
  \bibfield  {author} {\bibinfo {author} {\bibfnamefont {M.~W.}\ \bibnamefont {Hossain}}\ and\ \bibinfo {author} {\bibfnamefont {A.}~\bibnamefont {Maqsood}},\ }\bibfield  {title} {\bibinfo {title} {{Cosmological implications of Tracker scalar fields as dynamical dark energy}},\ }\href@noop {} {\  (\bibinfo {year} {2025})},\ \Eprint {https://arxiv.org/abs/2502.19274} {arXiv:2502.19274 [astro-ph.CO]} \BibitemShut {NoStop}%
\bibitem [{\citenamefont {Wolf}\ \emph {et~al.}(2025{\natexlab{a}})\citenamefont {Wolf}, \citenamefont {Ferreira},\ and\ \citenamefont {Garc\'\i{}a-Garc\'\i{}a}}]{Wolf:2024stt}%
  \BibitemOpen
  \bibfield  {author} {\bibinfo {author} {\bibfnamefont {W.~J.}\ \bibnamefont {Wolf}}, \bibinfo {author} {\bibfnamefont {P.~G.}\ \bibnamefont {Ferreira}},\ and\ \bibinfo {author} {\bibfnamefont {C.}~\bibnamefont {Garc\'\i{}a-Garc\'\i{}a}},\ }\bibfield  {title} {\bibinfo {title} {{Matching current observational constraints with nonminimally coupled dark energy}},\ }\href {https://doi.org/10.1103/PhysRevD.111.L041303} {\bibfield  {journal} {\bibinfo  {journal} {Phys. Rev. D}\ }\textbf {\bibinfo {volume} {111}},\ \bibinfo {pages} {L041303} (\bibinfo {year} {2025}{\natexlab{a}})},\ \Eprint {https://arxiv.org/abs/2409.17019} {arXiv:2409.17019 [astro-ph.CO]} \BibitemShut {NoStop}%
\bibitem [{\citenamefont {Wolf}\ \emph {et~al.}(2025{\natexlab{b}})\citenamefont {Wolf}, \citenamefont {Garc{\'\i}a-Garc{\'\i}a}, \citenamefont {Anton},\ and\ \citenamefont {Ferreira}}]{Wolf:2025jed}%
  \BibitemOpen
  \bibfield  {author} {\bibinfo {author} {\bibfnamefont {W.~J.}\ \bibnamefont {Wolf}}, \bibinfo {author} {\bibfnamefont {C.}~\bibnamefont {Garc{\'\i}a-Garc{\'\i}a}}, \bibinfo {author} {\bibfnamefont {T.}~\bibnamefont {Anton}},\ and\ \bibinfo {author} {\bibfnamefont {P.~G.}\ \bibnamefont {Ferreira}},\ }\bibfield  {title} {\bibinfo {title} {{Assessing Cosmological Evidence for Nonminimal Coupling}},\ }\href {https://doi.org/10.1103/jysf-k72m} {\bibfield  {journal} {\bibinfo  {journal} {Phys. Rev. Lett.}\ }\textbf {\bibinfo {volume} {135}},\ \bibinfo {pages} {081001} (\bibinfo {year} {2025}{\natexlab{b}})},\ \Eprint {https://arxiv.org/abs/2504.07679} {arXiv:2504.07679 [astro-ph.CO]} \BibitemShut {NoStop}%
\bibitem [{\citenamefont {Ye}\ \emph {et~al.}(2025)\citenamefont {Ye}, \citenamefont {Martinelli}, \citenamefont {Hu},\ and\ \citenamefont {Silvestri}}]{Ye:2024ywg}%
  \BibitemOpen
  \bibfield  {author} {\bibinfo {author} {\bibfnamefont {G.}~\bibnamefont {Ye}}, \bibinfo {author} {\bibfnamefont {M.}~\bibnamefont {Martinelli}}, \bibinfo {author} {\bibfnamefont {B.}~\bibnamefont {Hu}},\ and\ \bibinfo {author} {\bibfnamefont {A.}~\bibnamefont {Silvestri}},\ }\bibfield  {title} {\bibinfo {title} {{Hints of Nonminimally Coupled Gravity in DESI 2024 Baryon Acoustic Oscillation Measurements}},\ }\href {https://doi.org/10.1103/PhysRevLett.134.181002} {\bibfield  {journal} {\bibinfo  {journal} {Phys. Rev. Lett.}\ }\textbf {\bibinfo {volume} {134}},\ \bibinfo {pages} {181002} (\bibinfo {year} {2025})},\ \Eprint {https://arxiv.org/abs/2407.15832} {arXiv:2407.15832 [astro-ph.CO]} \BibitemShut {NoStop}%
\bibitem [{\citenamefont {Chudaykin}\ and\ \citenamefont {Kunz}(2024)}]{Chudaykin:2024gol}%
  \BibitemOpen
  \bibfield  {author} {\bibinfo {author} {\bibfnamefont {A.}~\bibnamefont {Chudaykin}}\ and\ \bibinfo {author} {\bibfnamefont {M.}~\bibnamefont {Kunz}},\ }\bibfield  {title} {\bibinfo {title} {{Modified gravity interpretation of the evolving dark energy in light of DESI data}},\ }\href {https://doi.org/10.1103/PhysRevD.110.123524} {\bibfield  {journal} {\bibinfo  {journal} {Phys. Rev. D}\ }\textbf {\bibinfo {volume} {110}},\ \bibinfo {pages} {123524} (\bibinfo {year} {2024})},\ \Eprint {https://arxiv.org/abs/2407.02558} {arXiv:2407.02558 [astro-ph.CO]} \BibitemShut {NoStop}%
\bibitem [{\citenamefont {Ye}\ and\ \citenamefont {Cai}(2025)}]{Ye:2025ulq}%
  \BibitemOpen
  \bibfield  {author} {\bibinfo {author} {\bibfnamefont {G.}~\bibnamefont {Ye}}\ and\ \bibinfo {author} {\bibfnamefont {Y.}~\bibnamefont {Cai}},\ }\bibfield  {title} {\bibinfo {title} {{NEC violation and ''beyond Horndeski'' physics in light of DESI DR2}},\ }\href@noop {} {\  (\bibinfo {year} {2025})},\ \Eprint {https://arxiv.org/abs/2503.22515} {arXiv:2503.22515 [gr-qc]} \BibitemShut {NoStop}%
\bibitem [{\citenamefont {Pan}\ and\ \citenamefont {Ye}(2025)}]{Pan:2025psn}%
  \BibitemOpen
  \bibfield  {author} {\bibinfo {author} {\bibfnamefont {J.}~\bibnamefont {Pan}}\ and\ \bibinfo {author} {\bibfnamefont {G.}~\bibnamefont {Ye}},\ }\bibfield  {title} {\bibinfo {title} {{Non-minimally coupled gravity constraints from DESI DR2 data}},\ }\href@noop {} {\  (\bibinfo {year} {2025})},\ \Eprint {https://arxiv.org/abs/2503.19898} {arXiv:2503.19898 [astro-ph.CO]} \BibitemShut {NoStop}%
\bibitem [{\citenamefont {Goldstein}\ \emph {et~al.}(2025)\citenamefont {Goldstein}, \citenamefont {Celoria},\ and\ \citenamefont {Schmidt}}]{Goldstein:2025epp}%
  \BibitemOpen
  \bibfield  {author} {\bibinfo {author} {\bibfnamefont {S.}~\bibnamefont {Goldstein}}, \bibinfo {author} {\bibfnamefont {M.}~\bibnamefont {Celoria}},\ and\ \bibinfo {author} {\bibfnamefont {F.}~\bibnamefont {Schmidt}},\ }\bibfield  {title} {\bibinfo {title} {{Monodromic Dark Energy and DESI}},\ }\href@noop {} {\  (\bibinfo {year} {2025})},\ \Eprint {https://arxiv.org/abs/2507.16970} {arXiv:2507.16970 [astro-ph.CO]} \BibitemShut {NoStop}%
\bibitem [{\citenamefont {Cai}\ \emph {et~al.}(2025)\citenamefont {Cai}, \citenamefont {Ren}, \citenamefont {Qiu}, \citenamefont {Li},\ and\ \citenamefont {Zhang}}]{Cai:2025mas}%
  \BibitemOpen
  \bibfield  {author} {\bibinfo {author} {\bibfnamefont {Y.}~\bibnamefont {Cai}}, \bibinfo {author} {\bibfnamefont {X.}~\bibnamefont {Ren}}, \bibinfo {author} {\bibfnamefont {T.}~\bibnamefont {Qiu}}, \bibinfo {author} {\bibfnamefont {M.}~\bibnamefont {Li}},\ and\ \bibinfo {author} {\bibfnamefont {X.}~\bibnamefont {Zhang}},\ }\bibfield  {title} {\bibinfo {title} {{The Quintom theory of dark energy after DESI DR2}},\ }\href@noop {} {\  (\bibinfo {year} {2025})},\ \Eprint {https://arxiv.org/abs/2505.24732} {arXiv:2505.24732 [astro-ph.CO]} \BibitemShut {NoStop}%
\bibitem [{\citenamefont {Adam}\ \emph {et~al.}(2025)\citenamefont {Adam}, \citenamefont {Hertzberg}, \citenamefont {Jim{\'e}nez-Aguilar},\ and\ \citenamefont {Khan}}]{Adam:2025kve}%
  \BibitemOpen
  \bibfield  {author} {\bibinfo {author} {\bibfnamefont {H.}~\bibnamefont {Adam}}, \bibinfo {author} {\bibfnamefont {M.~P.}\ \bibnamefont {Hertzberg}}, \bibinfo {author} {\bibfnamefont {D.}~\bibnamefont {Jim{\'e}nez-Aguilar}},\ and\ \bibinfo {author} {\bibfnamefont {I.}~\bibnamefont {Khan}},\ }\bibfield  {title} {\bibinfo {title} {{Comparing Minimal and Non-Minimal Quintessence Models to 2025 DESI Data}},\ }\href@noop {} {\  (\bibinfo {year} {2025})},\ \Eprint {https://arxiv.org/abs/2509.13302} {arXiv:2509.13302 [astro-ph.CO]} \BibitemShut {NoStop}%
\bibitem [{\citenamefont {Lu}\ \emph {et~al.}(2025)\citenamefont {Lu}, \citenamefont {Simon},\ and\ \citenamefont {Zhang}}]{Lu:2025gki}%
  \BibitemOpen
  \bibfield  {author} {\bibinfo {author} {\bibfnamefont {Z.}~\bibnamefont {Lu}}, \bibinfo {author} {\bibfnamefont {T.}~\bibnamefont {Simon}},\ and\ \bibinfo {author} {\bibfnamefont {P.}~\bibnamefont {Zhang}},\ }\bibfield  {title} {\bibinfo {title} {{Preference for evolving dark energy in light of the galaxy bispectrum}},\ }\href@noop {} {\  (\bibinfo {year} {2025})},\ \Eprint {https://arxiv.org/abs/2503.04602} {arXiv:2503.04602 [astro-ph.CO]} \BibitemShut {NoStop}%
\bibitem [{\citenamefont {Cataneo}\ and\ \citenamefont {Koyama}(2025)}]{Cataneo:2025vae}%
  \BibitemOpen
  \bibfield  {author} {\bibinfo {author} {\bibfnamefont {M.}~\bibnamefont {Cataneo}}\ and\ \bibinfo {author} {\bibfnamefont {K.}~\bibnamefont {Koyama}},\ }\bibfield  {title} {\bibinfo {title} {{Non-parametric exploration of minimally coupled gravity with phantom crossing}},\ }\href@noop {} {\  (\bibinfo {year} {2025})},\ \Eprint {https://arxiv.org/abs/2512.13691} {arXiv:2512.13691 [astro-ph.CO]} \BibitemShut {NoStop}%
\bibitem [{\citenamefont {Khoury}\ \emph {et~al.}(2025)\citenamefont {Khoury}, \citenamefont {Lin},\ and\ \citenamefont {Trodden}}]{Khoury:2025txd}%
  \BibitemOpen
  \bibfield  {author} {\bibinfo {author} {\bibfnamefont {J.}~\bibnamefont {Khoury}}, \bibinfo {author} {\bibfnamefont {M.-X.}\ \bibnamefont {Lin}},\ and\ \bibinfo {author} {\bibfnamefont {M.}~\bibnamefont {Trodden}},\ }\bibfield  {title} {\bibinfo {title} {{Apparent $w<-1$ and a Lower $S_8$ from Dark Axion and Dark Baryons Interactions}},\ }\href@noop {} {\  (\bibinfo {year} {2025})},\ \Eprint {https://arxiv.org/abs/2503.16415} {arXiv:2503.16415 [astro-ph.CO]} \BibitemShut {NoStop}%
\bibitem [{\citenamefont {Chakraborty}\ \emph {et~al.}(2025)\citenamefont {Chakraborty}, \citenamefont {Chanda}, \citenamefont {Das},\ and\ \citenamefont {Dutta}}]{Chakraborty:2025syu}%
  \BibitemOpen
  \bibfield  {author} {\bibinfo {author} {\bibfnamefont {A.}~\bibnamefont {Chakraborty}}, \bibinfo {author} {\bibfnamefont {P.~K.}\ \bibnamefont {Chanda}}, \bibinfo {author} {\bibfnamefont {S.}~\bibnamefont {Das}},\ and\ \bibinfo {author} {\bibfnamefont {K.}~\bibnamefont {Dutta}},\ }\bibfield  {title} {\bibinfo {title} {{DESI results: Hint towards coupled dark matter and dark energy}},\ }\href@noop {} {\  (\bibinfo {year} {2025})},\ \Eprint {https://arxiv.org/abs/2503.10806} {arXiv:2503.10806 [astro-ph.CO]} \BibitemShut {NoStop}%
\bibitem [{\citenamefont {van~der Westhuizen}\ \emph {et~al.}(2025{\natexlab{a}})\citenamefont {van~der Westhuizen}, \citenamefont {Abebe},\ and\ \citenamefont {Di~Valentino}}]{vanderWesthuizen:2025vcb}%
  \BibitemOpen
  \bibfield  {author} {\bibinfo {author} {\bibfnamefont {M.}~\bibnamefont {van~der Westhuizen}}, \bibinfo {author} {\bibfnamefont {A.}~\bibnamefont {Abebe}},\ and\ \bibinfo {author} {\bibfnamefont {E.}~\bibnamefont {Di~Valentino}},\ }\bibfield  {title} {\bibinfo {title} {{I. Linear Interacting Dark Energy: Analytical Solutions and Theoretical Pathologies}},\ }\href@noop {} {\  (\bibinfo {year} {2025}{\natexlab{a}})},\ \Eprint {https://arxiv.org/abs/2509.04495} {arXiv:2509.04495 [gr-qc]} \BibitemShut {NoStop}%
\bibitem [{\citenamefont {van~der Westhuizen}\ \emph {et~al.}(2025{\natexlab{b}})\citenamefont {van~der Westhuizen}, \citenamefont {Abebe},\ and\ \citenamefont {Di~Valentino}}]{vanderWesthuizen:2025mnw}%
  \BibitemOpen
  \bibfield  {author} {\bibinfo {author} {\bibfnamefont {M.}~\bibnamefont {van~der Westhuizen}}, \bibinfo {author} {\bibfnamefont {A.}~\bibnamefont {Abebe}},\ and\ \bibinfo {author} {\bibfnamefont {E.}~\bibnamefont {Di~Valentino}},\ }\bibfield  {title} {\bibinfo {title} {{II. Non-Linear Interacting Dark Energy: Analytical Solutions and Theoretical Pathologies}},\ }\href@noop {} {\  (\bibinfo {year} {2025}{\natexlab{b}})},\ \Eprint {https://arxiv.org/abs/2509.04494} {arXiv:2509.04494 [gr-qc]} \BibitemShut {NoStop}%
\bibitem [{\citenamefont {Shah}\ \emph {et~al.}(2025{\natexlab{a}})\citenamefont {Shah}, \citenamefont {Mukherjee},\ and\ \citenamefont {Pal}}]{Shah:2025ayl}%
  \BibitemOpen
  \bibfield  {author} {\bibinfo {author} {\bibfnamefont {R.}~\bibnamefont {Shah}}, \bibinfo {author} {\bibfnamefont {P.}~\bibnamefont {Mukherjee}},\ and\ \bibinfo {author} {\bibfnamefont {S.}~\bibnamefont {Pal}},\ }\bibfield  {title} {\bibinfo {title} {{Interacting dark sectors in light of DESI DR2}},\ }\href@noop {} {\  (\bibinfo {year} {2025}{\natexlab{a}})},\ \Eprint {https://arxiv.org/abs/2503.21652} {arXiv:2503.21652 [astro-ph.CO]} \BibitemShut {NoStop}%
\bibitem [{\citenamefont {Braglia}\ \emph {et~al.}(2025)\citenamefont {Braglia}, \citenamefont {Chen},\ and\ \citenamefont {Loeb}}]{Braglia:2025gdo}%
  \BibitemOpen
  \bibfield  {author} {\bibinfo {author} {\bibfnamefont {M.}~\bibnamefont {Braglia}}, \bibinfo {author} {\bibfnamefont {X.}~\bibnamefont {Chen}},\ and\ \bibinfo {author} {\bibfnamefont {A.}~\bibnamefont {Loeb}},\ }\bibfield  {title} {\bibinfo {title} {{Exotic Dark Matter and the DESI Anomaly}},\ }\href@noop {} {\  (\bibinfo {year} {2025})},\ \Eprint {https://arxiv.org/abs/2507.13925} {arXiv:2507.13925 [astro-ph.CO]} \BibitemShut {NoStop}%
\bibitem [{\citenamefont {Chen}\ \emph {et~al.}(2025)\citenamefont {Chen}, \citenamefont {Cline}, \citenamefont {Muralidharan},\ and\ \citenamefont {Salewicz}}]{Chen:2025ywv}%
  \BibitemOpen
  \bibfield  {author} {\bibinfo {author} {\bibfnamefont {R.}~\bibnamefont {Chen}}, \bibinfo {author} {\bibfnamefont {J.~M.}\ \bibnamefont {Cline}}, \bibinfo {author} {\bibfnamefont {V.}~\bibnamefont {Muralidharan}},\ and\ \bibinfo {author} {\bibfnamefont {B.}~\bibnamefont {Salewicz}},\ }\bibfield  {title} {\bibinfo {title} {{Quintessential dark energy crossing the phantom divide}},\ }\href@noop {} {\  (\bibinfo {year} {2025})},\ \Eprint {https://arxiv.org/abs/2508.19101} {arXiv:2508.19101 [astro-ph.CO]} \BibitemShut {NoStop}%
\bibitem [{\citenamefont {van~der Westhuizen}\ \emph {et~al.}(2025{\natexlab{c}})\citenamefont {van~der Westhuizen}, \citenamefont {Abebe},\ and\ \citenamefont {Di~Valentino}}]{vanderWesthuizen:2025rip}%
  \BibitemOpen
  \bibfield  {author} {\bibinfo {author} {\bibfnamefont {M.}~\bibnamefont {van~der Westhuizen}}, \bibinfo {author} {\bibfnamefont {A.}~\bibnamefont {Abebe}},\ and\ \bibinfo {author} {\bibfnamefont {E.}~\bibnamefont {Di~Valentino}},\ }\bibfield  {title} {\bibinfo {title} {{III. Interacting Dark Energy: Summary of Models, Pathologies, and Constraints}},\ }\href@noop {} {\  (\bibinfo {year} {2025}{\natexlab{c}})},\ \Eprint {https://arxiv.org/abs/2509.04496} {arXiv:2509.04496 [gr-qc]} \BibitemShut {NoStop}%
\bibitem [{\citenamefont {Andriot}(2025)}]{Andriot:2025los}%
  \BibitemOpen
  \bibfield  {author} {\bibinfo {author} {\bibfnamefont {D.}~\bibnamefont {Andriot}},\ }\bibfield  {title} {\bibinfo {title} {{Phantom matters}},\ }\href {https://doi.org/10.1016/j.dark.2025.102000} {\bibfield  {journal} {\bibinfo  {journal} {Phys. Dark Univ.}\ }\textbf {\bibinfo {volume} {49}},\ \bibinfo {pages} {102000} (\bibinfo {year} {2025})},\ \Eprint {https://arxiv.org/abs/2505.10410} {arXiv:2505.10410 [hep-th]} \BibitemShut {NoStop}%
\bibitem [{\citenamefont {Li}\ \emph {et~al.}(2024)\citenamefont {Li}, \citenamefont {Wu}, \citenamefont {Du}, \citenamefont {Jin}, \citenamefont {Li}, \citenamefont {Zhang},\ and\ \citenamefont {Zhang}}]{Li:2024qso}%
  \BibitemOpen
  \bibfield  {author} {\bibinfo {author} {\bibfnamefont {T.-N.}\ \bibnamefont {Li}}, \bibinfo {author} {\bibfnamefont {P.-J.}\ \bibnamefont {Wu}}, \bibinfo {author} {\bibfnamefont {G.-H.}\ \bibnamefont {Du}}, \bibinfo {author} {\bibfnamefont {S.-J.}\ \bibnamefont {Jin}}, \bibinfo {author} {\bibfnamefont {H.-L.}\ \bibnamefont {Li}}, \bibinfo {author} {\bibfnamefont {J.-F.}\ \bibnamefont {Zhang}},\ and\ \bibinfo {author} {\bibfnamefont {X.}~\bibnamefont {Zhang}},\ }\bibfield  {title} {\bibinfo {title} {{Constraints on Interacting Dark Energy Models from the DESI Baryon Acoustic Oscillation and DES Supernovae Data}},\ }\href {https://doi.org/10.3847/1538-4357/ad87f0} {\bibfield  {journal} {\bibinfo  {journal} {Astrophys. J.}\ }\textbf {\bibinfo {volume} {976}},\ \bibinfo {pages} {1} (\bibinfo {year} {2024})},\ \Eprint {https://arxiv.org/abs/2407.14934} {arXiv:2407.14934 [astro-ph.CO]} \BibitemShut {NoStop}%
\bibitem [{\citenamefont {Wolf}\ \emph {et~al.}(2025{\natexlab{c}})\citenamefont {Wolf}, \citenamefont {Garc\'\i{}a-Garc\'\i{}a},\ and\ \citenamefont {Ferreira}}]{Wolf:2025jlc}%
  \BibitemOpen
  \bibfield  {author} {\bibinfo {author} {\bibfnamefont {W.~J.}\ \bibnamefont {Wolf}}, \bibinfo {author} {\bibfnamefont {C.}~\bibnamefont {Garc\'\i{}a-Garc\'\i{}a}},\ and\ \bibinfo {author} {\bibfnamefont {P.~G.}\ \bibnamefont {Ferreira}},\ }\bibfield  {title} {\bibinfo {title} {{Robustness of dark energy phenomenology across different parameterizations}},\ }\href {https://doi.org/10.1088/1475-7516/2025/05/034} {\bibfield  {journal} {\bibinfo  {journal} {JCAP}\ }\textbf {\bibinfo {volume} {05}},\ \bibinfo {pages} {034}},\ \Eprint {https://arxiv.org/abs/2502.04929} {arXiv:2502.04929 [astro-ph.CO]} \BibitemShut {NoStop}%
\bibitem [{\citenamefont {Shlivko}\ \emph {et~al.}(2025)\citenamefont {Shlivko}, \citenamefont {Steinhardt},\ and\ \citenamefont {Steinhardt}}]{Shlivko:2025fgv}%
  \BibitemOpen
  \bibfield  {author} {\bibinfo {author} {\bibfnamefont {D.}~\bibnamefont {Shlivko}}, \bibinfo {author} {\bibfnamefont {P.~J.}\ \bibnamefont {Steinhardt}},\ and\ \bibinfo {author} {\bibfnamefont {C.~L.}\ \bibnamefont {Steinhardt}},\ }\bibfield  {title} {\bibinfo {title} {{Optimal parameterizations for observational constraints on thawing dark energy}},\ }\href@noop {} {\  (\bibinfo {year} {2025})},\ \Eprint {https://arxiv.org/abs/2504.02028} {arXiv:2504.02028 [astro-ph.CO]} \BibitemShut {NoStop}%
\bibitem [{\citenamefont {Giar\`e}\ \emph {et~al.}(2024)\citenamefont {Giar\`e}, \citenamefont {Najafi}, \citenamefont {Pan}, \citenamefont {Di~Valentino},\ and\ \citenamefont {Firouzjaee}}]{Giare:2024gpk}%
  \BibitemOpen
  \bibfield  {author} {\bibinfo {author} {\bibfnamefont {W.}~\bibnamefont {Giar\`e}}, \bibinfo {author} {\bibfnamefont {M.}~\bibnamefont {Najafi}}, \bibinfo {author} {\bibfnamefont {S.}~\bibnamefont {Pan}}, \bibinfo {author} {\bibfnamefont {E.}~\bibnamefont {Di~Valentino}},\ and\ \bibinfo {author} {\bibfnamefont {J.~T.}\ \bibnamefont {Firouzjaee}},\ }\bibfield  {title} {\bibinfo {title} {{Robust preference for Dynamical Dark Energy in DESI BAO and SN measurements}},\ }\href {https://doi.org/10.1088/1475-7516/2024/10/035} {\bibfield  {journal} {\bibinfo  {journal} {JCAP}\ }\textbf {\bibinfo {volume} {10}},\ \bibinfo {pages} {035}},\ \Eprint {https://arxiv.org/abs/2407.16689} {arXiv:2407.16689 [astro-ph.CO]} \BibitemShut {NoStop}%
\bibitem [{\citenamefont {Giani}\ \emph {et~al.}(2025)\citenamefont {Giani}, \citenamefont {Von~Marttens},\ and\ \citenamefont {Piattella}}]{Giani:2025hhs}%
  \BibitemOpen
  \bibfield  {author} {\bibinfo {author} {\bibfnamefont {L.}~\bibnamefont {Giani}}, \bibinfo {author} {\bibfnamefont {R.}~\bibnamefont {Von~Marttens}},\ and\ \bibinfo {author} {\bibfnamefont {O.~F.}\ \bibnamefont {Piattella}},\ }\bibfield  {title} {\bibinfo {title} {{The matter with(in) CPL}},\ }\href@noop {} {\  (\bibinfo {year} {2025})},\ \Eprint {https://arxiv.org/abs/2505.08467} {arXiv:2505.08467 [astro-ph.CO]} \BibitemShut {NoStop}%
\bibitem [{\citenamefont {Dinda}\ and\ \citenamefont {Maartens}(2025{\natexlab{a}})}]{Dinda:2025iaq}%
  \BibitemOpen
  \bibfield  {author} {\bibinfo {author} {\bibfnamefont {B.~R.}\ \bibnamefont {Dinda}}\ and\ \bibinfo {author} {\bibfnamefont {R.}~\bibnamefont {Maartens}},\ }\bibfield  {title} {\bibinfo {title} {{Physical vs phantom dark energy after DESI: thawing quintessence in a curved background}}\ }\href {https://doi.org/10.1093/mnrasl/slaf063} {10.1093/mnrasl/slaf063} (\bibinfo {year} {2025}{\natexlab{a}}),\ \Eprint {https://arxiv.org/abs/2504.15190} {arXiv:2504.15190 [astro-ph.CO]} \BibitemShut {NoStop}%
\bibitem [{\citenamefont {Wu}(2025)}]{Wu:2025wyk}%
  \BibitemOpen
  \bibfield  {author} {\bibinfo {author} {\bibfnamefont {P.-J.}\ \bibnamefont {Wu}},\ }\bibfield  {title} {\bibinfo {title} {{Comparison of dark energy models using late-universe observations}},\ }\href {https://doi.org/10.1103/mcgb-ntwr} {\bibfield  {journal} {\bibinfo  {journal} {Phys. Rev. D}\ }\textbf {\bibinfo {volume} {112}},\ \bibinfo {pages} {043527} (\bibinfo {year} {2025})},\ \Eprint {https://arxiv.org/abs/2504.09054} {arXiv:2504.09054 [astro-ph.CO]} \BibitemShut {NoStop}%
\bibitem [{\citenamefont {Yang}(2025)}]{Yang:2025oax}%
  \BibitemOpen
  \bibfield  {author} {\bibinfo {author} {\bibfnamefont {Y.}~\bibnamefont {Yang}},\ }\bibfield  {title} {\bibinfo {title} {{Constraining deviations from $\Lambda$CDM in the Hubble expansion rate}},\ }\href@noop {} {\  (\bibinfo {year} {2025})},\ \Eprint {https://arxiv.org/abs/2508.17848} {arXiv:2508.17848 [astro-ph.CO]} \BibitemShut {NoStop}%
\bibitem [{\citenamefont {Lee}\ \emph {et~al.}(2025)\citenamefont {Lee}, \citenamefont {Yang}, \citenamefont {Di~Valentino}, \citenamefont {Pan},\ and\ \citenamefont {van~de Bruck}}]{Lee:2025pzo}%
  \BibitemOpen
  \bibfield  {author} {\bibinfo {author} {\bibfnamefont {D.~H.}\ \bibnamefont {Lee}}, \bibinfo {author} {\bibfnamefont {W.}~\bibnamefont {Yang}}, \bibinfo {author} {\bibfnamefont {E.}~\bibnamefont {Di~Valentino}}, \bibinfo {author} {\bibfnamefont {S.}~\bibnamefont {Pan}},\ and\ \bibinfo {author} {\bibfnamefont {C.}~\bibnamefont {van~de Bruck}},\ }\bibfield  {title} {\bibinfo {title} {{The Shape of Dark Energy: Constraining Its Evolution with a General Parametrization}},\ }\href@noop {} {\  (\bibinfo {year} {2025})},\ \Eprint {https://arxiv.org/abs/2507.11432} {arXiv:2507.11432 [astro-ph.CO]} \BibitemShut {NoStop}%
\bibitem [{\citenamefont {Li}\ \emph {et~al.}(2025)\citenamefont {Li}, \citenamefont {Du}, \citenamefont {Zhou}, \citenamefont {Li}, \citenamefont {Zhang},\ and\ \citenamefont {Zhang}}]{Li:2025vuh}%
  \BibitemOpen
  \bibfield  {author} {\bibinfo {author} {\bibfnamefont {T.-N.}\ \bibnamefont {Li}}, \bibinfo {author} {\bibfnamefont {G.-H.}\ \bibnamefont {Du}}, \bibinfo {author} {\bibfnamefont {S.-H.}\ \bibnamefont {Zhou}}, \bibinfo {author} {\bibfnamefont {Y.-H.}\ \bibnamefont {Li}}, \bibinfo {author} {\bibfnamefont {J.-F.}\ \bibnamefont {Zhang}},\ and\ \bibinfo {author} {\bibfnamefont {X.}~\bibnamefont {Zhang}},\ }\bibfield  {title} {\bibinfo {title} {{Robust evidence for dynamical dark energy in light of DESI DR2 and joint ACT, SPT, and Planck data}},\ }\href@noop {} {\  (\bibinfo {year} {2025})},\ \Eprint {https://arxiv.org/abs/2511.22512} {arXiv:2511.22512 [astro-ph.CO]} \BibitemShut {NoStop}%
\bibitem [{\citenamefont {Dinda}\ and\ \citenamefont {Maartens}(2025{\natexlab{b}})}]{Dinda:2024ktd}%
  \BibitemOpen
  \bibfield  {author} {\bibinfo {author} {\bibfnamefont {B.~R.}\ \bibnamefont {Dinda}}\ and\ \bibinfo {author} {\bibfnamefont {R.}~\bibnamefont {Maartens}},\ }\bibfield  {title} {\bibinfo {title} {{Model-agnostic assessment of dark energy after DESI DR1 BAO}},\ }\href {https://doi.org/10.1088/1475-7516/2025/01/120} {\bibfield  {journal} {\bibinfo  {journal} {JCAP}\ }\textbf {\bibinfo {volume} {01}},\ \bibinfo {pages} {120}},\ \Eprint {https://arxiv.org/abs/2407.17252} {arXiv:2407.17252 [astro-ph.CO]} \BibitemShut {NoStop}%
\bibitem [{\citenamefont {Calderon}\ \emph {et~al.}(2024)\citenamefont {Calderon} \emph {et~al.}}]{DESI:2024aqx}%
  \BibitemOpen
  \bibfield  {author} {\bibinfo {author} {\bibfnamefont {R.}~\bibnamefont {Calderon}} \emph {et~al.} (\bibinfo {collaboration} {DESI}),\ }\bibfield  {title} {\bibinfo {title} {{DESI 2024: Reconstructing Dark Energy using Crossing Statistics with DESI DR1 BAO data}},\ }\href@noop {} {\  (\bibinfo {year} {2024})},\ \Eprint {https://arxiv.org/abs/2405.04216} {arXiv:2405.04216 [astro-ph.CO]} \BibitemShut {NoStop}%
\bibitem [{\citenamefont {Jiang}\ \emph {et~al.}(2024)\citenamefont {Jiang}, \citenamefont {Pedrotti}, \citenamefont {da~Costa},\ and\ \citenamefont {Vagnozzi}}]{Jiang:2024xnu}%
  \BibitemOpen
  \bibfield  {author} {\bibinfo {author} {\bibfnamefont {J.-Q.}\ \bibnamefont {Jiang}}, \bibinfo {author} {\bibfnamefont {D.}~\bibnamefont {Pedrotti}}, \bibinfo {author} {\bibfnamefont {S.~S.}\ \bibnamefont {da~Costa}},\ and\ \bibinfo {author} {\bibfnamefont {S.}~\bibnamefont {Vagnozzi}},\ }\bibfield  {title} {\bibinfo {title} {{Nonparametric late-time expansion history reconstruction and implications for the Hubble tension in light of recent DESI and type Ia supernovae data}},\ }\href {https://doi.org/10.1103/PhysRevD.110.123519} {\bibfield  {journal} {\bibinfo  {journal} {Phys. Rev. D}\ }\textbf {\bibinfo {volume} {110}},\ \bibinfo {pages} {123519} (\bibinfo {year} {2024})},\ \Eprint {https://arxiv.org/abs/2408.02365} {arXiv:2408.02365 [astro-ph.CO]} \BibitemShut {NoStop}%
\bibitem [{\citenamefont {Berti}\ \emph {et~al.}(2025)\citenamefont {Berti}, \citenamefont {Bellini}, \citenamefont {Bonvin}, \citenamefont {Kunz}, \citenamefont {Viel},\ and\ \citenamefont {Zumalacarregui}}]{Berti:2025phi}%
  \BibitemOpen
  \bibfield  {author} {\bibinfo {author} {\bibfnamefont {M.}~\bibnamefont {Berti}}, \bibinfo {author} {\bibfnamefont {E.}~\bibnamefont {Bellini}}, \bibinfo {author} {\bibfnamefont {C.}~\bibnamefont {Bonvin}}, \bibinfo {author} {\bibfnamefont {M.}~\bibnamefont {Kunz}}, \bibinfo {author} {\bibfnamefont {M.}~\bibnamefont {Viel}},\ and\ \bibinfo {author} {\bibfnamefont {M.}~\bibnamefont {Zumalacarregui}},\ }\bibfield  {title} {\bibinfo {title} {{Reconstructing the dark energy density in light of DESI BAO observations}},\ }\href@noop {} {\  (\bibinfo {year} {2025})},\ \Eprint {https://arxiv.org/abs/2503.13198} {arXiv:2503.13198 [astro-ph.CO]} \BibitemShut {NoStop}%
\bibitem [{\citenamefont {Li}\ and\ \citenamefont {Wang}(2025)}]{Li:2025ops}%
  \BibitemOpen
  \bibfield  {author} {\bibinfo {author} {\bibfnamefont {J.-X.}\ \bibnamefont {Li}}\ and\ \bibinfo {author} {\bibfnamefont {S.}~\bibnamefont {Wang}},\ }\bibfield  {title} {\bibinfo {title} {{Reconstructing dark energy with model independent methods after DESI DR2 BAO}},\ }\href@noop {} {\  (\bibinfo {year} {2025})},\ \Eprint {https://arxiv.org/abs/2506.22953} {arXiv:2506.22953 [astro-ph.CO]} \BibitemShut {NoStop}%
\bibitem [{\citenamefont {Efstathiou}(2024)}]{Efstathiou_2024}%
  \BibitemOpen
  \bibfield  {author} {\bibinfo {author} {\bibfnamefont {G.}~\bibnamefont {Efstathiou}},\ }\href {https://arxiv.org/abs/2408.07175} {\bibinfo {title} {Evolving dark energy or supernovae systematics?}} (\bibinfo {year} {2024}),\ \Eprint {https://arxiv.org/abs/2408.07175} {arXiv:2408.07175 [astro-ph.CO]} \BibitemShut {NoStop}%
\bibitem [{\citenamefont {Efstathiou}(2025)}]{Efstathiou:2025tie}%
  \BibitemOpen
  \bibfield  {author} {\bibinfo {author} {\bibfnamefont {G.}~\bibnamefont {Efstathiou}},\ }\bibfield  {title} {\bibinfo {title} {{Baryon Acoustic Oscillations from a Different Angle}},\ }\href@noop {} {\  (\bibinfo {year} {2025})},\ \Eprint {https://arxiv.org/abs/2505.02658} {arXiv:2505.02658 [astro-ph.CO]} \BibitemShut {NoStop}%
\bibitem [{\citenamefont {Carloni}\ \emph {et~al.}(2025)\citenamefont {Carloni}, \citenamefont {Luongo},\ and\ \citenamefont {Muccino}}]{Carloni:2024zpl}%
  \BibitemOpen
  \bibfield  {author} {\bibinfo {author} {\bibfnamefont {Y.}~\bibnamefont {Carloni}}, \bibinfo {author} {\bibfnamefont {O.}~\bibnamefont {Luongo}},\ and\ \bibinfo {author} {\bibfnamefont {M.}~\bibnamefont {Muccino}},\ }\bibfield  {title} {\bibinfo {title} {{Does dark energy really revive using DESI 2024 data?}},\ }\href {https://doi.org/10.1103/PhysRevD.111.023512} {\bibfield  {journal} {\bibinfo  {journal} {Phys. Rev. D}\ }\textbf {\bibinfo {volume} {111}},\ \bibinfo {pages} {023512} (\bibinfo {year} {2025})},\ \Eprint {https://arxiv.org/abs/2404.12068} {arXiv:2404.12068 [astro-ph.CO]} \BibitemShut {NoStop}%
\bibitem [{\citenamefont {Vincenzi}\ \emph {et~al.}(2025)\citenamefont {Vincenzi} \emph {et~al.}}]{DES:2025tir}%
  \BibitemOpen
  \bibfield  {author} {\bibinfo {author} {\bibfnamefont {M.}~\bibnamefont {Vincenzi}} \emph {et~al.} (\bibinfo {collaboration} {DES}),\ }\bibfield  {title} {\bibinfo {title} {{Comparing the DES-SN5YR and Pantheon+ SN cosmology analyses: Investigation based on ''Evolving Dark Energy or Supernovae systematics?''}},\ }\href@noop {} {\  (\bibinfo {year} {2025})},\ \Eprint {https://arxiv.org/abs/2501.06664} {arXiv:2501.06664 [astro-ph.CO]} \BibitemShut {NoStop}%
\bibitem [{\citenamefont {Cort{\^e}s}\ and\ \citenamefont {Liddle}(2025)}]{Cortes:2025joz}%
  \BibitemOpen
  \bibfield  {author} {\bibinfo {author} {\bibfnamefont {M.}~\bibnamefont {Cort{\^e}s}}\ and\ \bibinfo {author} {\bibfnamefont {A.~R.}\ \bibnamefont {Liddle}},\ }\bibfield  {title} {\bibinfo {title} {{On DESI's DR2 exclusion of $\Lambda$CDM}},\ }\href@noop {} {\  (\bibinfo {year} {2025})},\ \Eprint {https://arxiv.org/abs/2504.15336} {arXiv:2504.15336 [astro-ph.CO]} \BibitemShut {NoStop}%
\bibitem [{\citenamefont {Wang}\ and\ \citenamefont {Piao}(2024)}]{Wang:2024dka}%
  \BibitemOpen
  \bibfield  {author} {\bibinfo {author} {\bibfnamefont {H.}~\bibnamefont {Wang}}\ and\ \bibinfo {author} {\bibfnamefont {Y.-S.}\ \bibnamefont {Piao}},\ }\bibfield  {title} {\bibinfo {title} {{Dark energy in light of recent DESI BAO and Hubble tension}},\ }\href@noop {} {\  (\bibinfo {year} {2024})},\ \Eprint {https://arxiv.org/abs/2404.18579} {arXiv:2404.18579 [astro-ph.CO]} \BibitemShut {NoStop}%
\bibitem [{\citenamefont {Huang}\ \emph {et~al.}(2025)\citenamefont {Huang}, \citenamefont {Cai},\ and\ \citenamefont {Wang}}]{Huang:2025som}%
  \BibitemOpen
  \bibfield  {author} {\bibinfo {author} {\bibfnamefont {L.}~\bibnamefont {Huang}}, \bibinfo {author} {\bibfnamefont {R.-G.}\ \bibnamefont {Cai}},\ and\ \bibinfo {author} {\bibfnamefont {S.-J.}\ \bibnamefont {Wang}},\ }\bibfield  {title} {\bibinfo {title} {{The DESI 2024 hint for dynamical dark energy is biased by low-redshift supernovae}},\ }\href@noop {} {\  (\bibinfo {year} {2025})},\ \Eprint {https://arxiv.org/abs/2502.04212} {arXiv:2502.04212 [astro-ph.CO]} \BibitemShut {NoStop}%
\bibitem [{\citenamefont {Gialamas}\ \emph {et~al.}(2025{\natexlab{b}})\citenamefont {Gialamas}, \citenamefont {H\"utsi}, \citenamefont {Kannike}, \citenamefont {Racioppi}, \citenamefont {Raidal}, \citenamefont {Vasar},\ and\ \citenamefont {Veerm\"ae}}]{Gialamas:2024lyw}%
  \BibitemOpen
  \bibfield  {author} {\bibinfo {author} {\bibfnamefont {I.~D.}\ \bibnamefont {Gialamas}}, \bibinfo {author} {\bibfnamefont {G.}~\bibnamefont {H\"utsi}}, \bibinfo {author} {\bibfnamefont {K.}~\bibnamefont {Kannike}}, \bibinfo {author} {\bibfnamefont {A.}~\bibnamefont {Racioppi}}, \bibinfo {author} {\bibfnamefont {M.}~\bibnamefont {Raidal}}, \bibinfo {author} {\bibfnamefont {M.}~\bibnamefont {Vasar}},\ and\ \bibinfo {author} {\bibfnamefont {H.}~\bibnamefont {Veerm\"ae}},\ }\bibfield  {title} {\bibinfo {title} {{Interpreting DESI 2024 BAO: Late-time dynamical dark energy or a local effect?}},\ }\href {https://doi.org/10.1103/PhysRevD.111.043540} {\bibfield  {journal} {\bibinfo  {journal} {Phys. Rev. D}\ }\textbf {\bibinfo {volume} {111}},\ \bibinfo {pages} {043540} (\bibinfo {year} {2025}{\natexlab{b}})},\ \Eprint {https://arxiv.org/abs/2406.07533} {arXiv:2406.07533 [astro-ph.CO]} \BibitemShut {NoStop}%
\bibitem [{\citenamefont {Dinda}\ \emph {et~al.}(2025)\citenamefont {Dinda}, \citenamefont {Maartens}, \citenamefont {Saito},\ and\ \citenamefont {Clarkson}}]{Dinda:2025svh}%
  \BibitemOpen
  \bibfield  {author} {\bibinfo {author} {\bibfnamefont {B.~R.}\ \bibnamefont {Dinda}}, \bibinfo {author} {\bibfnamefont {R.}~\bibnamefont {Maartens}}, \bibinfo {author} {\bibfnamefont {S.}~\bibnamefont {Saito}},\ and\ \bibinfo {author} {\bibfnamefont {C.}~\bibnamefont {Clarkson}},\ }\bibfield  {title} {\bibinfo {title} {{Improved null tests of $\Lambda$CDM and FLRW in light of DESI DR2}},\ }\href@noop {} {\  (\bibinfo {year} {2025})},\ \Eprint {https://arxiv.org/abs/2504.09681} {arXiv:2504.09681 [astro-ph.CO]} \BibitemShut {NoStop}%
\bibitem [{\citenamefont {Ghosh}\ and\ \citenamefont {Bengaly}(2024)}]{Ghosh:2024kyd}%
  \BibitemOpen
  \bibfield  {author} {\bibinfo {author} {\bibfnamefont {B.}~\bibnamefont {Ghosh}}\ and\ \bibinfo {author} {\bibfnamefont {C.}~\bibnamefont {Bengaly}},\ }\bibfield  {title} {\bibinfo {title} {{Consistency tests between SDSS and DESI BAO measurements}},\ }\href {https://doi.org/10.1016/j.dark.2024.101699} {\bibfield  {journal} {\bibinfo  {journal} {Phys. Dark Univ.}\ }\textbf {\bibinfo {volume} {46}},\ \bibinfo {pages} {101699} (\bibinfo {year} {2024})},\ \Eprint {https://arxiv.org/abs/2408.04432} {arXiv:2408.04432 [astro-ph.CO]} \BibitemShut {NoStop}%
\bibitem [{\citenamefont {Bansal}\ and\ \citenamefont {Huterer}(2025)}]{Bansal:2025ipo}%
  \BibitemOpen
  \bibfield  {author} {\bibinfo {author} {\bibfnamefont {P.}~\bibnamefont {Bansal}}\ and\ \bibinfo {author} {\bibfnamefont {D.}~\bibnamefont {Huterer}},\ }\bibfield  {title} {\bibinfo {title} {{Expansion-history preferences of DESI and external data}},\ }\href@noop {} {\  (\bibinfo {year} {2025})},\ \Eprint {https://arxiv.org/abs/2502.07185} {arXiv:2502.07185 [astro-ph.CO]} \BibitemShut {NoStop}%
\bibitem [{\citenamefont {Park}\ \emph {et~al.}(2024{\natexlab{a}})\citenamefont {Park}, \citenamefont {de~Cruz~P\'erez},\ and\ \citenamefont {Ratra}}]{Chan-GyungPark:2024mlx}%
  \BibitemOpen
  \bibfield  {author} {\bibinfo {author} {\bibfnamefont {C.-G.}\ \bibnamefont {Park}}, \bibinfo {author} {\bibfnamefont {J.}~\bibnamefont {de~Cruz~P\'erez}},\ and\ \bibinfo {author} {\bibfnamefont {B.}~\bibnamefont {Ratra}},\ }\bibfield  {title} {\bibinfo {title} {{Using non-DESI data to confirm and strengthen the DESI 2024 spatially flat w0waCDM cosmological parametrization result}},\ }\href {https://doi.org/10.1103/PhysRevD.110.123533} {\bibfield  {journal} {\bibinfo  {journal} {Phys. Rev. D}\ }\textbf {\bibinfo {volume} {110}},\ \bibinfo {pages} {123533} (\bibinfo {year} {2024}{\natexlab{a}})},\ \Eprint {https://arxiv.org/abs/2405.00502} {arXiv:2405.00502 [astro-ph.CO]} \BibitemShut {NoStop}%
\bibitem [{\citenamefont {Park}\ \emph {et~al.}(2024{\natexlab{b}})\citenamefont {Park}, \citenamefont {de~Cruz~Perez},\ and\ \citenamefont {Ratra}}]{Chan-GyungPark:2024brx}%
  \BibitemOpen
  \bibfield  {author} {\bibinfo {author} {\bibfnamefont {C.-G.}\ \bibnamefont {Park}}, \bibinfo {author} {\bibfnamefont {J.}~\bibnamefont {de~Cruz~Perez}},\ and\ \bibinfo {author} {\bibfnamefont {B.}~\bibnamefont {Ratra}},\ }\bibfield  {title} {\bibinfo {title} {{Is the $w_0w_a$CDM cosmological parameterization evidence for dark energy dynamics partially caused by the excess smoothing of Planck CMB anisotropy data?}},\ }\href@noop {} {\  (\bibinfo {year} {2024}{\natexlab{b}})},\ \Eprint {https://arxiv.org/abs/2410.13627} {arXiv:2410.13627 [astro-ph.CO]} \BibitemShut {NoStop}%
\bibitem [{\citenamefont {Qiang}\ \emph {et~al.}(2025)\citenamefont {Qiang}, \citenamefont {Jia},\ and\ \citenamefont {Wei}}]{Qiang:2025cxp}%
  \BibitemOpen
  \bibfield  {author} {\bibinfo {author} {\bibfnamefont {D.-C.}\ \bibnamefont {Qiang}}, \bibinfo {author} {\bibfnamefont {J.-Y.}\ \bibnamefont {Jia}},\ and\ \bibinfo {author} {\bibfnamefont {H.}~\bibnamefont {Wei}},\ }\bibfield  {title} {\bibinfo {title} {{New Insights into Dark Energy from DESI DR2 with CMB and SNIa}},\ }\href@noop {} {\  (\bibinfo {year} {2025})},\ \Eprint {https://arxiv.org/abs/2507.09981} {arXiv:2507.09981 [astro-ph.CO]} \BibitemShut {NoStop}%
\bibitem [{\citenamefont {Chaudhary}\ \emph {et~al.}(2025{\natexlab{a}})\citenamefont {Chaudhary}, \citenamefont {Capozziello}, \citenamefont {Sharma},\ and\ \citenamefont {Mustafa}}]{Chaudhary:2025pcc}%
  \BibitemOpen
  \bibfield  {author} {\bibinfo {author} {\bibfnamefont {H.}~\bibnamefont {Chaudhary}}, \bibinfo {author} {\bibfnamefont {S.}~\bibnamefont {Capozziello}}, \bibinfo {author} {\bibfnamefont {V.~K.}\ \bibnamefont {Sharma}},\ and\ \bibinfo {author} {\bibfnamefont {G.}~\bibnamefont {Mustafa}},\ }\bibfield  {title} {\bibinfo {title} {{Does DESI DR2 challenge $\Lambda$CDM paradigm?}},\ }\href@noop {} {\  (\bibinfo {year} {2025}{\natexlab{a}})},\ \Eprint {https://arxiv.org/abs/2507.21607} {arXiv:2507.21607 [astro-ph.CO]} \BibitemShut {NoStop}%
\bibitem [{\citenamefont {Camarena}\ \emph {et~al.}(2025)\citenamefont {Camarena}, \citenamefont {Greene}, \citenamefont {Houghteling},\ and\ \citenamefont {Cyr-Racine}}]{Camarena:2025upt}%
  \BibitemOpen
  \bibfield  {author} {\bibinfo {author} {\bibfnamefont {D.}~\bibnamefont {Camarena}}, \bibinfo {author} {\bibfnamefont {K.}~\bibnamefont {Greene}}, \bibinfo {author} {\bibfnamefont {J.}~\bibnamefont {Houghteling}},\ and\ \bibinfo {author} {\bibfnamefont {F.-Y.}\ \bibnamefont {Cyr-Racine}},\ }\bibfield  {title} {\bibinfo {title} {{DESIgning concordant distances in the age of precision cosmology: the impact of density fluctuations}},\ }\href@noop {} {\  (\bibinfo {year} {2025})},\ \Eprint {https://arxiv.org/abs/2507.17969} {arXiv:2507.17969 [astro-ph.CO]} \BibitemShut {NoStop}%
\bibitem [{\citenamefont {Luongo}\ and\ \citenamefont {Muccino}(2024)}]{Luongo:2024fww}%
  \BibitemOpen
  \bibfield  {author} {\bibinfo {author} {\bibfnamefont {O.}~\bibnamefont {Luongo}}\ and\ \bibinfo {author} {\bibfnamefont {M.}~\bibnamefont {Muccino}},\ }\bibfield  {title} {\bibinfo {title} {{Model independent cosmographic constraints from DESI 2024}},\ }\href@noop {} {\  (\bibinfo {year} {2024})},\ \Eprint {https://arxiv.org/abs/2404.07070} {arXiv:2404.07070 [astro-ph.CO]} \BibitemShut {NoStop}%
\bibitem [{\citenamefont {Luongo}\ and\ \citenamefont {Muccino}(2025)}]{Luongo:2024zhc}%
  \BibitemOpen
  \bibfield  {author} {\bibinfo {author} {\bibfnamefont {O.}~\bibnamefont {Luongo}}\ and\ \bibinfo {author} {\bibfnamefont {M.}~\bibnamefont {Muccino}},\ }\bibfield  {title} {\bibinfo {title} {{Dark energy reconstructions combining baryonic acoustic oscillation data with galaxy clusters and intermediate-redshift catalogs}},\ }\href {https://doi.org/10.1051/0004-6361/202452973} {\bibfield  {journal} {\bibinfo  {journal} {Astron. Astrophys.}\ }\textbf {\bibinfo {volume} {693}},\ \bibinfo {pages} {A187} (\bibinfo {year} {2025})},\ \Eprint {https://arxiv.org/abs/2411.04901} {arXiv:2411.04901 [astro-ph.CO]} \BibitemShut {NoStop}%
\bibitem [{\citenamefont {Sousa-Neto}\ \emph {et~al.}(2025)\citenamefont {Sousa-Neto}, \citenamefont {Bengaly}, \citenamefont {Gonzalez},\ and\ \citenamefont {Alcaniz}}]{Sousa-Neto:2025gpj}%
  \BibitemOpen
  \bibfield  {author} {\bibinfo {author} {\bibfnamefont {A.}~\bibnamefont {Sousa-Neto}}, \bibinfo {author} {\bibfnamefont {C.}~\bibnamefont {Bengaly}}, \bibinfo {author} {\bibfnamefont {J.~E.}\ \bibnamefont {Gonzalez}},\ and\ \bibinfo {author} {\bibfnamefont {J.}~\bibnamefont {Alcaniz}},\ }\bibfield  {title} {\bibinfo {title} {{Evidence for dynamical dark energy from DESI-DR2 and SN data? A symbolic regression analysis}},\ }\href@noop {} {\  (\bibinfo {year} {2025})},\ \Eprint {https://arxiv.org/abs/2502.10506} {arXiv:2502.10506 [astro-ph.CO]} \BibitemShut {NoStop}%
\bibitem [{\citenamefont {Chaudhary}\ \emph {et~al.}(2025{\natexlab{b}})\citenamefont {Chaudhary}, \citenamefont {Capozziello}, \citenamefont {Sharma}, \citenamefont {G{\'o}mez-Vargas},\ and\ \citenamefont {Mustafa}}]{Chaudhary:2025vzy}%
  \BibitemOpen
  \bibfield  {author} {\bibinfo {author} {\bibfnamefont {H.}~\bibnamefont {Chaudhary}}, \bibinfo {author} {\bibfnamefont {S.}~\bibnamefont {Capozziello}}, \bibinfo {author} {\bibfnamefont {V.~K.}\ \bibnamefont {Sharma}}, \bibinfo {author} {\bibfnamefont {I.}~\bibnamefont {G{\'o}mez-Vargas}},\ and\ \bibinfo {author} {\bibfnamefont {G.}~\bibnamefont {Mustafa}},\ }\bibfield  {title} {\bibinfo {title} {{Evidence for Evolving Dark Energy from LRG1 and Low-$z$ SNe Ia Data}},\ }\href@noop {} {\  (\bibinfo {year} {2025}{\natexlab{b}})},\ \Eprint {https://arxiv.org/abs/2508.10514} {arXiv:2508.10514 [astro-ph.CO]} \BibitemShut {NoStop}%
\bibitem [{\citenamefont {Afroz}\ and\ \citenamefont {Mukherjee}(2025)}]{Afroz:2025iwo}%
  \BibitemOpen
  \bibfield  {author} {\bibinfo {author} {\bibfnamefont {S.}~\bibnamefont {Afroz}}\ and\ \bibinfo {author} {\bibfnamefont {S.}~\bibnamefont {Mukherjee}},\ }\bibfield  {title} {\bibinfo {title} {{Hint towards inconsistency between BAO and Supernovae Dataset: The Evidence of Redshift Evolving Dark Energy from DESI DR2 is Absent}},\ }\href@noop {} {\  (\bibinfo {year} {2025})},\ \Eprint {https://arxiv.org/abs/2504.16868} {arXiv:2504.16868 [astro-ph.CO]} \BibitemShut {NoStop}%
\bibitem [{\citenamefont {Toomey}\ \emph {et~al.}(2025)\citenamefont {Toomey}, \citenamefont {Montefalcone}, \citenamefont {McDonough},\ and\ \citenamefont {Freese}}]{Toomey:2025xyo}%
  \BibitemOpen
  \bibfield  {author} {\bibinfo {author} {\bibfnamefont {M.~W.}\ \bibnamefont {Toomey}}, \bibinfo {author} {\bibfnamefont {G.}~\bibnamefont {Montefalcone}}, \bibinfo {author} {\bibfnamefont {E.}~\bibnamefont {McDonough}},\ and\ \bibinfo {author} {\bibfnamefont {K.}~\bibnamefont {Freese}},\ }\bibfield  {title} {\bibinfo {title} {{How Theory-Informed Priors Affect DESI Evidence for Evolving Dark Energy}},\ }\href@noop {} {\  (\bibinfo {year} {2025})},\ \Eprint {https://arxiv.org/abs/2509.13318} {arXiv:2509.13318 [astro-ph.CO]} \BibitemShut {NoStop}%
\bibitem [{\citenamefont {Adi}(2025)}]{Adi:2025hyj}%
  \BibitemOpen
  \bibfield  {author} {\bibinfo {author} {\bibfnamefont {T.}~\bibnamefont {Adi}},\ }\bibfield  {title} {\bibinfo {title} {{Lowering the Horizon on Dark Energy: A Late-Time Response to Early Solutions for the Hubble Tension}},\ }\href@noop {} {\  (\bibinfo {year} {2025})},\ \Eprint {https://arxiv.org/abs/2509.12331} {arXiv:2509.12331 [astro-ph.CO]} \BibitemShut {NoStop}%
\bibitem [{\citenamefont {Roy~Choudhury}\ and\ \citenamefont {Okumura}(2024)}]{RoyChoudhury:2024wri}%
  \BibitemOpen
  \bibfield  {author} {\bibinfo {author} {\bibfnamefont {S.}~\bibnamefont {Roy~Choudhury}}\ and\ \bibinfo {author} {\bibfnamefont {T.}~\bibnamefont {Okumura}},\ }\bibfield  {title} {\bibinfo {title} {{Updated cosmological constraints in extended parameter space with Planck PR4, DESI BAO, and SN: dynamical dark energy, neutrino masses, lensing anomaly, and the Hubble tension}},\ }\href@noop {} {\  (\bibinfo {year} {2024})},\ \Eprint {https://arxiv.org/abs/2409.13022} {arXiv:2409.13022 [astro-ph.CO]} \BibitemShut {NoStop}%
\bibitem [{\citenamefont {Roy~Choudhury}(2025)}]{RoyChoudhury:2025dhe}%
  \BibitemOpen
  \bibfield  {author} {\bibinfo {author} {\bibfnamefont {S.}~\bibnamefont {Roy~Choudhury}},\ }\bibfield  {title} {\bibinfo {title} {{Cosmology in Extended Parameter Space with DESI Data Release 2 Baryon Acoustic Oscillations: A 2{\ensuremath{\sigma}}+ Detection of Nonzero Neutrino Masses with an Update on Dynamical Dark Energy and Lensing Anomaly}},\ }\href {https://doi.org/10.3847/2041-8213/ade1cc} {\bibfield  {journal} {\bibinfo  {journal} {Astrophys. J. Lett.}\ }\textbf {\bibinfo {volume} {986}},\ \bibinfo {pages} {L31} (\bibinfo {year} {2025})},\ \Eprint {https://arxiv.org/abs/2504.15340} {arXiv:2504.15340 [astro-ph.CO]} \BibitemShut {NoStop}%
\bibitem [{\citenamefont {Linder}(2025)}]{Linder:2025zxb}%
  \BibitemOpen
  \bibfield  {author} {\bibinfo {author} {\bibfnamefont {E.~V.}\ \bibnamefont {Linder}},\ }\bibfield  {title} {\bibinfo {title} {{Uplifting, Depressing, and Tilting Dark Energy}},\ }\href@noop {} {\  (\bibinfo {year} {2025})},\ \Eprint {https://arxiv.org/abs/2506.02122} {arXiv:2506.02122 [astro-ph.CO]} \BibitemShut {NoStop}%
\bibitem [{\citenamefont {{\"O}z{\"u}lker}\ \emph {et~al.}(2025)\citenamefont {{\"O}z{\"u}lker}, \citenamefont {Di~Valentino},\ and\ \citenamefont {Giar{\`e}}}]{Ozulker:2025ehg}%
  \BibitemOpen
  \bibfield  {author} {\bibinfo {author} {\bibfnamefont {E.}~\bibnamefont {{\"O}z{\"u}lker}}, \bibinfo {author} {\bibfnamefont {E.}~\bibnamefont {Di~Valentino}},\ and\ \bibinfo {author} {\bibfnamefont {W.}~\bibnamefont {Giar{\`e}}},\ }\bibfield  {title} {\bibinfo {title} {{Dark Energy Crosses the Line: Quantifying and Testing the Evidence for Phantom Crossing}},\ }\href@noop {} {\  (\bibinfo {year} {2025})},\ \Eprint {https://arxiv.org/abs/2506.19053} {arXiv:2506.19053 [astro-ph.CO]} \BibitemShut {NoStop}%
\bibitem [{\citenamefont {Scherer}\ \emph {et~al.}(2025)\citenamefont {Scherer}, \citenamefont {Sabogal}, \citenamefont {Nunes},\ and\ \citenamefont {De~Felice}}]{Scherer:2025esj}%
  \BibitemOpen
  \bibfield  {author} {\bibinfo {author} {\bibfnamefont {M.}~\bibnamefont {Scherer}}, \bibinfo {author} {\bibfnamefont {M.~A.}\ \bibnamefont {Sabogal}}, \bibinfo {author} {\bibfnamefont {R.~C.}\ \bibnamefont {Nunes}},\ and\ \bibinfo {author} {\bibfnamefont {A.}~\bibnamefont {De~Felice}},\ }\bibfield  {title} {\bibinfo {title} {{Challenging the {\ensuremath{\Lambda}}CDM model: 5{\ensuremath{\sigma}} evidence for a dynamical dark energy late-time transition}},\ }\href {https://doi.org/10.1103/n86r-sjgm} {\bibfield  {journal} {\bibinfo  {journal} {Phys. Rev. D}\ }\textbf {\bibinfo {volume} {112}},\ \bibinfo {pages} {043513} (\bibinfo {year} {2025})},\ \Eprint {https://arxiv.org/abs/2504.20664} {arXiv:2504.20664 [astro-ph.CO]} \BibitemShut {NoStop}%
\bibitem [{\citenamefont {Keeley}\ \emph {et~al.}(2025)\citenamefont {Keeley}, \citenamefont {Shafieloo},\ and\ \citenamefont {Matthewson}}]{Keeley:2025rlg}%
  \BibitemOpen
  \bibfield  {author} {\bibinfo {author} {\bibfnamefont {R.~E.}\ \bibnamefont {Keeley}}, \bibinfo {author} {\bibfnamefont {A.}~\bibnamefont {Shafieloo}},\ and\ \bibinfo {author} {\bibfnamefont {W.~L.}\ \bibnamefont {Matthewson}},\ }\bibfield  {title} {\bibinfo {title} {{Could We Be Fooled about Phantom Crossing?}},\ }\href@noop {} {\  (\bibinfo {year} {2025})},\ \Eprint {https://arxiv.org/abs/2506.15091} {arXiv:2506.15091 [astro-ph.CO]} \BibitemShut {NoStop}%
\bibitem [{\citenamefont {Liu}\ \emph {et~al.}(2025)\citenamefont {Liu}, \citenamefont {Zhu}, \citenamefont {Hu},\ and\ \citenamefont {Miranda}}]{Liu:2025bss}%
  \BibitemOpen
  \bibfield  {author} {\bibinfo {author} {\bibfnamefont {R.}~\bibnamefont {Liu}}, \bibinfo {author} {\bibfnamefont {Y.}~\bibnamefont {Zhu}}, \bibinfo {author} {\bibfnamefont {W.}~\bibnamefont {Hu}},\ and\ \bibinfo {author} {\bibfnamefont {V.}~\bibnamefont {Miranda}},\ }\bibfield  {title} {\bibinfo {title} {{Phantom Mirage from Axion Dark Energy}},\ }\href@noop {} {\  (\bibinfo {year} {2025})},\ \Eprint {https://arxiv.org/abs/2510.14957} {arXiv:2510.14957 [astro-ph.CO]} \BibitemShut {NoStop}%
\bibitem [{\citenamefont {Park}\ \emph {et~al.}(2010)\citenamefont {Park}, \citenamefont {Zurek},\ and\ \citenamefont {Watson}}]{Park:2010cw}%
  \BibitemOpen
  \bibfield  {author} {\bibinfo {author} {\bibfnamefont {M.}~\bibnamefont {Park}}, \bibinfo {author} {\bibfnamefont {K.~M.}\ \bibnamefont {Zurek}},\ and\ \bibinfo {author} {\bibfnamefont {S.}~\bibnamefont {Watson}},\ }\bibfield  {title} {\bibinfo {title} {{A Unified Approach to Cosmic Acceleration}},\ }\href {https://doi.org/10.1103/PhysRevD.81.124008} {\bibfield  {journal} {\bibinfo  {journal} {Phys. Rev. D}\ }\textbf {\bibinfo {volume} {81}},\ \bibinfo {pages} {124008} (\bibinfo {year} {2010})},\ \Eprint {https://arxiv.org/abs/1003.1722} {arXiv:1003.1722 [hep-th]} \BibitemShut {NoStop}%
\bibitem [{\citenamefont {Caldwell}\ \emph {et~al.}(1998)\citenamefont {Caldwell}, \citenamefont {Dave},\ and\ \citenamefont {Steinhardt}}]{Caldwell:1997ii}%
  \BibitemOpen
  \bibfield  {author} {\bibinfo {author} {\bibfnamefont {R.~R.}\ \bibnamefont {Caldwell}}, \bibinfo {author} {\bibfnamefont {R.}~\bibnamefont {Dave}},\ and\ \bibinfo {author} {\bibfnamefont {P.~J.}\ \bibnamefont {Steinhardt}},\ }\bibfield  {title} {\bibinfo {title} {{Cosmological imprint of an energy component with general equation of state}},\ }\href {https://doi.org/10.1103/PhysRevLett.80.1582} {\bibfield  {journal} {\bibinfo  {journal} {Phys. Rev. Lett.}\ }\textbf {\bibinfo {volume} {80}},\ \bibinfo {pages} {1582} (\bibinfo {year} {1998})},\ \Eprint {https://arxiv.org/abs/astro-ph/9708069} {arXiv:astro-ph/9708069} \BibitemShut {NoStop}%
\bibitem [{\citenamefont {Tsujikawa}(2013)}]{Tsujikawa:2013fta}%
  \BibitemOpen
  \bibfield  {author} {\bibinfo {author} {\bibfnamefont {S.}~\bibnamefont {Tsujikawa}},\ }\bibfield  {title} {\bibinfo {title} {{Quintessence: A Review}},\ }\href {https://doi.org/10.1088/0264-9381/30/21/214003} {\bibfield  {journal} {\bibinfo  {journal} {Class. Quant. Grav.}\ }\textbf {\bibinfo {volume} {30}},\ \bibinfo {pages} {214003} (\bibinfo {year} {2013})},\ \Eprint {https://arxiv.org/abs/1304.1961} {arXiv:1304.1961 [gr-qc]} \BibitemShut {NoStop}%
\bibitem [{\citenamefont {Ratra}\ and\ \citenamefont {Peebles}(1988)}]{Ratra:1987rm}%
  \BibitemOpen
  \bibfield  {author} {\bibinfo {author} {\bibfnamefont {B.}~\bibnamefont {Ratra}}\ and\ \bibinfo {author} {\bibfnamefont {P.~J.~E.}\ \bibnamefont {Peebles}},\ }\bibfield  {title} {\bibinfo {title} {{Cosmological Consequences of a Rolling Homogeneous Scalar Field}},\ }\href {https://doi.org/10.1103/PhysRevD.37.3406} {\bibfield  {journal} {\bibinfo  {journal} {Phys. Rev. D}\ }\textbf {\bibinfo {volume} {37}},\ \bibinfo {pages} {3406} (\bibinfo {year} {1988})}\BibitemShut {NoStop}%
\bibitem [{\citenamefont {Copeland}\ \emph {et~al.}(2006)\citenamefont {Copeland}, \citenamefont {Sami},\ and\ \citenamefont {Tsujikawa}}]{Copeland:2006wr}%
  \BibitemOpen
  \bibfield  {author} {\bibinfo {author} {\bibfnamefont {E.~J.}\ \bibnamefont {Copeland}}, \bibinfo {author} {\bibfnamefont {M.}~\bibnamefont {Sami}},\ and\ \bibinfo {author} {\bibfnamefont {S.}~\bibnamefont {Tsujikawa}},\ }\bibfield  {title} {\bibinfo {title} {{Dynamics of dark energy}},\ }\href {https://doi.org/10.1142/S021827180600942X} {\bibfield  {journal} {\bibinfo  {journal} {Int. J. Mod. Phys. D}\ }\textbf {\bibinfo {volume} {15}},\ \bibinfo {pages} {1753} (\bibinfo {year} {2006})},\ \Eprint {https://arxiv.org/abs/hep-th/0603057} {arXiv:hep-th/0603057} \BibitemShut {NoStop}%
\bibitem [{\citenamefont {Linder}(2003)}]{Linder:2002et}%
  \BibitemOpen
  \bibfield  {author} {\bibinfo {author} {\bibfnamefont {E.~V.}\ \bibnamefont {Linder}},\ }\bibfield  {title} {\bibinfo {title} {{Exploring the expansion history of the universe}},\ }\href {https://doi.org/10.1103/PhysRevLett.90.091301} {\bibfield  {journal} {\bibinfo  {journal} {Phys. Rev. Lett.}\ }\textbf {\bibinfo {volume} {90}},\ \bibinfo {pages} {091301} (\bibinfo {year} {2003})},\ \Eprint {https://arxiv.org/abs/astro-ph/0208512} {arXiv:astro-ph/0208512} \BibitemShut {NoStop}%
\bibitem [{\citenamefont {Chevallier}\ and\ \citenamefont {Polarski}(2001)}]{Chevallier:2000qy}%
  \BibitemOpen
  \bibfield  {author} {\bibinfo {author} {\bibfnamefont {M.}~\bibnamefont {Chevallier}}\ and\ \bibinfo {author} {\bibfnamefont {D.}~\bibnamefont {Polarski}},\ }\bibfield  {title} {\bibinfo {title} {{Accelerating universes with scaling dark matter}},\ }\href {https://doi.org/10.1142/S0218271801000822} {\bibfield  {journal} {\bibinfo  {journal} {Int. J. Mod. Phys. D}\ }\textbf {\bibinfo {volume} {10}},\ \bibinfo {pages} {213} (\bibinfo {year} {2001})},\ \Eprint {https://arxiv.org/abs/gr-qc/0009008} {arXiv:gr-qc/0009008} \BibitemShut {NoStop}%
\bibitem [{\citenamefont {Gannouji}\ \emph {et~al.}(2006)\citenamefont {Gannouji}, \citenamefont {Polarski}, \citenamefont {Ranquet},\ and\ \citenamefont {Starobinsky}}]{Gannouji:2006jm}%
  \BibitemOpen
  \bibfield  {author} {\bibinfo {author} {\bibfnamefont {R.}~\bibnamefont {Gannouji}}, \bibinfo {author} {\bibfnamefont {D.}~\bibnamefont {Polarski}}, \bibinfo {author} {\bibfnamefont {A.}~\bibnamefont {Ranquet}},\ and\ \bibinfo {author} {\bibfnamefont {A.~A.}\ \bibnamefont {Starobinsky}},\ }\bibfield  {title} {\bibinfo {title} {{Scalar-Tensor Models of Normal and Phantom Dark Energy}},\ }\href {https://doi.org/10.1088/1475-7516/2006/09/016} {\bibfield  {journal} {\bibinfo  {journal} {JCAP}\ }\textbf {\bibinfo {volume} {09}},\ \bibinfo {pages} {016}},\ \Eprint {https://arxiv.org/abs/astro-ph/0606287} {arXiv:astro-ph/0606287} \BibitemShut {NoStop}%
\bibitem [{\citenamefont {Boisseau}\ \emph {et~al.}(2000)\citenamefont {Boisseau}, \citenamefont {Esposito-Farese}, \citenamefont {Polarski},\ and\ \citenamefont {Starobinsky}}]{Boisseau:2000pr}%
  \BibitemOpen
  \bibfield  {author} {\bibinfo {author} {\bibfnamefont {B.}~\bibnamefont {Boisseau}}, \bibinfo {author} {\bibfnamefont {G.}~\bibnamefont {Esposito-Farese}}, \bibinfo {author} {\bibfnamefont {D.}~\bibnamefont {Polarski}},\ and\ \bibinfo {author} {\bibfnamefont {A.~A.}\ \bibnamefont {Starobinsky}},\ }\bibfield  {title} {\bibinfo {title} {{Reconstruction of a scalar tensor theory of gravity in an accelerating universe}},\ }\href {https://doi.org/10.1103/PhysRevLett.85.2236} {\bibfield  {journal} {\bibinfo  {journal} {Phys. Rev. Lett.}\ }\textbf {\bibinfo {volume} {85}},\ \bibinfo {pages} {2236} (\bibinfo {year} {2000})},\ \Eprint {https://arxiv.org/abs/gr-qc/0001066} {arXiv:gr-qc/0001066} \BibitemShut {NoStop}%
\bibitem [{\citenamefont {Caldwell}(2002)}]{Caldwell:1999ew}%
  \BibitemOpen
  \bibfield  {author} {\bibinfo {author} {\bibfnamefont {R.~R.}\ \bibnamefont {Caldwell}},\ }\bibfield  {title} {\bibinfo {title} {{A Phantom menace?}},\ }\href {https://doi.org/10.1016/S0370-2693(02)02589-3} {\bibfield  {journal} {\bibinfo  {journal} {Phys. Lett. B}\ }\textbf {\bibinfo {volume} {545}},\ \bibinfo {pages} {23} (\bibinfo {year} {2002})},\ \Eprint {https://arxiv.org/abs/astro-ph/9908168} {arXiv:astro-ph/9908168} \BibitemShut {NoStop}%
\bibitem [{\citenamefont {Torres}(2002)}]{Torres:2002pe}%
  \BibitemOpen
  \bibfield  {author} {\bibinfo {author} {\bibfnamefont {D.~F.}\ \bibnamefont {Torres}},\ }\bibfield  {title} {\bibinfo {title} {{Quintessence, superquintessence and observable quantities in Brans-Dicke and nonminimally coupled theories}},\ }\href {https://doi.org/10.1103/PhysRevD.66.043522} {\bibfield  {journal} {\bibinfo  {journal} {Phys. Rev. D}\ }\textbf {\bibinfo {volume} {66}},\ \bibinfo {pages} {043522} (\bibinfo {year} {2002})},\ \Eprint {https://arxiv.org/abs/astro-ph/0204504} {arXiv:astro-ph/0204504} \BibitemShut {NoStop}%
\bibitem [{\citenamefont {Deffayet}\ \emph {et~al.}(2009)\citenamefont {Deffayet}, \citenamefont {Esposito-Farese},\ and\ \citenamefont {Vikman}}]{Deffayet:2009wt}%
  \BibitemOpen
  \bibfield  {author} {\bibinfo {author} {\bibfnamefont {C.}~\bibnamefont {Deffayet}}, \bibinfo {author} {\bibfnamefont {G.}~\bibnamefont {Esposito-Farese}},\ and\ \bibinfo {author} {\bibfnamefont {A.}~\bibnamefont {Vikman}},\ }\bibfield  {title} {\bibinfo {title} {{Covariant Galileon}},\ }\href {https://doi.org/10.1103/PhysRevD.79.084003} {\bibfield  {journal} {\bibinfo  {journal} {Phys. Rev. D}\ }\textbf {\bibinfo {volume} {79}},\ \bibinfo {pages} {084003} (\bibinfo {year} {2009})},\ \Eprint {https://arxiv.org/abs/0901.1314} {arXiv:0901.1314 [hep-th]} \BibitemShut {NoStop}%
\bibitem [{\citenamefont {Deffayet}\ \emph {et~al.}(2010)\citenamefont {Deffayet}, \citenamefont {Pujolas}, \citenamefont {Sawicki},\ and\ \citenamefont {Vikman}}]{Deffayet:2010qz}%
  \BibitemOpen
  \bibfield  {author} {\bibinfo {author} {\bibfnamefont {C.}~\bibnamefont {Deffayet}}, \bibinfo {author} {\bibfnamefont {O.}~\bibnamefont {Pujolas}}, \bibinfo {author} {\bibfnamefont {I.}~\bibnamefont {Sawicki}},\ and\ \bibinfo {author} {\bibfnamefont {A.}~\bibnamefont {Vikman}},\ }\bibfield  {title} {\bibinfo {title} {{Imperfect Dark Energy from Kinetic Gravity Braiding}},\ }\href {https://doi.org/10.1088/1475-7516/2010/10/026} {\bibfield  {journal} {\bibinfo  {journal} {JCAP}\ }\textbf {\bibinfo {volume} {10}},\ \bibinfo {pages} {026}},\ \Eprint {https://arxiv.org/abs/1008.0048} {arXiv:1008.0048 [hep-th]} \BibitemShut {NoStop}%
\bibitem [{\citenamefont {Nicolis}\ \emph {et~al.}(2009)\citenamefont {Nicolis}, \citenamefont {Rattazzi},\ and\ \citenamefont {Trincherini}}]{Nicolis:2008in}%
  \BibitemOpen
  \bibfield  {author} {\bibinfo {author} {\bibfnamefont {A.}~\bibnamefont {Nicolis}}, \bibinfo {author} {\bibfnamefont {R.}~\bibnamefont {Rattazzi}},\ and\ \bibinfo {author} {\bibfnamefont {E.}~\bibnamefont {Trincherini}},\ }\bibfield  {title} {\bibinfo {title} {{The Galileon as a local modification of gravity}},\ }\href {https://doi.org/10.1103/PhysRevD.79.064036} {\bibfield  {journal} {\bibinfo  {journal} {Phys. Rev. D}\ }\textbf {\bibinfo {volume} {79}},\ \bibinfo {pages} {064036} (\bibinfo {year} {2009})},\ \Eprint {https://arxiv.org/abs/0811.2197} {arXiv:0811.2197 [hep-th]} \BibitemShut {NoStop}%
\bibitem [{\citenamefont {Traykova}\ \emph {et~al.}(2021)\citenamefont {Traykova}, \citenamefont {Bellini}, \citenamefont {Ferreira}, \citenamefont {Garc\'\i{}a-Garc\'\i{}a}, \citenamefont {Noller},\ and\ \citenamefont {Zumalac\'arregui}}]{Traykova:2021hbr}%
  \BibitemOpen
  \bibfield  {author} {\bibinfo {author} {\bibfnamefont {D.}~\bibnamefont {Traykova}}, \bibinfo {author} {\bibfnamefont {E.}~\bibnamefont {Bellini}}, \bibinfo {author} {\bibfnamefont {P.~G.}\ \bibnamefont {Ferreira}}, \bibinfo {author} {\bibfnamefont {C.}~\bibnamefont {Garc\'\i{}a-Garc\'\i{}a}}, \bibinfo {author} {\bibfnamefont {J.}~\bibnamefont {Noller}},\ and\ \bibinfo {author} {\bibfnamefont {M.}~\bibnamefont {Zumalac\'arregui}},\ }\bibfield  {title} {\bibinfo {title} {{Theoretical priors in scalar-tensor cosmologies: Shift-symmetric Horndeski models}},\ }\href {https://doi.org/10.1103/PhysRevD.104.083502} {\bibfield  {journal} {\bibinfo  {journal} {Phys. Rev. D}\ }\textbf {\bibinfo {volume} {104}},\ \bibinfo {pages} {083502} (\bibinfo {year} {2021})},\ \Eprint {https://arxiv.org/abs/2103.11195} {arXiv:2103.11195 [astro-ph.CO]} \BibitemShut {NoStop}%
\bibitem [{\citenamefont {De~Felice}\ and\ \citenamefont {Tsujikawa}(2010)}]{DeFelice:2010pv}%
  \BibitemOpen
  \bibfield  {author} {\bibinfo {author} {\bibfnamefont {A.}~\bibnamefont {De~Felice}}\ and\ \bibinfo {author} {\bibfnamefont {S.}~\bibnamefont {Tsujikawa}},\ }\bibfield  {title} {\bibinfo {title} {{Cosmology of a covariant Galileon field}},\ }\href {https://doi.org/10.1103/PhysRevLett.105.111301} {\bibfield  {journal} {\bibinfo  {journal} {Phys. Rev. Lett.}\ }\textbf {\bibinfo {volume} {105}},\ \bibinfo {pages} {111301} (\bibinfo {year} {2010})},\ \Eprint {https://arxiv.org/abs/1007.2700} {arXiv:1007.2700 [astro-ph.CO]} \BibitemShut {NoStop}%
\bibitem [{\citenamefont {Ye}\ and\ \citenamefont {Silvestri}(2025)}]{Ye:2024kus}%
  \BibitemOpen
  \bibfield  {author} {\bibinfo {author} {\bibfnamefont {G.}~\bibnamefont {Ye}}\ and\ \bibinfo {author} {\bibfnamefont {A.}~\bibnamefont {Silvestri}},\ }\bibfield  {title} {\bibinfo {title} {{Cubic Galileon gravity effects in the CMB}},\ }\href {https://doi.org/10.1103/PhysRevD.111.023502} {\bibfield  {journal} {\bibinfo  {journal} {Phys. Rev. D}\ }\textbf {\bibinfo {volume} {111}},\ \bibinfo {pages} {023502} (\bibinfo {year} {2025})},\ \Eprint {https://arxiv.org/abs/2407.02471} {arXiv:2407.02471 [astro-ph.CO]} \BibitemShut {NoStop}%
\bibitem [{\citenamefont {Zumalacarregui}(2020)}]{Zumalacarregui:2020cjh}%
  \BibitemOpen
  \bibfield  {author} {\bibinfo {author} {\bibfnamefont {M.}~\bibnamefont {Zumalacarregui}},\ }\bibfield  {title} {\bibinfo {title} {{Gravity in the Era of Equality: Towards solutions to the Hubble problem without fine-tuned initial conditions}},\ }\href {https://doi.org/10.1103/PhysRevD.102.023523} {\bibfield  {journal} {\bibinfo  {journal} {Phys. Rev. D}\ }\textbf {\bibinfo {volume} {102}},\ \bibinfo {pages} {023523} (\bibinfo {year} {2020})},\ \Eprint {https://arxiv.org/abs/2003.06396} {arXiv:2003.06396 [astro-ph.CO]} \BibitemShut {NoStop}%
\bibitem [{\citenamefont {Brahma}\ and\ \citenamefont {Hossain}(2021)}]{Brahma:2020eqd}%
  \BibitemOpen
  \bibfield  {author} {\bibinfo {author} {\bibfnamefont {S.}~\bibnamefont {Brahma}}\ and\ \bibinfo {author} {\bibfnamefont {M.~W.}\ \bibnamefont {Hossain}},\ }\bibfield  {title} {\bibinfo {title} {{Consistency of Cubic Galileon Cosmology: Model-Independent Bounds from Background Expansion and Perturbative Analyses}},\ }\href {https://doi.org/10.3390/universe7060167} {\bibfield  {journal} {\bibinfo  {journal} {Universe}\ }\textbf {\bibinfo {volume} {7}},\ \bibinfo {pages} {167} (\bibinfo {year} {2021})},\ \Eprint {https://arxiv.org/abs/2007.06425} {arXiv:2007.06425 [astro-ph.CO]} \BibitemShut {NoStop}%
\bibitem [{\citenamefont {Barreira}\ \emph {et~al.}(2012)\citenamefont {Barreira}, \citenamefont {Li}, \citenamefont {Baugh},\ and\ \citenamefont {Pascoli}}]{Barreira:2012kk}%
  \BibitemOpen
  \bibfield  {author} {\bibinfo {author} {\bibfnamefont {A.}~\bibnamefont {Barreira}}, \bibinfo {author} {\bibfnamefont {B.}~\bibnamefont {Li}}, \bibinfo {author} {\bibfnamefont {C.~M.}\ \bibnamefont {Baugh}},\ and\ \bibinfo {author} {\bibfnamefont {S.}~\bibnamefont {Pascoli}},\ }\bibfield  {title} {\bibinfo {title} {{Linear perturbations in Galileon gravity models}},\ }\href {https://doi.org/10.1103/PhysRevD.86.124016} {\bibfield  {journal} {\bibinfo  {journal} {Phys. Rev. D}\ }\textbf {\bibinfo {volume} {86}},\ \bibinfo {pages} {124016} (\bibinfo {year} {2012})},\ \Eprint {https://arxiv.org/abs/1208.0600} {arXiv:1208.0600 [astro-ph.CO]} \BibitemShut {NoStop}%
\bibitem [{\citenamefont {Barreira}\ \emph {et~al.}(2014)\citenamefont {Barreira}, \citenamefont {Li}, \citenamefont {Baugh},\ and\ \citenamefont {Pascoli}}]{Barreira:2014jha}%
  \BibitemOpen
  \bibfield  {author} {\bibinfo {author} {\bibfnamefont {A.}~\bibnamefont {Barreira}}, \bibinfo {author} {\bibfnamefont {B.}~\bibnamefont {Li}}, \bibinfo {author} {\bibfnamefont {C.}~\bibnamefont {Baugh}},\ and\ \bibinfo {author} {\bibfnamefont {S.}~\bibnamefont {Pascoli}},\ }\bibfield  {title} {\bibinfo {title} {{The observational status of Galileon gravity after Planck}},\ }\href {https://doi.org/10.1088/1475-7516/2014/08/059} {\bibfield  {journal} {\bibinfo  {journal} {JCAP}\ }\textbf {\bibinfo {volume} {08}},\ \bibinfo {pages} {059}},\ \Eprint {https://arxiv.org/abs/1406.0485} {arXiv:1406.0485 [astro-ph.CO]} \BibitemShut {NoStop}%
\bibitem [{\citenamefont {Leloup}\ \emph {et~al.}(2019)\citenamefont {Leloup}, \citenamefont {Ruhlmann-Kleider}, \citenamefont {Neveu},\ and\ \citenamefont {De~Mattia}}]{Leloup:2019fas}%
  \BibitemOpen
  \bibfield  {author} {\bibinfo {author} {\bibfnamefont {C.}~\bibnamefont {Leloup}}, \bibinfo {author} {\bibfnamefont {V.}~\bibnamefont {Ruhlmann-Kleider}}, \bibinfo {author} {\bibfnamefont {J.}~\bibnamefont {Neveu}},\ and\ \bibinfo {author} {\bibfnamefont {A.}~\bibnamefont {De~Mattia}},\ }\bibfield  {title} {\bibinfo {title} {{Observational status of the Galileon model general solution from cosmological data and gravitational waves}},\ }\href {https://doi.org/10.1088/1475-7516/2019/05/011} {\bibfield  {journal} {\bibinfo  {journal} {JCAP}\ }\textbf {\bibinfo {volume} {05}},\ \bibinfo {pages} {011}},\ \Eprint {https://arxiv.org/abs/1902.07065} {arXiv:1902.07065 [astro-ph.CO]} \BibitemShut {NoStop}%
\bibitem [{\citenamefont {Renk}\ \emph {et~al.}(2017)\citenamefont {Renk}, \citenamefont {Zumalac\'arregui}, \citenamefont {Montanari},\ and\ \citenamefont {Barreira}}]{Renk:2017rzu}%
  \BibitemOpen
  \bibfield  {author} {\bibinfo {author} {\bibfnamefont {J.}~\bibnamefont {Renk}}, \bibinfo {author} {\bibfnamefont {M.}~\bibnamefont {Zumalac\'arregui}}, \bibinfo {author} {\bibfnamefont {F.}~\bibnamefont {Montanari}},\ and\ \bibinfo {author} {\bibfnamefont {A.}~\bibnamefont {Barreira}},\ }\bibfield  {title} {\bibinfo {title} {{Galileon gravity in light of ISW, CMB, BAO and H$_0$ data}},\ }\href {https://doi.org/10.1088/1475-7516/2017/10/020} {\bibfield  {journal} {\bibinfo  {journal} {JCAP}\ }\textbf {\bibinfo {volume} {10}},\ \bibinfo {pages} {020}},\ \Eprint {https://arxiv.org/abs/1707.02263} {arXiv:1707.02263 [astro-ph.CO]} \BibitemShut {NoStop}%
\bibitem [{\citenamefont {Peirone}\ \emph {et~al.}(2019)\citenamefont {Peirone}, \citenamefont {Benevento}, \citenamefont {Frusciante},\ and\ \citenamefont {Tsujikawa}}]{Peirone:2019aua}%
  \BibitemOpen
  \bibfield  {author} {\bibinfo {author} {\bibfnamefont {S.}~\bibnamefont {Peirone}}, \bibinfo {author} {\bibfnamefont {G.}~\bibnamefont {Benevento}}, \bibinfo {author} {\bibfnamefont {N.}~\bibnamefont {Frusciante}},\ and\ \bibinfo {author} {\bibfnamefont {S.}~\bibnamefont {Tsujikawa}},\ }\bibfield  {title} {\bibinfo {title} {{Cosmological data favor Galileon ghost condensate over $\Lambda$CDM}},\ }\href {https://doi.org/10.1103/PhysRevD.100.063540} {\bibfield  {journal} {\bibinfo  {journal} {Phys. Rev. D}\ }\textbf {\bibinfo {volume} {100}},\ \bibinfo {pages} {063540} (\bibinfo {year} {2019})},\ \Eprint {https://arxiv.org/abs/1905.05166} {arXiv:1905.05166 [astro-ph.CO]} \BibitemShut {NoStop}%
\bibitem [{\citenamefont {Frusciante}\ \emph {et~al.}(2020)\citenamefont {Frusciante}, \citenamefont {Peirone}, \citenamefont {Atayde},\ and\ \citenamefont {De~Felice}}]{Frusciante:2019puu}%
  \BibitemOpen
  \bibfield  {author} {\bibinfo {author} {\bibfnamefont {N.}~\bibnamefont {Frusciante}}, \bibinfo {author} {\bibfnamefont {S.}~\bibnamefont {Peirone}}, \bibinfo {author} {\bibfnamefont {L.}~\bibnamefont {Atayde}},\ and\ \bibinfo {author} {\bibfnamefont {A.}~\bibnamefont {De~Felice}},\ }\bibfield  {title} {\bibinfo {title} {{Phenomenology of the generalized cubic covariant Galileon model and cosmological bounds}},\ }\href {https://doi.org/10.1103/PhysRevD.101.064001} {\bibfield  {journal} {\bibinfo  {journal} {Phys. Rev. D}\ }\textbf {\bibinfo {volume} {101}},\ \bibinfo {pages} {064001} (\bibinfo {year} {2020})},\ \Eprint {https://arxiv.org/abs/1912.07586} {arXiv:1912.07586 [astro-ph.CO]} \BibitemShut {NoStop}%
\bibitem [{\citenamefont {De~Felice}\ and\ \citenamefont {Tsujikawa}(2012)}]{DeFelice:2011bh}%
  \BibitemOpen
  \bibfield  {author} {\bibinfo {author} {\bibfnamefont {A.}~\bibnamefont {De~Felice}}\ and\ \bibinfo {author} {\bibfnamefont {S.}~\bibnamefont {Tsujikawa}},\ }\bibfield  {title} {\bibinfo {title} {{Conditions for the cosmological viability of the most general scalar-tensor theories and their applications to extended Galileon dark energy models}},\ }\href {https://doi.org/10.1088/1475-7516/2012/02/007} {\bibfield  {journal} {\bibinfo  {journal} {JCAP}\ }\textbf {\bibinfo {volume} {02}},\ \bibinfo {pages} {007}},\ \Eprint {https://arxiv.org/abs/1110.3878} {arXiv:1110.3878 [gr-qc]} \BibitemShut {NoStop}%
\bibitem [{\citenamefont {Bartlett}\ \emph {et~al.}(2021)\citenamefont {Bartlett}, \citenamefont {Desmond},\ and\ \citenamefont {Ferreira}}]{Bartlett:2020tjd}%
  \BibitemOpen
  \bibfield  {author} {\bibinfo {author} {\bibfnamefont {D.~J.}\ \bibnamefont {Bartlett}}, \bibinfo {author} {\bibfnamefont {H.}~\bibnamefont {Desmond}},\ and\ \bibinfo {author} {\bibfnamefont {P.~G.}\ \bibnamefont {Ferreira}},\ }\bibfield  {title} {\bibinfo {title} {{Constraints on galileons from the positions of supermassive black holes}},\ }\href {https://doi.org/10.1103/PhysRevD.103.023523} {\bibfield  {journal} {\bibinfo  {journal} {Phys. Rev. D}\ }\textbf {\bibinfo {volume} {103}},\ \bibinfo {pages} {023523} (\bibinfo {year} {2021})},\ \Eprint {https://arxiv.org/abs/2010.05811} {arXiv:2010.05811 [astro-ph.CO]} \BibitemShut {NoStop}%
\bibitem [{\citenamefont {Barreira}\ \emph {et~al.}(2013)\citenamefont {Barreira}, \citenamefont {Li}, \citenamefont {Sanchez}, \citenamefont {Baugh},\ and\ \citenamefont {Pascoli}}]{Barreira:2013jma}%
  \BibitemOpen
  \bibfield  {author} {\bibinfo {author} {\bibfnamefont {A.}~\bibnamefont {Barreira}}, \bibinfo {author} {\bibfnamefont {B.}~\bibnamefont {Li}}, \bibinfo {author} {\bibfnamefont {A.}~\bibnamefont {Sanchez}}, \bibinfo {author} {\bibfnamefont {C.~M.}\ \bibnamefont {Baugh}},\ and\ \bibinfo {author} {\bibfnamefont {S.}~\bibnamefont {Pascoli}},\ }\bibfield  {title} {\bibinfo {title} {{Parameter space in Galileon gravity models}},\ }\href {https://doi.org/10.1103/PhysRevD.87.103511} {\bibfield  {journal} {\bibinfo  {journal} {Phys. Rev. D}\ }\textbf {\bibinfo {volume} {87}},\ \bibinfo {pages} {103511} (\bibinfo {year} {2013})},\ \Eprint {https://arxiv.org/abs/1302.6241} {arXiv:1302.6241 [astro-ph.CO]} \BibitemShut {NoStop}%
\bibitem [{\citenamefont {Tsujikawa}(2025)}]{Tsujikawa:2025wca}%
  \BibitemOpen
  \bibfield  {author} {\bibinfo {author} {\bibfnamefont {S.}~\bibnamefont {Tsujikawa}},\ }\bibfield  {title} {\bibinfo {title} {{Crossing the phantom divide in scalar-tensor and vector-tensor theories}},\ }\href@noop {} {\  (\bibinfo {year} {2025})},\ \Eprint {https://arxiv.org/abs/2508.17231} {arXiv:2508.17231 [astro-ph.CO]} \BibitemShut {NoStop}%
\bibitem [{\citenamefont {Samaddar}\ and\ \citenamefont {Singh}(2025)}]{Samaddar:2025xak}%
  \BibitemOpen
  \bibfield  {author} {\bibinfo {author} {\bibfnamefont {A.}~\bibnamefont {Samaddar}}\ and\ \bibinfo {author} {\bibfnamefont {S.~S.}\ \bibnamefont {Singh}},\ }\bibfield  {title} {\bibinfo {title} {{CPL-parametrized cosmic expansion in Galileon gravity: Constraints from recent data}},\ }\href@noop {} {\  (\bibinfo {year} {2025})},\ \Eprint {https://arxiv.org/abs/2508.14392} {arXiv:2508.14392 [astro-ph.CO]} \BibitemShut {NoStop}%
\bibitem [{\citenamefont {Bellini}\ and\ \citenamefont {Sawicki}(2014)}]{Bellini:2014fua}%
  \BibitemOpen
  \bibfield  {author} {\bibinfo {author} {\bibfnamefont {E.}~\bibnamefont {Bellini}}\ and\ \bibinfo {author} {\bibfnamefont {I.}~\bibnamefont {Sawicki}},\ }\bibfield  {title} {\bibinfo {title} {{Maximal freedom at minimum cost: linear large-scale structure in general modifications of gravity}},\ }\href {https://doi.org/10.1088/1475-7516/2014/07/050} {\bibfield  {journal} {\bibinfo  {journal} {JCAP}\ }\textbf {\bibinfo {volume} {07}},\ \bibinfo {pages} {050}},\ \Eprint {https://arxiv.org/abs/1404.3713} {arXiv:1404.3713 [astro-ph.CO]} \BibitemShut {NoStop}%
\bibitem [{\citenamefont {Zumalac\'arregui}\ \emph {et~al.}(2017)\citenamefont {Zumalac\'arregui}, \citenamefont {Bellini}, \citenamefont {Sawicki}, \citenamefont {Lesgourgues},\ and\ \citenamefont {Ferreira}}]{hi_class1}%
  \BibitemOpen
  \bibfield  {author} {\bibinfo {author} {\bibfnamefont {M.}~\bibnamefont {Zumalac\'arregui}}, \bibinfo {author} {\bibfnamefont {E.}~\bibnamefont {Bellini}}, \bibinfo {author} {\bibfnamefont {I.}~\bibnamefont {Sawicki}}, \bibinfo {author} {\bibfnamefont {J.}~\bibnamefont {Lesgourgues}},\ and\ \bibinfo {author} {\bibfnamefont {P.~G.}\ \bibnamefont {Ferreira}},\ }\bibfield  {title} {\bibinfo {title} {{hi\_class: Horndeski in the Cosmic Linear Anisotropy Solving System}},\ }\href {https://doi.org/10.1088/1475-7516/2017/08/019} {\bibfield  {journal} {\bibinfo  {journal} {JCAP}\ }\textbf {\bibinfo {volume} {08}},\ \bibinfo {pages} {019}},\ \Eprint {https://arxiv.org/abs/1605.06102} {arXiv:1605.06102 [astro-ph.CO]} \BibitemShut {NoStop}%
\bibitem [{\citenamefont {Bellini}\ \emph {et~al.}(2020)\citenamefont {Bellini}, \citenamefont {Sawicki},\ and\ \citenamefont {Zumalac\'arregui}}]{hi_class2}%
  \BibitemOpen
  \bibfield  {author} {\bibinfo {author} {\bibfnamefont {E.}~\bibnamefont {Bellini}}, \bibinfo {author} {\bibfnamefont {I.}~\bibnamefont {Sawicki}},\ and\ \bibinfo {author} {\bibfnamefont {M.}~\bibnamefont {Zumalac\'arregui}},\ }\bibfield  {title} {\bibinfo {title} {{hi\_class: Background Evolution, Initial Conditions and Approximation Schemes}},\ }\href {https://doi.org/10.1088/1475-7516/2020/02/008} {\bibfield  {journal} {\bibinfo  {journal} {JCAP}\ }\textbf {\bibinfo {volume} {02}},\ \bibinfo {pages} {008}},\ \Eprint {https://arxiv.org/abs/1909.01828} {arXiv:1909.01828 [astro-ph.CO]} \BibitemShut {NoStop}%
\bibitem [{\citenamefont {Garc\'\i{}a-Garc\'\i{}a}\ \emph {et~al.}(2020)\citenamefont {Garc\'\i{}a-Garc\'\i{}a}, \citenamefont {Bellini}, \citenamefont {Ferreira}, \citenamefont {Traykova},\ and\ \citenamefont {Zumalac\'arregui}}]{Garcia-Garcia:2019cvr}%
  \BibitemOpen
  \bibfield  {author} {\bibinfo {author} {\bibfnamefont {C.}~\bibnamefont {Garc\'\i{}a-Garc\'\i{}a}}, \bibinfo {author} {\bibfnamefont {E.}~\bibnamefont {Bellini}}, \bibinfo {author} {\bibfnamefont {P.~G.}\ \bibnamefont {Ferreira}}, \bibinfo {author} {\bibfnamefont {D.}~\bibnamefont {Traykova}},\ and\ \bibinfo {author} {\bibfnamefont {M.}~\bibnamefont {Zumalac\'arregui}},\ }\bibfield  {title} {\bibinfo {title} {{Theoretical priors in scalar-tensor cosmologies: Thawing quintessence}},\ }\href {https://doi.org/10.1103/PhysRevD.101.063508} {\bibfield  {journal} {\bibinfo  {journal} {Phys. Rev. D}\ }\textbf {\bibinfo {volume} {101}},\ \bibinfo {pages} {063508} (\bibinfo {year} {2020})},\ \Eprint {https://arxiv.org/abs/1911.02868} {arXiv:1911.02868 [astro-ph.CO]} \BibitemShut {NoStop}%
\bibitem [{\citenamefont {Scolnic}\ \emph {et~al.}(2022)\citenamefont {Scolnic} \emph {et~al.}}]{Scolnic:2021amr}%
  \BibitemOpen
  \bibfield  {author} {\bibinfo {author} {\bibfnamefont {D.}~\bibnamefont {Scolnic}} \emph {et~al.},\ }\bibfield  {title} {\bibinfo {title} {{The Pantheon+ Analysis: The Full Data Set and Light-curve Release}},\ }\href {https://doi.org/10.3847/1538-4357/ac8b7a} {\bibfield  {journal} {\bibinfo  {journal} {Astrophys. J.}\ }\textbf {\bibinfo {volume} {938}},\ \bibinfo {pages} {113} (\bibinfo {year} {2022})},\ \Eprint {https://arxiv.org/abs/2112.03863} {arXiv:2112.03863 [astro-ph.CO]} \BibitemShut {NoStop}%
\bibitem [{\citenamefont {Rubin}\ \emph {et~al.}(2023)\citenamefont {Rubin} \emph {et~al.}}]{Rubin:2023ovl}%
  \BibitemOpen
  \bibfield  {author} {\bibinfo {author} {\bibfnamefont {D.}~\bibnamefont {Rubin}} \emph {et~al.},\ }\bibfield  {title} {\bibinfo {title} {{Union Through UNITY: Cosmology with 2,000 SNe Using a Unified Bayesian Framework}},\ }\href@noop {} {\  (\bibinfo {year} {2023})},\ \Eprint {https://arxiv.org/abs/2311.12098} {arXiv:2311.12098 [astro-ph.CO]} \BibitemShut {NoStop}%
\bibitem [{\citenamefont {Abbott}\ \emph {et~al.}(2024)\citenamefont {Abbott} \emph {et~al.}}]{DES:2024jxu}%
  \BibitemOpen
  \bibfield  {author} {\bibinfo {author} {\bibfnamefont {T.~M.~C.}\ \bibnamefont {Abbott}} \emph {et~al.} (\bibinfo {collaboration} {DES}),\ }\bibfield  {title} {\bibinfo {title} {{The Dark Energy Survey: Cosmology Results with \ensuremath{\sim}1500 New High-redshift Type Ia Supernovae Using the Full 5 yr Data Set}},\ }\href {https://doi.org/10.3847/2041-8213/ad6f9f} {\bibfield  {journal} {\bibinfo  {journal} {Astrophys. J. Lett.}\ }\textbf {\bibinfo {volume} {973}},\ \bibinfo {pages} {L14} (\bibinfo {year} {2024})},\ \Eprint {https://arxiv.org/abs/2401.02929} {arXiv:2401.02929 [astro-ph.CO]} \BibitemShut {NoStop}%
\bibitem [{\citenamefont {Aghanim}\ \emph {et~al.}(2020{\natexlab{a}})\citenamefont {Aghanim} \emph {et~al.}}]{Planck:2019nip}%
  \BibitemOpen
  \bibfield  {author} {\bibinfo {author} {\bibfnamefont {N.}~\bibnamefont {Aghanim}} \emph {et~al.} (\bibinfo {collaboration} {Planck}),\ }\bibfield  {title} {\bibinfo {title} {{Planck 2018 results. V. CMB power spectra and likelihoods}},\ }\href {https://doi.org/10.1051/0004-6361/201936386} {\bibfield  {journal} {\bibinfo  {journal} {Astron. Astrophys.}\ }\textbf {\bibinfo {volume} {641}},\ \bibinfo {pages} {A5} (\bibinfo {year} {2020}{\natexlab{a}})},\ \Eprint {https://arxiv.org/abs/1907.12875} {arXiv:1907.12875 [astro-ph.CO]} \BibitemShut {NoStop}%
\bibitem [{\citenamefont {Aghanim}\ \emph {et~al.}(2020{\natexlab{b}})\citenamefont {Aghanim} \emph {et~al.}}]{Planck:2018vyg}%
  \BibitemOpen
  \bibfield  {author} {\bibinfo {author} {\bibfnamefont {N.}~\bibnamefont {Aghanim}} \emph {et~al.} (\bibinfo {collaboration} {Planck}),\ }\bibfield  {title} {\bibinfo {title} {{Planck 2018 results. VI. Cosmological parameters}},\ }\href {https://doi.org/10.1051/0004-6361/201833910} {\bibfield  {journal} {\bibinfo  {journal} {Astron. Astrophys.}\ }\textbf {\bibinfo {volume} {641}},\ \bibinfo {pages} {A6} (\bibinfo {year} {2020}{\natexlab{b}})},\ \bibinfo {note} {[Erratum: Astron.Astrophys. 652, C4 (2021)]},\ \Eprint {https://arxiv.org/abs/1807.06209} {arXiv:1807.06209 [astro-ph.CO]} \BibitemShut {NoStop}%
\bibitem [{\citenamefont {Qu}\ \emph {et~al.}(2024)\citenamefont {Qu} \emph {et~al.}}]{ACT:2023dou}%
  \BibitemOpen
  \bibfield  {author} {\bibinfo {author} {\bibfnamefont {F.~J.}\ \bibnamefont {Qu}} \emph {et~al.} (\bibinfo {collaboration} {ACT}),\ }\bibfield  {title} {\bibinfo {title} {{The Atacama Cosmology Telescope: A Measurement of the DR6 CMB Lensing Power Spectrum and Its Implications for Structure Growth}},\ }\href {https://doi.org/10.3847/1538-4357/acfe06} {\bibfield  {journal} {\bibinfo  {journal} {Astrophys. J.}\ }\textbf {\bibinfo {volume} {962}},\ \bibinfo {pages} {112} (\bibinfo {year} {2024})},\ \Eprint {https://arxiv.org/abs/2304.05202} {arXiv:2304.05202 [astro-ph.CO]} \BibitemShut {NoStop}%
\bibitem [{\citenamefont {Madhavacheril}\ \emph {et~al.}(2024)\citenamefont {Madhavacheril} \emph {et~al.}}]{ACT:2023kun}%
  \BibitemOpen
  \bibfield  {author} {\bibinfo {author} {\bibfnamefont {M.~S.}\ \bibnamefont {Madhavacheril}} \emph {et~al.} (\bibinfo {collaboration} {ACT}),\ }\bibfield  {title} {\bibinfo {title} {{The Atacama Cosmology Telescope: DR6 Gravitational Lensing Map and Cosmological Parameters}},\ }\href {https://doi.org/10.3847/1538-4357/acff5f} {\bibfield  {journal} {\bibinfo  {journal} {Astrophys. J.}\ }\textbf {\bibinfo {volume} {962}},\ \bibinfo {pages} {113} (\bibinfo {year} {2024})},\ \Eprint {https://arxiv.org/abs/2304.05203} {arXiv:2304.05203 [astro-ph.CO]} \BibitemShut {NoStop}%
\bibitem [{\citenamefont {Aghanim}\ \emph {et~al.}(2020{\natexlab{c}})\citenamefont {Aghanim} \emph {et~al.}}]{Aghanim:2019ame}%
  \BibitemOpen
  \bibfield  {author} {\bibinfo {author} {\bibfnamefont {N.}~\bibnamefont {Aghanim}} \emph {et~al.} (\bibinfo {collaboration} {Planck}),\ }\bibfield  {title} {\bibinfo {title} {{Planck 2018 results. V. CMB power spectra and likelihoods}},\ }\href {https://doi.org/10.1051/0004-6361/201936386} {\bibfield  {journal} {\bibinfo  {journal} {Astron. Astrophys.}\ }\textbf {\bibinfo {volume} {641}},\ \bibinfo {pages} {A5} (\bibinfo {year} {2020}{\natexlab{c}})},\ \Eprint {https://arxiv.org/abs/1907.12875} {arXiv:1907.12875 [astro-ph.CO]} \BibitemShut {NoStop}%
\bibitem [{\citenamefont {Torrado}\ and\ \citenamefont {Lewis}(2021)}]{Cobaya}%
  \BibitemOpen
  \bibfield  {author} {\bibinfo {author} {\bibfnamefont {J.}~\bibnamefont {Torrado}}\ and\ \bibinfo {author} {\bibfnamefont {A.}~\bibnamefont {Lewis}},\ }\bibfield  {title} {\bibinfo {title} {{Cobaya: Code for Bayesian Analysis of hierarchical physical models}},\ }\href {https://doi.org/10.1088/1475-7516/2021/05/057} {\bibfield  {journal} {\bibinfo  {journal} {JCAP}\ }\textbf {\bibinfo {volume} {05}},\ \bibinfo {pages} {057}},\ \Eprint {https://arxiv.org/abs/2005.05290} {arXiv:2005.05290 [astro-ph.IM]} \BibitemShut {NoStop}%
\bibitem [{\citenamefont {Handley}\ \emph {et~al.}(2015)\citenamefont {Handley}, \citenamefont {Hobson},\ and\ \citenamefont {Lasenby}}]{Handley:2015fda}%
  \BibitemOpen
  \bibfield  {author} {\bibinfo {author} {\bibfnamefont {W.~J.}\ \bibnamefont {Handley}}, \bibinfo {author} {\bibfnamefont {M.~P.}\ \bibnamefont {Hobson}},\ and\ \bibinfo {author} {\bibfnamefont {A.~N.}\ \bibnamefont {Lasenby}},\ }\bibfield  {title} {\bibinfo {title} {{PolyChord: nested sampling for cosmology}},\ }\href {https://doi.org/10.1093/mnrasl/slv047} {\bibfield  {journal} {\bibinfo  {journal} {Mon. Not. Roy. Astron. Soc.}\ }\textbf {\bibinfo {volume} {450}},\ \bibinfo {pages} {L61} (\bibinfo {year} {2015})},\ \Eprint {https://arxiv.org/abs/1502.01856} {arXiv:1502.01856 [astro-ph.CO]} \BibitemShut {NoStop}%
\bibitem [{\citenamefont {Lewis}\ and\ \citenamefont {Bridle}(2002)}]{Lewis:2002ah}%
  \BibitemOpen
  \bibfield  {author} {\bibinfo {author} {\bibfnamefont {A.}~\bibnamefont {Lewis}}\ and\ \bibinfo {author} {\bibfnamefont {S.}~\bibnamefont {Bridle}},\ }\bibfield  {title} {\bibinfo {title} {{Cosmological parameters from CMB and other data: A Monte Carlo approach}},\ }\href {https://doi.org/10.1103/PhysRevD.66.103511} {\bibfield  {journal} {\bibinfo  {journal} {Phys. Rev.}\ }\textbf {\bibinfo {volume} {D66}},\ \bibinfo {pages} {103511} (\bibinfo {year} {2002})},\ \Eprint {https://arxiv.org/abs/astro-ph/0205436} {arXiv:astro-ph/0205436 [astro-ph]} \BibitemShut {NoStop}%
\bibitem [{\citenamefont {Jeffreys}(1939)}]{Jeffreys1939}%
  \BibitemOpen
  \bibfield  {author} {\bibinfo {author} {\bibfnamefont {H.}~\bibnamefont {Jeffreys}},\ }\href@noop {} {\emph {\bibinfo {title} {The Theory of Probability}}},\ \bibinfo {edition} {1st}\ ed.,\ Oxford Classic Texts in the Physical Sciences\ (\bibinfo  {publisher} {Oxford University Press},\ \bibinfo {address} {Oxford},\ \bibinfo {year} {1939})\BibitemShut {NoStop}%
\bibitem [{\citenamefont {Trotta}(2007)}]{Trotta:2005ar}%
  \BibitemOpen
  \bibfield  {author} {\bibinfo {author} {\bibfnamefont {R.}~\bibnamefont {Trotta}},\ }\bibfield  {title} {\bibinfo {title} {{Applications of Bayesian model selection to cosmological parameters}},\ }\href {https://doi.org/10.1111/j.1365-2966.2007.11738.x} {\bibfield  {journal} {\bibinfo  {journal} {Mon. Not. Roy. Astron. Soc.}\ }\textbf {\bibinfo {volume} {378}},\ \bibinfo {pages} {72} (\bibinfo {year} {2007})},\ \Eprint {https://arxiv.org/abs/astro-ph/0504022} {arXiv:astro-ph/0504022} \BibitemShut {NoStop}%
\bibitem [{\citenamefont {Trotta}(2008)}]{Trotta:2008qt}%
  \BibitemOpen
  \bibfield  {author} {\bibinfo {author} {\bibfnamefont {R.}~\bibnamefont {Trotta}},\ }\bibfield  {title} {\bibinfo {title} {{Bayes in the sky: Bayesian inference and model selection in cosmology}},\ }\href {https://doi.org/10.1080/00107510802066753} {\bibfield  {journal} {\bibinfo  {journal} {Contemp. Phys.}\ }\textbf {\bibinfo {volume} {49}},\ \bibinfo {pages} {71} (\bibinfo {year} {2008})},\ \Eprint {https://arxiv.org/abs/0803.4089} {arXiv:0803.4089 [astro-ph]} \BibitemShut {NoStop}%
\bibitem [{\citenamefont {Burnham}\ and\ \citenamefont {Anderson}(2002)}]{Burnham2002}%
  \BibitemOpen
  \bibfield  {author} {\bibinfo {author} {\bibfnamefont {K.~P.}\ \bibnamefont {Burnham}}\ and\ \bibinfo {author} {\bibfnamefont {D.~R.}\ \bibnamefont {Anderson}},\ }\bibfield  {title} {\bibinfo {title} {Advanced issues and deeper insights},\ }in\ \href {https://doi.org/10.1007/978-0-387-22456-5_6} {\emph {\bibinfo {booktitle} {Model Selection and Multimodel Inference}}}\ (\bibinfo  {publisher} {Springer},\ \bibinfo {address} {New York, NY},\ \bibinfo {year} {2002})\ \bibinfo {edition} {2nd}\ ed.\BibitemShut {Stop}%
\bibitem [{\citenamefont {Liddle}(2007)}]{Liddle:2007fy}%
  \BibitemOpen
  \bibfield  {author} {\bibinfo {author} {\bibfnamefont {A.~R.}\ \bibnamefont {Liddle}},\ }\bibfield  {title} {\bibinfo {title} {{Information criteria for astrophysical model selection}},\ }\href {https://doi.org/10.1111/j.1745-3933.2007.00306.x} {\bibfield  {journal} {\bibinfo  {journal} {Mon. Not. Roy. Astron. Soc.}\ }\textbf {\bibinfo {volume} {377}},\ \bibinfo {pages} {L74} (\bibinfo {year} {2007})},\ \Eprint {https://arxiv.org/abs/astro-ph/0701113} {arXiv:astro-ph/0701113} \BibitemShut {NoStop}%
\bibitem [{\citenamefont {Ferreira}\ \emph {et~al.}(2025)\citenamefont {Ferreira}, \citenamefont {Wolf},\ and\ \citenamefont {Read}}]{Ferreira:2025fpn}%
  \BibitemOpen
  \bibfield  {author} {\bibinfo {author} {\bibfnamefont {P.~G.}\ \bibnamefont {Ferreira}}, \bibinfo {author} {\bibfnamefont {W.~J.}\ \bibnamefont {Wolf}},\ and\ \bibinfo {author} {\bibfnamefont {J.}~\bibnamefont {Read}},\ }\bibfield  {title} {\bibinfo {title} {{The Spectre of Underdetermination in Modern Cosmology}},\ }\href {https://doi.org/10.31389/pop.218} {\bibfield  {journal} {\bibinfo  {journal} {Phil. Phys.}\ }\textbf {\bibinfo {volume} {3}},\ \bibinfo {pages} {1} (\bibinfo {year} {2025})},\ \Eprint {https://arxiv.org/abs/2501.06095} {arXiv:2501.06095 [physics.hist-ph]} \BibitemShut {NoStop}%
\bibitem [{\citenamefont {Shah}\ \emph {et~al.}(2025{\natexlab{b}})\citenamefont {Shah}, \citenamefont {Koyama},\ and\ \citenamefont {Noller}}]{Shah:2025vnt}%
  \BibitemOpen
  \bibfield  {author} {\bibinfo {author} {\bibfnamefont {N.}~\bibnamefont {Shah}}, \bibinfo {author} {\bibfnamefont {K.}~\bibnamefont {Koyama}},\ and\ \bibinfo {author} {\bibfnamefont {J.}~\bibnamefont {Noller}},\ }\bibfield  {title} {\bibinfo {title} {{Dark energy constraints in light of theoretical priors}},\ }\href@noop {} {\  (\bibinfo {year} {2025}{\natexlab{b}})},\ \Eprint {https://arxiv.org/abs/2507.19450} {arXiv:2507.19450 [astro-ph.CO]} \BibitemShut {NoStop}%
\bibitem [{\citenamefont {Ishak}\ \emph {et~al.}(2024)\citenamefont {Ishak} \emph {et~al.}}]{Ishak:2024jhs}%
  \BibitemOpen
  \bibfield  {author} {\bibinfo {author} {\bibfnamefont {M.}~\bibnamefont {Ishak}} \emph {et~al.},\ }\bibfield  {title} {\bibinfo {title} {{Modified Gravity Constraints from the Full Shape Modeling of Clustering Measurements from DESI 2024}},\ }\href@noop {} {\  (\bibinfo {year} {2024})},\ \Eprint {https://arxiv.org/abs/2411.12026} {arXiv:2411.12026 [astro-ph.CO]} \BibitemShut {NoStop}%
\bibitem [{\citenamefont {Uzan}(2024)}]{Uzan:2024ded}%
  \BibitemOpen
  \bibfield  {author} {\bibinfo {author} {\bibfnamefont {J.-P.}\ \bibnamefont {Uzan}},\ }\bibfield  {title} {\bibinfo {title} {{Fundamental constants: from measurement to the universe, a window on gravitation and cosmology}},\ }\href@noop {} {\  (\bibinfo {year} {2024})},\ \Eprint {https://arxiv.org/abs/2410.07281} {arXiv:2410.07281 [astro-ph.CO]} \BibitemShut {NoStop}%
\bibitem [{\citenamefont {Will}(2014)}]{Will:2014kxa}%
  \BibitemOpen
  \bibfield  {author} {\bibinfo {author} {\bibfnamefont {C.~M.}\ \bibnamefont {Will}},\ }\bibfield  {title} {\bibinfo {title} {{The Confrontation between General Relativity and Experiment}},\ }\href {https://doi.org/10.12942/lrr-2014-4} {\bibfield  {journal} {\bibinfo  {journal} {Living Rev. Rel.}\ }\textbf {\bibinfo {volume} {17}},\ \bibinfo {pages} {4} (\bibinfo {year} {2014})},\ \Eprint {https://arxiv.org/abs/1403.7377} {arXiv:1403.7377 [gr-qc]} \BibitemShut {NoStop}%
\bibitem [{\citenamefont {Ke}\ \emph {et~al.}(2021)\citenamefont {Ke}, \citenamefont {Luo}, \citenamefont {Shao}, \citenamefont {Tan}, \citenamefont {Tan},\ and\ \citenamefont {Yang}}]{Ke:2021jtj}%
  \BibitemOpen
  \bibfield  {author} {\bibinfo {author} {\bibfnamefont {J.}~\bibnamefont {Ke}}, \bibinfo {author} {\bibfnamefont {J.}~\bibnamefont {Luo}}, \bibinfo {author} {\bibfnamefont {C.-G.}\ \bibnamefont {Shao}}, \bibinfo {author} {\bibfnamefont {Y.-J.}\ \bibnamefont {Tan}}, \bibinfo {author} {\bibfnamefont {W.-H.}\ \bibnamefont {Tan}},\ and\ \bibinfo {author} {\bibfnamefont {S.-Q.}\ \bibnamefont {Yang}},\ }\bibfield  {title} {\bibinfo {title} {{Combined Test of the Gravitational Inverse-Square Law at the Centimeter Range}},\ }\href {https://doi.org/10.1103/PhysRevLett.126.211101} {\bibfield  {journal} {\bibinfo  {journal} {Phys. Rev. Lett.}\ }\textbf {\bibinfo {volume} {126}},\ \bibinfo {pages} {211101} (\bibinfo {year} {2021})}\BibitemShut {NoStop}%
\bibitem [{\citenamefont {Bassi}\ \emph {et~al.}(2022)\citenamefont {Bassi}, \citenamefont {Cacciapuoti}, \citenamefont {Capozziello},\ and\ \citenamefont {et~al.}}]{Bassi2022}%
  \BibitemOpen
  \bibfield  {author} {\bibinfo {author} {\bibfnamefont {A.}~\bibnamefont {Bassi}}, \bibinfo {author} {\bibfnamefont {L.}~\bibnamefont {Cacciapuoti}}, \bibinfo {author} {\bibfnamefont {S.}~\bibnamefont {Capozziello}},\ and\ \bibinfo {author} {\bibnamefont {et~al.}},\ }\bibfield  {title} {\bibinfo {title} {A way forward for fundamental physics in space},\ }\href {https://doi.org/10.1038/s41526-022-00229-0} {\bibfield  {journal} {\bibinfo  {journal} {npj Microgravity}\ }\textbf {\bibinfo {volume} {8}},\ \bibinfo {pages} {49} (\bibinfo {year} {2022})}\BibitemShut {NoStop}%
\bibitem [{\citenamefont {Babichev}\ and\ \citenamefont {Deffayet}(2013)}]{Babichev:2013usa}%
  \BibitemOpen
  \bibfield  {author} {\bibinfo {author} {\bibfnamefont {E.}~\bibnamefont {Babichev}}\ and\ \bibinfo {author} {\bibfnamefont {C.}~\bibnamefont {Deffayet}},\ }\bibfield  {title} {\bibinfo {title} {{An introduction to the Vainshtein mechanism}},\ }\href {https://doi.org/10.1088/0264-9381/30/18/184001} {\bibfield  {journal} {\bibinfo  {journal} {Class. Quant. Grav.}\ }\textbf {\bibinfo {volume} {30}},\ \bibinfo {pages} {184001} (\bibinfo {year} {2013})},\ \Eprint {https://arxiv.org/abs/1304.7240} {arXiv:1304.7240 [gr-qc]} \BibitemShut {NoStop}%
\bibitem [{\citenamefont {Burrage}\ \emph {et~al.}(2021)\citenamefont {Burrage}, \citenamefont {Coltman}, \citenamefont {Padilla}, \citenamefont {Saadeh},\ and\ \citenamefont {Wilson}}]{Burrage_2021}%
  \BibitemOpen
  \bibfield  {author} {\bibinfo {author} {\bibfnamefont {C.}~\bibnamefont {Burrage}}, \bibinfo {author} {\bibfnamefont {B.}~\bibnamefont {Coltman}}, \bibinfo {author} {\bibfnamefont {A.}~\bibnamefont {Padilla}}, \bibinfo {author} {\bibfnamefont {D.}~\bibnamefont {Saadeh}},\ and\ \bibinfo {author} {\bibfnamefont {T.}~\bibnamefont {Wilson}},\ }\bibfield  {title} {\bibinfo {title} {Massive galileons and vainshtein screening},\ }\href {https://doi.org/10.1088/1475-7516/2021/02/050} {\bibfield  {journal} {\bibinfo  {journal} {Journal of Cosmology and Astroparticle Physics}\ }\textbf {\bibinfo {volume} {2021}}\bibinfo  {number} { (02)},\ \bibinfo {pages} {050–050}}\BibitemShut {NoStop}%
\bibitem [{\citenamefont {Belgacem}\ \emph {et~al.}(2019)\citenamefont {Belgacem} \emph {et~al.}}]{LISACosmologyWorkingGroup:2019mwx}%
  \BibitemOpen
\bibfield  {number} {  }\bibfield  {author} {\bibinfo {author} {\bibfnamefont {E.}~\bibnamefont {Belgacem}} \emph {et~al.} (\bibinfo {collaboration} {LISA Cosmology Working Group}),\ }\bibfield  {title} {\bibinfo {title} {{Testing modified gravity at cosmological distances with LISA standard sirens}},\ }\href {https://doi.org/10.1088/1475-7516/2019/07/024} {\bibfield  {journal} {\bibinfo  {journal} {JCAP}\ }\textbf {\bibinfo {volume} {07}},\ \bibinfo {pages} {024}},\ \Eprint {https://arxiv.org/abs/1906.01593} {arXiv:1906.01593 [astro-ph.CO]} \BibitemShut {NoStop}%
\bibitem [{\citenamefont {Lagos}\ \emph {et~al.}(2019)\citenamefont {Lagos}, \citenamefont {Fishbach}, \citenamefont {Landry},\ and\ \citenamefont {Holz}}]{Lagos:2019kds}%
  \BibitemOpen
  \bibfield  {author} {\bibinfo {author} {\bibfnamefont {M.}~\bibnamefont {Lagos}}, \bibinfo {author} {\bibfnamefont {M.}~\bibnamefont {Fishbach}}, \bibinfo {author} {\bibfnamefont {P.}~\bibnamefont {Landry}},\ and\ \bibinfo {author} {\bibfnamefont {D.~E.}\ \bibnamefont {Holz}},\ }\bibfield  {title} {\bibinfo {title} {{Standard sirens with a running Planck mass}},\ }\href {https://doi.org/10.1103/PhysRevD.99.083504} {\bibfield  {journal} {\bibinfo  {journal} {Phys. Rev. D}\ }\textbf {\bibinfo {volume} {99}},\ \bibinfo {pages} {083504} (\bibinfo {year} {2019})},\ \Eprint {https://arxiv.org/abs/1901.03321} {arXiv:1901.03321 [astro-ph.CO]} \BibitemShut {NoStop}%
\bibitem [{\citenamefont {Wolf}\ and\ \citenamefont {Lagos}(2020)}]{Wolf:2019hun}%
  \BibitemOpen
  \bibfield  {author} {\bibinfo {author} {\bibfnamefont {W.~J.}\ \bibnamefont {Wolf}}\ and\ \bibinfo {author} {\bibfnamefont {M.}~\bibnamefont {Lagos}},\ }\bibfield  {title} {\bibinfo {title} {{Standard Sirens as a Novel Probe of Dark Energy}},\ }\href {https://doi.org/10.1103/PhysRevLett.124.061101} {\bibfield  {journal} {\bibinfo  {journal} {Phys. Rev. Lett.}\ }\textbf {\bibinfo {volume} {124}},\ \bibinfo {pages} {061101} (\bibinfo {year} {2020})},\ \Eprint {https://arxiv.org/abs/1910.10580} {arXiv:1910.10580 [gr-qc]} \BibitemShut {NoStop}%
\bibitem [{\citenamefont {Kable}\ \emph {et~al.}(2022)\citenamefont {Kable}, \citenamefont {Benevento}, \citenamefont {Frusciante}, \citenamefont {De~Felice},\ and\ \citenamefont {Tsujikawa}}]{Kable:2021yws}%
  \BibitemOpen
  \bibfield  {author} {\bibinfo {author} {\bibfnamefont {J.~A.}\ \bibnamefont {Kable}}, \bibinfo {author} {\bibfnamefont {G.}~\bibnamefont {Benevento}}, \bibinfo {author} {\bibfnamefont {N.}~\bibnamefont {Frusciante}}, \bibinfo {author} {\bibfnamefont {A.}~\bibnamefont {De~Felice}},\ and\ \bibinfo {author} {\bibfnamefont {S.}~\bibnamefont {Tsujikawa}},\ }\bibfield  {title} {\bibinfo {title} {{Probing modified gravity with integrated Sachs-Wolfe CMB and galaxy cross-correlations}},\ }\href {https://doi.org/10.1088/1475-7516/2022/09/002} {\bibfield  {journal} {\bibinfo  {journal} {JCAP}\ }\textbf {\bibinfo {volume} {09}},\ \bibinfo {pages} {002}},\ \Eprint {https://arxiv.org/abs/2111.10432} {arXiv:2111.10432 [astro-ph.CO]} \BibitemShut {NoStop}%
\bibitem [{\citenamefont {Yao}\ \emph {et~al.}(2025)\citenamefont {Yao}, \citenamefont {Ye},\ and\ \citenamefont {Silvestri}}]{Yao:2025wlx}%
  \BibitemOpen
  \bibfield  {author} {\bibinfo {author} {\bibfnamefont {Z.}~\bibnamefont {Yao}}, \bibinfo {author} {\bibfnamefont {G.}~\bibnamefont {Ye}},\ and\ \bibinfo {author} {\bibfnamefont {A.}~\bibnamefont {Silvestri}},\ }\bibfield  {title} {\bibinfo {title} {{A General Model for Dark Energy Crossing the Phantom Divide}},\ }\href@noop {} {\  (\bibinfo {year} {2025})},\ \Eprint {https://arxiv.org/abs/2508.01378} {arXiv:2508.01378 [gr-qc]} \BibitemShut {NoStop}%
\bibitem [{\citenamefont {Horndeski}(1974)}]{Horndeski:1974wa}%
  \BibitemOpen
  \bibfield  {author} {\bibinfo {author} {\bibfnamefont {G.~W.}\ \bibnamefont {Horndeski}},\ }\bibfield  {title} {\bibinfo {title} {{Second-order scalar-tensor field equations in a four-dimensional space}},\ }\href {https://doi.org/10.1007/BF01807638} {\bibfield  {journal} {\bibinfo  {journal} {Int. J. Theor. Phys.}\ }\textbf {\bibinfo {volume} {10}},\ \bibinfo {pages} {363} (\bibinfo {year} {1974})}\BibitemShut {NoStop}%
\bibitem [{\citenamefont {Charmousis}\ \emph {et~al.}(2012)\citenamefont {Charmousis}, \citenamefont {Copeland}, \citenamefont {Padilla},\ and\ \citenamefont {Saffin}}]{Charmousis:2011bf}%
  \BibitemOpen
  \bibfield  {author} {\bibinfo {author} {\bibfnamefont {C.}~\bibnamefont {Charmousis}}, \bibinfo {author} {\bibfnamefont {E.~J.}\ \bibnamefont {Copeland}}, \bibinfo {author} {\bibfnamefont {A.}~\bibnamefont {Padilla}},\ and\ \bibinfo {author} {\bibfnamefont {P.~M.}\ \bibnamefont {Saffin}},\ }\bibfield  {title} {\bibinfo {title} {{General second order scalar-tensor theory, self tuning, and the Fab Four}},\ }\href {https://doi.org/10.1103/PhysRevLett.108.051101} {\bibfield  {journal} {\bibinfo  {journal} {Phys. Rev. Lett.}\ }\textbf {\bibinfo {volume} {108}},\ \bibinfo {pages} {051101} (\bibinfo {year} {2012})},\ \Eprint {https://arxiv.org/abs/1106.2000} {arXiv:1106.2000 [hep-th]} \BibitemShut {NoStop}%
\end{thebibliography}%

\end{document}